\documentclass[preprint]{aastex}
\usepackage{graphicx}
\usepackage{epsfig}
\usepackage{lscape}
\usepackage{graphicx}
\usepackage{amsmath}%
\usepackage[english]{babel}
\usepackage[titletoc,title]{appendix}
\usepackage{txfonts}
\usepackage{times}
\usepackage{natbib}
\begin{document}
\title{Outflows in the Discs of Active Galaxies} 
\author{N. Menci$^1$, F. Fiore$^2$, C. Feruglio$^2$,  A. Lamastra$^{1,3}$, F. Shankar$^{4}$, E. Piconcelli$^1$, E. Giallongo$^1$, A. Grazian$^1$}
\affil{$^1$INAF - Osservatorio Astronomico di Roma, via Frascati 33, I-00078 Monteporzio, Italy}
\affil{$^2$INAF – Osservatorio Astronomico di Trieste, via Tiepolo 11, I-34143 Trieste, Italy}
\affil{$^3$Space Science Data Center - ASI, via del Politecnico SNC, I-00133 Roma, Italy}
\affil{$^4$Department of Physics and Astronomy, University of Southampton, Highfield SO17 1BJ, UK}

\begin{abstract}
Recent advances in observations have provided a wealth of measurements of the expansions of outflows in galactic discs out to large radii in a variety of galactic hosts. To provide an updated baseline for the interpretation of such data, and to assess to what extent the present status of the modeling is consistent with the existing observations, 
we provide a compact two-dimensional description for the expansion of AGN-driven shocks in realistic galactic discs with 
exponential gas density profiles in a disc geometry. We derive solutions for the outflow expansion and the mass outflow rates in different directions with respect to the plane of the disc. These are expressed  in terms of the global  properties of the host galaxy and of the central AGN to allow for an easy and direct comparison with existing observations in a variety of galactic hosts with measured properties, and out to distances $\sim 10$ kpc from the centre.  
The results are compared with a state-of-the-art compilation of observed outflows in 19 galaxies with different measured gas and dynamical mass, allowing for a detailed, one-by-one comparison with the model predictions. The  agreement we obtain for a wide range of host galaxy gas mass ($10^{9}\,M_{\odot}\lesssim M_{gas}\lesssim 10^{12}\,M_{\odot}$) and AGN bolometric luminosity
 ($10^{43}\,{\rm erg\,s^{-1}}\lesssim L_{AGN}\lesssim 10^{47}{\rm erg\,s^{-1}}$) provides a  quantitative systematic test  for the modeling of  AGN-driven outflows in galactic discs. We also consider a larger sample of 48 objects in galaxies with no reliable measurements of the gas and dynamical mass. In this case  we perform a comparison of the model predictions for different bins of AGN luminosities assuming different reference values for the  gas mass and dynamical mass derived from average scaling relations. 
Finally, we reconsider the AGN wind scaling laws empirically derived by many authors  in light of the results from our updated models.
The encouraging, quantitative agreement of the model predictions with a wide set of existing observations constitutes a baseline for the interpretation of forthcoming data, and for a more detailed treatment of AGN feedback in galaxy formation models. 
\end{abstract}

\section{Introduction} 

In the last decade a wealth of observations has provided an increasingly detailed characterization of galaxy-scale outflows driven by active galactic nuclei (AGN). The early observations of ultra-fast AGN-driven winds on small scales (from the accretion disc scale up to the dusty torus) with velocities 
$\approx 0.1\cdot c$ (see King \& Pounds 2015 for a review)  through blue-shifted absorption lines in the X-ray spectra in a substantial fraction ($\approx 40\%$) of AGNs (e.g. Piconcelli et al. 2005; Tombesi et al. 2010; Gofford et al 2013) 
have been recently complemented with a wide set of measurements of fast (velocities of the order of 1000 km $s^{-1}$), massive flows of ionized, neutral and molecular gas, extended on kpc scales. These have been performed through deep optical/NIR spectroscopy (Nesvadba et al. 2006, 2008; Alexander et al. 2010; Rupke \& Veilleux 2011; Riffel \& Storchi-Bergmann 2011; Cano-Diaz et al. 2012; Greene et al. 2012; Harrison et al. 2012, 2014; Liu et al. 2013a,b; Cimatti et al. 2013; Tadhunter et al. 2014; Genzel et al. 2014; Brusa et al. 2015a,b; Cresci et al. 2015; Carniani et al. 2015; Perna et al. 2015a,b; Zakamska et al. 2016ò Bischetti et al. 2017), through interferometric observations in the (sub)millimetre domain (e.g. Feruglio et al. 2010, 2013, 2015; Alatalo et al. 2011; Aalto et al. 2012; Cicone et al. 2012, 2014, 2015; Maiolino et al. 2012, Krips et al. 2011; Morganti et al. 2013a,b; Combes et al. 2013; Garcia-Burillo et al. 2014), and through far-infrared spectroscopy from Herschel (e.g. Fischer et al. 2010; Sturm et al. 2011; Veilleux et al. 2013; Spoon et al. 2013; Stone et al. 2016; Gonzalez-Alfonso et al. 2017). These observations have enabled to determine the detailed physical properties of the outflows (velocities, mass outflow rate, kinetic energy rate) for a number of sources with different AGN luminosity and host galaxy properties (gas mass, circular velocity). 
Recent works by Cicone et al. (2014) and Fiore et al. (2017) have allowed to assemble samples with more than a hundred outflow measurements with detected massive winds at different scales (sub-pc to kpc) and with different molecular/ion compositions. For several molecular outflows the complementary measurement of the host galaxy circular velocity and  gas mass has been used to constrain the relationships between wind parameters, AGN parameters and host galaxy parameters. 

Parallel  theoretical work (Silk \& Rees 1998; Cavaliere, Lapi, Menci 2002; King 2003; Lapi, Cavaliere, Menci 2005; Granato et al. 2004; Silk \& Nusser 2010; King, Zubovas \& Power 2011; Zubovas \& King 2012;  Faucher-Giguere \& Quataert 2012; King \& Pounds 2015) 
 has focused on capturing the main features of the outflows and on pinning down their main expansion and cooling properties, mainly through the implementation of models based on shocks expanding into the inter-stellar medium (ISM) approximated as a sphere with a power-law density profile $\rho\sim R^{-\alpha}$ (for the extension to exponential discs see Hartwick, Volonteri \& Dashyan 2018). Within the large uncertainties and approximations,  energy conserving shock models are consistent with present measurements that indicate that AGN-driven, galaxy-scale outflows may commonly have momentum fluxes $10\,L_{AGN}/c$ (in terms of the AGN bolometric luminosity $L_{AGN}$). These models allowed to derive scaling laws for the run of the shock velocity $V_s$, for the associated mass outflow rate $\dot M_{S,\theta}$, and for their dependence on the AGN luminosity; e.g., for the case of an isothermal sphere ($\alpha=-2$) models including cooling predict mass outflow rates $\dot M_s\sim L_{AGN}^{1/3}$, while the energy-conserving model by Lapi et al. (2005) yields a slightly steeper dependence $\dot M_s\sim L_{AGN}^{1/2}$. The observed steeper dependencies $\dot M_s\sim L_{AGN}^{0.8}$ for molecular winds, and $\dot M_s\sim L_{AGN}^{1.3}$ for ionized winds, see Fiore et al. (2017) then point toward a medium where the density profile is flatter than the isothermal case (Faucher-Giguere,  \& Quataert, 2012), although such conclusions may be affected by biases in the observational results. 
 
Although refined, recent shock models have started to compare with the distribution of observational outflow measurements (see, e.g., Richings \& Faucher-Giguere 2018b), the increasing wealth of data concerning the physical properties of a large  number of AGN-driven outflows calls for a more detailed and quantitative comparison with models, starting with a "one-by-one"comparison of models predictions with the measured outflow properties in well studied objects, residing in galaxies with different measured gas and dynamical mass. 
 
Toward this aim, we extend the shock model for AGN outflow to include realistic exponential density profiles for the ISM, where the normalization of the gas density  is related to the global gas content of the host galaxy disc, and the disc scale radius is related to the total host galaxy mass. This allows us to compare the shock model results with the most recent compilations of data concerning AGN outflows with different AGN luminosity and host galaxy gas and dark matter (DM) mass, measured at different distances from the host galaxy centre. 
Our goal is to incorporate most previous advances into a single yet manageable
analytic framework so as to describe the expansion of AGN-driven shocks for realistic exponential density profiles for the interstellar medium in a disc geometry, and to derive  solutions in terms  of the global  properties of the host galaxy and of the central AGN. This allows us to perform a direct comparison with existing observations in a variety of galactic hosts with measured properties, and out to distance $\sim 10$ kpc from the centre. The goal is to provide an observationally-based test ground for the current description of AGN-driven shocks in realistic galactic hosts, and to assess to what extent the present status of the modeling is consistent with the existing observational distribution of the expansion and mass outflow rates. 

While the main observables we compare with, i.e. expansion and mass outflow rates, can be reliably computed using
the analytical formalism we adopt (as found for in earlier works comparing simulations with analytical computations in well studied cases, e.g., Richings
\& Faucher-Giguere 2018), our analytical approach is complementary to numerical simulations. E.g., 
while a precise description of the position-dependent molecular, ionization and chemical 
 properties of the shocked shell requires numerical simulations to account for the effects of Rayleigh-Taylor instabilities (Richings \& Faucher-Giguere 2018a,b, see also Zubovas \& King 2014),  our treatment effectively follows the expansion velocity of the shock and the mass outflow rate out to large radii where the assumption of power-law gas density profiles (adopted in such simulations) fails, and where the disc is the dominant component with respect to the rapidly-declining bulge component (the shock expansion in this case is treated, e.g., in King, Zubovas, Power 2011). 
Also, our analytical model allows us to easily explore the dependence of the AGN-driven outflows on a variety of quantities (including the AGN luminosity, the gas mass fraction and the total mass of the host galaxy) over a huge range of values. 
 In addition, our computation allows to describe the two-dimensional structure of the outflow, as opposite to the isotropic situations considered in the most simulations (for simulations of outflows for a non-spherical, elliptical distribution of gas see Zubovas and Nayakshin 2014). In this sense, our approach is similar to that adopted in 
 Hartwick, Volonteri \& Dashyan (2018) but focused on the exploration of a wide set of properties of galactic hosts (including the gas mass fraction) and on the systematic comparison with the most recent compilation of observational data encompassing a wide range of properties of the host galaxy and of the central AGN.
 
The plan of the paper is as follows. We provide the basic equations governing the expansion of AGN-driven shocks in Sect. 2. In Sect. 3 we first derive solutions for an isotropic distribution of gas with power-law density profile (Sect. 3.1), to compare with previous studies. Then we derive solutions in the 
case of exponential gas density profiles along the plane of the disc (3.2.1) and in the other directions (3.2.2). Sect. 4. is devoted to a detailed comparison with existing samples of observed AGN-driven outflows. In Sect. 4.1 we compare with outflows in galaxies with measured gas and total mass, allowing for a one-by-one quantitative comparison with the model predictions, while in Sect. 4.2 we consider a large sample of observed outflows in galaxies where measurements of gas and total mass are not available, so that the comparison has to be performed assuming observational scaling laws for the 
observed host galaxy properties. In Sect. 5 we reconsider the AGN wind scaling laws empirically derived by many authors in light of the results from our updated models. Sect. 6 is devoted to discussion and conclusion.

\section{The Model}

We adopt the standard shell approximation (see Cavaliere \& Messina 1976; Ostriker \& McKee 1988; Cavaliere, Lapi, Menci 2002; King et al. 2003; Lapi et al. 2005, Faucher-Giguere \& Quataert 2012; King 2010; Ishibashi \& Fabian 2014, 2015; King \& Pounds 2015; Hartwick, Volonteri \& Dashyan 2018) for the expansion of shocks into the ambient ISM of the host galaxy. We assume that nuclear winds with velocities $V_{in}\approx 3\,10^{4}$ km/s generated by the central AGN accelerate a forward shock expanding into the ambient medium.  In the general two-dimensional case (see fig. 1) the shock radius $R_{S,\theta}$ and 
 velocity $V_{S,\theta}=\dot R_{S,\theta}$ depend not only on time, but also on the angle $\theta$ between the direction of expansion and the plane of the disk. The shock expansion results into the formation of a shell of swept-up material defined by $R\approx R_{S,\theta}$, with a mass outflow rate 
 $\dot M_{S,\theta}$ in the considered direction, enclosing a bubble of hot shocked medium. We compute the expansion of the bubble and the properties of the shocked shell, in turn. 

\vspace{-0.4cm}
\begin{center}
\scalebox{0.55}[0.55]{\rotatebox{0}{\includegraphics{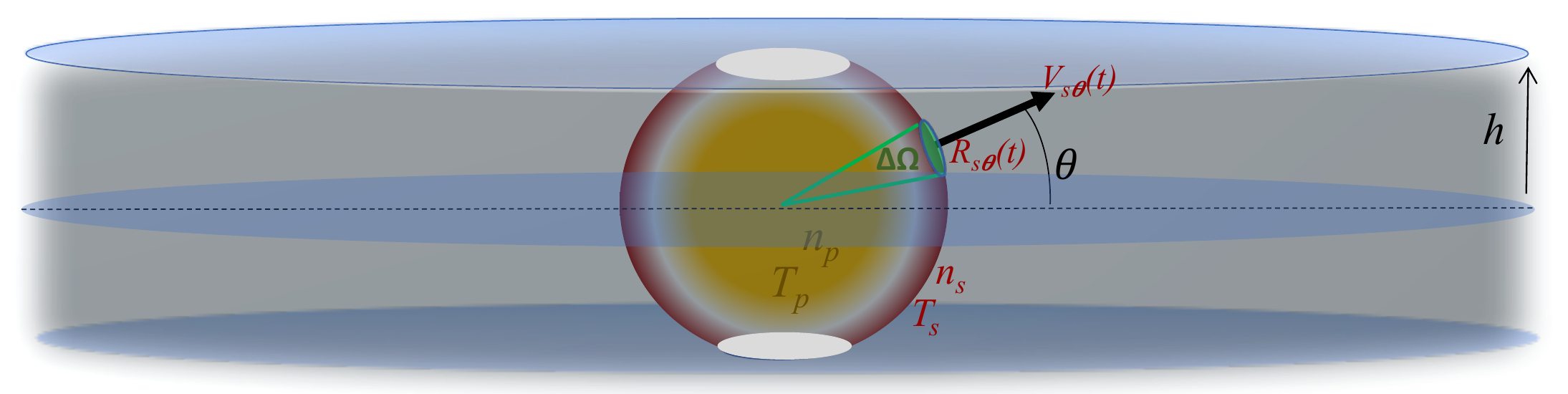}}}
\end{center}
\vspace{-0.4cm }
{\footnotesize Fig. 1. The disc geometry considered in the text. Within the vertical boundaries corresponding to a disc scale height $h$, the galactic gas density outside the shock depends only on the radial coordinate $R$. The external boundary of the red region corresponds to the shock position $R_{S,\theta}$. We also show the density $n_p$ and temperature $T_p$ of the hot bubble  inside the shock (yellow region) and the density $n_{S}$ and temperature $T_{S}$ of the shocked shell (red region).
}

\subsection{The Expansion of the Shock}

In the shell approximation, we consider the motion of a mass element $\Delta M_{S,\theta}$ of the swept-out gas at a shock radius $R_{s,\theta}$ 
 propagating with velocity $V_{S,\theta}=\dot R_{S,\theta}$ in the direction defined by the angle $\theta$ with respect to the plane of the disc. Within a solid angle  $\Delta \Omega=2\,\pi\,cos(\theta)\Delta \theta$,  the mass element $\Delta M_{S,\theta}=\Delta \Omega \,\int_0^{R_{S,\theta}}\,dR^{2}\,R^{2}\,\rho(R,\theta)$ is given by the initial galaxy gas mass (distributed  according to a density profile $\rho(R,\theta)$)  enclosed within the shock radius $R_{S,\theta}$ in the considered direction. We define the mass swept out per 
 unit solid angle $M'_{S,\theta}\equiv d M_{S,\theta}/d \Omega $, so that the total mass of the swept out gas 
  is $M_S=\int\,d\Omega M'_{S,\theta}$. 
   Then the expansion of the shock in the solid angle $\Delta \Omega$ along the considered direction $\theta$ is given by
\begin{equation}
{d\over dt}\Bigg(M'_{s,\theta}\Delta \Omega\,V_{s,\theta}\Bigg)=\Delta \Omega\,R_{s,\theta}^{2}\big(P_b-P_0)-{G\, M'_{s,\theta}\,\Delta \Omega \,M(<R_{S,\theta})\over R_{s,\theta}^{2}}
 \end{equation}
 The above equation accounts for the balance between the pressure term (fueling the expansion) acting on the surface element corresponding to the solid angle in the considered direction $\theta$, and the counter-acting gravitational term, determined 
 by the total mass $M$ (contributed by the DM and by the central Black Hole) within the shock radius $R_{s,\theta}$.  
 Here $P_0$ is the initial ambient pressure, while the pressure of the hot gas in the bubble is $P_b=E_b/(3/2)Q(t)$, where $Q(t)$ is the 
 volume enclosed by the bubble at the considered time $t$.  In turn, the thermal energy $E_b$ of the bubble  evolves according to 
\begin{equation}
{d\over dt}E_b=\epsilon\,L_{AGN}-\,P_b\int d\Omega {dS\over d\Omega}\,V_{S,\theta}- L_{cool}
 \end{equation}
Here $S(t)$ is the surface enclosing the hot bubble at the considered time $t$, $\epsilon \approx v_{in}/c$ is the efficiency for the AGN radiation to transfer energy to the ISM medium, while $L_{cool}$ is the cooling rate of the bubble $L_{cool}=Q(t)\,(\Lambda_{IC}+\Lambda_{ff})$, in turn related to the cooling functions for inverse Compton ($\Lambda_{IC}$) and free-free emission ($\Lambda_{ff})$. Notice that we have assumed  the inner boundary of the shocked wind bubble to be much smaller than its outer boundary $R_{s,\theta}$, an approximation which is known to impact the results by less than 5\% (see Richings \& 
Faucher-Giguere 2018b). In the following, we shall also neglect the initial pressure $P_0$ since it is found to be much smaller that the pressure $P_b$
 in view of the initial small temperature $\sim 10^{4}$ K of the unperturbed medium when compared to the temperatures $T_b\sim 10^{9}-10^{11}$ K of the bubble (see also Richings \& Faucher-Giguere 2018b). 
  
 We then define the geometrical factor $C(R_{S,\theta})\equiv Q(t)/(4\pi/3\,R_{S,\theta}^{3})$  which expresses the 
 deviation of the volume from the spherical case. 
 With the above notation, and within the approximations for the bubble volume and for the initial pressure of the ambient ISM discussed above, we can recast eq. 1 as
\begin{equation}
{d\,V_{S,\theta}\over dt}={2\over C(R_{S,\theta})}\,{E_b\over 4\,\pi\,M'_{S,\theta}\,R_{S,\theta}}-{M(<R_{S,\theta})\over M}\,{R_v\over R_{S,\theta}}\,{V_c^{2}\over R_{S,\theta}}-V_{S,\theta}\,{\dot M'_{S,\theta}\over M'_{S,\theta}}
\end{equation}

The last term on the right hand side can be readily computed for specific assumed density profile. We start from the  general form $\rho\equiv \rho_0\,g(R/R_v,\theta)$, where $R_v$ is the virial radius of the host galaxy. 
Defining the rescaled radius $x\equiv R/R_v$, the normalization $\rho_0$ is related to the total gas content of the galaxy by the relation 
 $\rho_0=M_{gas}/R_v^{3}\,I$ where the form factor $I$ is obtained integrating the density profile $g(x,\theta)$ over the volume occupied by the galactic  gas. Thus, $I=4\,\pi\int_0^{1}\,g(x)\,x^{2}dx$ in the case of a spherical distribution, while $I=4\,\pi\int_{0}^{arctg(h/x\,R_v)}\int_0^{1}\,g(x)\,x^{2}\,cos\theta d\theta\,dx$ in the case of an isotropic distribution inside a disc, with a sharp cutoff at a distance $h$ in the direction perpendicular to the disc. With such a notation, we get
\begin{equation}
{d\,V_{S,\theta}\over dt}={2\over C(R_{S,\theta})}\,{E_b\over 4\,\pi\,M'_{S,\theta}\,R_{S,\theta}}-{M(<R_{S,\theta})\over M}\,{R_v\over R_{S,\theta}}\,{V_c^{2}\over R_{S,\theta}}-
{ g(R_{S,\theta}/R_V,\theta)\over \int_{0}^{R_{S,\theta}/R_v}g(x,\theta)\,x^{2}\,dx}\,
\Bigg({R_{S,\theta}\over R_v}\Bigg)^{3}\,{V_{S,\theta}^{2}\over R_{S,\theta}}
\end{equation}

where we have defined the total mass (mainly contributed by DM) within the virial radius $M\equiv M(<R_v)$, and 
we have expressed the ratio $G\,M(<R_{S,\theta})/R_v=V_c^{2}\,[M(<R_{S,\theta})/M]$ in terms of the host galaxy circular galaxy velocity $V_c=G\,M/R_v$.  
In the following we shall express distances in units of $R_0=1$ kpc, velocities in units of $V_0=1000$ km/s, masses in units of $M_0=10^{12}\,M_{\odot}$. Correspondingly, energies are expressed in units of $E_0=M_0\,V_0^{2}$ and time in units $t_0=R_0/V_0$. 
 After defining $r\equiv R/R_0$, $r_v\equiv R_v/R_0$, $v=V/V_0$, $m\equiv M/M_0$, $e\equiv E/E_0$, and $\tilde{t}\equiv t/t_0$ the set of equations defining the expansion of the shock into the  ambient ISM are:
\begin{eqnarray}
{d\,v_{S,\theta}\over d\tilde{t}} & = & {2\over C(r_{S,\theta})}\,{e_b\over 4\,\pi\,m'_{S,\theta}\,r_{S,\theta}}-{m(<r_{S,\theta})\over m}\,{r_v\over r_{S,\theta}}\,{v_c^{2}\over r_{S,\theta}}-
{ g(r_{S,\theta}/r_V,\theta)\over \int_{0}^{r_{S,\theta}/r_v}g(x,\theta)\,x^{2}\,dx}\,
\,\Bigg({r_{S,\theta}\over r_v}\Bigg)^{3}\,{v_{S,\theta}^{2}\over r_{S,\theta}}\\
{d\,e_b\over d\tilde{t}} & = & 1.5\,10^{-4}\epsilon\,l_{AGN}-{2\,e_b\over 3}\,{1\over Q(t)}\,\int\,d\Omega {dS(t)\over d\Omega}\,v_{S,\theta}- l_{cool}(e_b,l_{AGN})\\
r_{S,\theta} & =& \int_{0}^{\tilde{t}} d\tilde{t}'\,v_{S,\theta}(\tilde{t}')\\
m'_{S,\theta} & = & m_{gas}\int_0^{r_{S,\theta}/r_v}\,g(x)\,x^{2}\,dx/I
\end{eqnarray}
 where all luminosities $l_{AGN}\equiv L/L_0$ and $l_{cool}\equiv l_{cool}/L_0$ are  expressed in units of $L_0=10^{45}$ erg/s, and the volume integral in eq. (7) is performed over the regions where the gas is initially distributed, and extends up to a rescaled radius $r_{S,\theta}/r_v$.
 For the computation of the fraction of total mass within the shock radius $m(<r_{S,\theta})/ m$, we include in $m(<r_{S,\theta})$ the contributions from both the central black hole mass $m_{BH}$ and the dark matter mass $m_{DM}$. For the latter we assume a Navarro, Frenk \& White (1996) form 
 $m_{DM}(r)/ m_{DM}(r_v)=[ln(1+cx)-cx/(1+cx)]/[ln(1+c)-c/(1+c)]$, where $c$ is the concentration for which we adopt the 
 expression given in Maccio', Dutton, van den Bosch (2008), and $x=r/r_v$ . The computation of the cooling term requires 
 the bubble temperature $T_b$ and densities $n_b$. These are computed after eq. 3.7 and 3.8 in Richings \& Faucher-Giguere (2018b) assuming 
a fully ionized plasma with mean molecular weight is $\mu$ = 14/23, to get  $T_b=28/69\,(m_p\,E_b\,V_{in}^{2}/k_b\,\epsilon \,L_{AGN}\,t)$ and $n_b=L_{AGN}\,t\,X_H/Q(t)\,V_{in}^{2}\,m_p$ (here $k_b$ is the Boltzmann constant, $m_p$ is the proton mass and $X_H=0.7$ is the hydrogen mass fraction); the associated electron density is taken to be $n_e=1.2\,n_b$. From these quantities, the Inverse Compton and free-free cooling functions entering the computation of $l_{cool}$ are computed after eqs. 3.4, 3.5, B7 and B8 in Richings \& Faucher-Giguere (2018b). 
 
\subsection{The proprties of the shocked ISM shell} 
To compute the properties of the shocked ISM shell we adopt the approach in Richings \& Faucher-Giguere (2018b). We first compute the 
evolution of the  energy in the layer as
\begin{equation}
{d\over dt} E_{s}=P_b\,\int\,d\Omega {dS\over d\Omega}\,V_{S,\theta}- \int d\Omega {G\,M(<R_{S,\theta})\,\,M'_{S,\theta}\over R_{S,\theta}^{2}}\,V_{S,\theta}-L_{cool, shell}(E_s, n_s)˜. 
\end{equation}

This accounts for the balance between the work done on the shocked ISM layer by the shocked wind bubble pressure, and the effects of the gravitational potential and of the cooling. 
The thermal energy of the shell is $E_{s,th}=E_{s}-(1/2)\,\int\,d\Omega M'_{S,\theta}\,V_{S,\theta}^{2}$ and the associated temperature is 
$T_s=(28/ 69)\,E_{s,th}\,m_p/M_{S}\,k_B$; it evolves according to eq. 8, where the bubble energy $E_b$ satisfies eq. 5. The associated hydrogen number density in the shocked shell is $n_s=(3/2)\,P_b\,M_{S}\,X_H/m_p\,E_{s,th}$ (eq. 3.17 in Richings \& Faucher-Giguere 2018b) which can be recast (in our usual units defined in sect. 2.1) as $n_s\approx 0.5\,(e_b/e_{s,th})\, [m_{S}/Q(t)]\,10^{4}$ cm$^{-3}$,  and the associated electron number density is 1.2 $n_s$. The above values of temperature $T_s$ and density $n_s$ are used to compute the cooling rate $L_{cool, shell}(E_s, n_s)$  from 
from free-free and line emission after  eqs. 3.11 and 3.12 in Richings \& Faucher-Giguere (2018b).

In sum, the following set of equations describes the energy evolution the shocked ISM shell, in terms of our rescaled variables:
\begin{eqnarray}
{d\,e_s\over d\tilde{t}} & = & e_b\,{2\over 3\,Q(t)}\,\int d\Omega\,{dS\over d\Omega}\,v_{s,\theta}-\int d \Omega {m(<r_{S,\theta})\over m}\,{r_v\over r_{S,\theta}}\,{v_c^{2}\over r_{S,\theta}}\,v_{S,\theta}\,m'_{S,\theta}-l_{cool,shell}(e_s,n_s)\\
e_{s,th}& = & e_s-(1/2)\,\int d\Omega\,m'_{S,\theta}\,v_{S,\theta}^{2}\\
n_s & = &  0.5\,(e_b/e_{s,th})\,[m_{S}/Q(t)]\,10^{4}\,{\rm cm}^{-3}\\
T_s & = & (e_{s,th}/m_{S})\,0.5\,10^{8}\,{\rm K}
\end{eqnarray}
where the volume $Q(t)$ is expressed in units of $R_{0}^{3}=$ 1 kpc$^{3}$. We note that this set of equations is coupled with those describing the expansion of the shock (eqs. 4-7) through the bubble energy $e_b$, the shock position $r_{S,\theta}$ and the shock velocity $v_{S,\theta}$. 

\section{Properties of Solutions}
The solutions of eqs.  4-7 and 9-12 depend on the assumed initial density distribution $\rho=\rho_0\,g(x,\theta)$ where $x=R/R_v$ and 
the normalization $\rho_0$ is related to the total gas content $M_{gas}$ of the host galaxy (see sect 2.).
Although we shall focus on exponential density profiles in a non-isotropic disc geometry, we first 
derive solutions for a spherical initial density distribution with a scale-free,  power-law dependence on the radius $R$, since in this case analytical solutions exist in the limit of energy-conserving shock.This allows us to test the reliability of our numerical solutions against analytical results. 

\subsection{Testing the Numerical Solutions: the Case of Power-Law Density Profiles} 

To test our solutions, we first consider a spherically symmetric initial gas density distribution with a power-law  profiles  $g(x)=x^{-\alpha}$, and the proper form factor  $I$ entering eq. 7 is simply $I=4\pi\,\int_0^{1}\,x^{2-\alpha}dx$. 
In figs. 1a-1c we show our results for different values of the power-law index $\alpha$, of the AGN bolometric luminosity $L_{AGN}$, 
and of the gas mass $M_{gas}$, for a host galaxy with DM mass $M=10^{12}\,M_{\odot}$.  Due to the spherical symmetry, in this case we 
have shock solutions which are independent of the inclination $\theta$, i.e.,  in all equations in Sect. 2.1 we have $R_{S,\theta}=R_s$, 
$V_{S,\theta}=V_S$, $C=1$, $Q(t)=(4\,\pi/3)\,R_s^{3}(t)$, and  $dS/d\Omega=R_S^{2}(t)$. Notice that in this case the equations can be written in terms  of the total swept-up mass $M_S=4\,\pi\,M'_S$. 

\vspace{0.1cm}
\scalebox{0.49}[0.49]{\rotatebox{-0}{\includegraphics{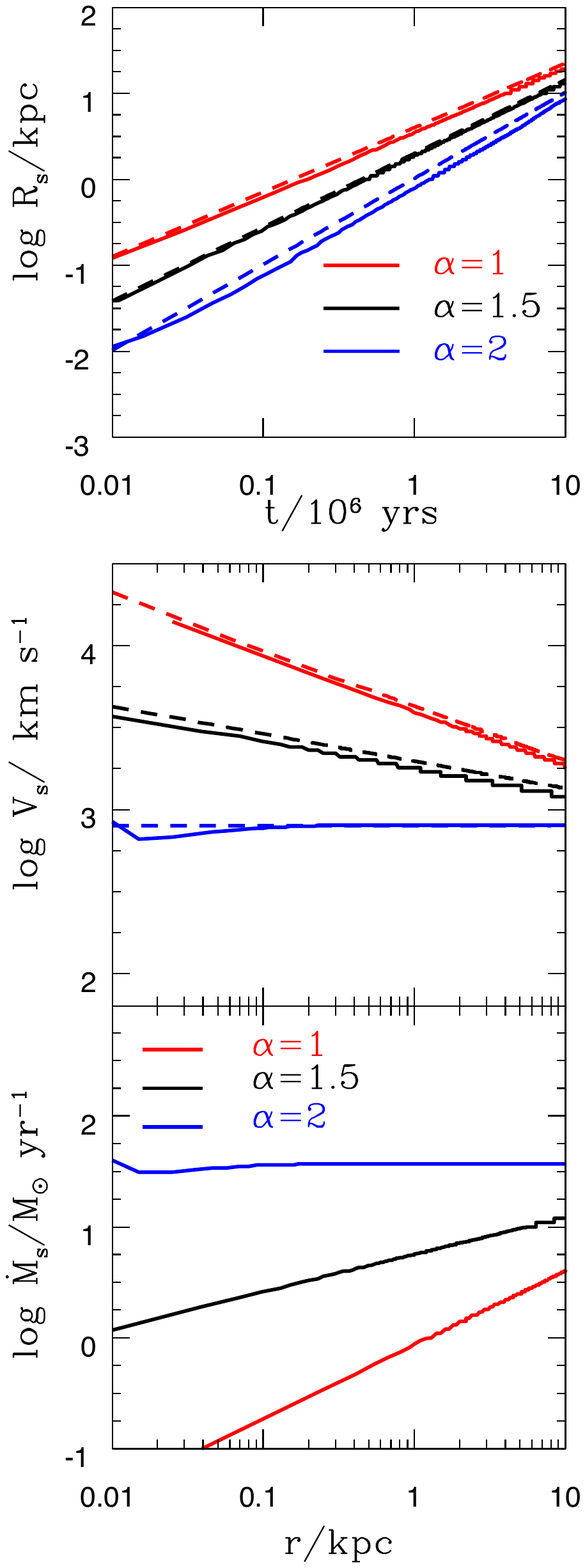}}}\hspace{0.5cm}
\scalebox{0.5}[0.5]{\rotatebox{-0}{\includegraphics{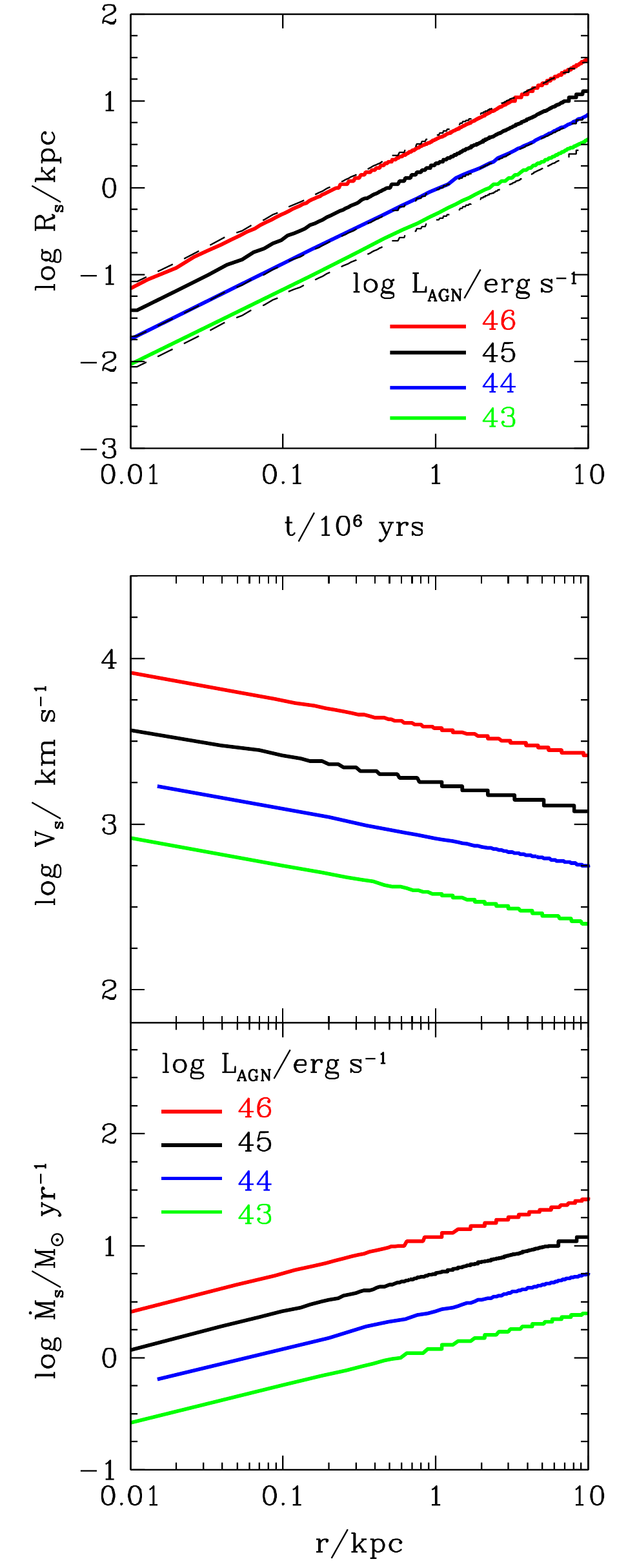}}}\hspace{0.5cm}
\scalebox{0.5}[0.5]{\rotatebox{-0}{\includegraphics{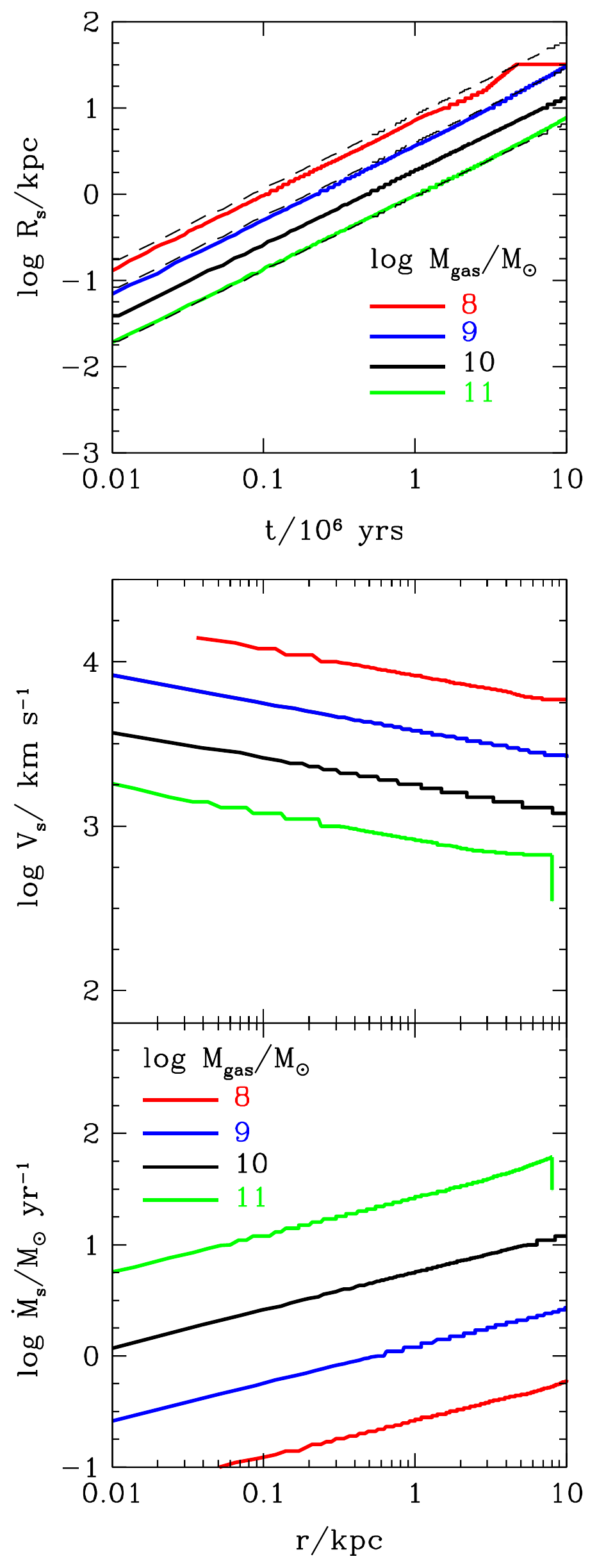}}}
\vspace{-0.1cm }
\newline
{\footnotesize Fig. 2. The dependence of our numerical solutions on the assumed logarithmic slope $\alpha$ of the density profile 
(left columns), on the input AGN bolometric luminosity $L_{AGN}$ (central panel), and on the total gas mass of the host galaxy $M_{gas}$ (right panel). 
A dark matter mass $M=10^{12}\,M_{\odot}$ has been assumed, and the black hole mass is derived from $L_{AGN}$ assuming Eddington emission. 
In all panels the black line corresponds to the reference case $\alpha=1.5$, $L_{AGN}=10^{45}$ erg/s, and $M_{gas}=10^{10}\,m_{\odot}$, and an 
initial wind velocity $V_{in}=3\,10^{4}$ km/s has been assumed in all cases.
The dashed lines correspond to the analytical self-similar solutions for $R_{S,\theta}$, as discussed in the text. 
}

We explore the dependence of our numerical solutions on the assumed value of $\alpha$ in the left panels of fig. 2. 
We note that decreasing $\alpha$ corresponds to a faster decline of the  velocity $V_S$ and to a steeper increase 
of the mass outflow  $\dot M_S$ as a function of the shock position.  This behavior was already found by earlier numerical and analytical works (see Cavaliere, Lapi, Menci 2002; Faucher Giguere 2012). Indeed, in the case of a scale-free, power-law density profile, self-similar analytical scalings can be derived for $R_S\sim t^{3/5-\alpha}$ (and hence for $V_{S}=\dot R_{S}$) in the limit of negligible cooling (energy conserving outflows). These self-similar solutions are shown as dashed lines, and provide a test for our numerical solutions. The matching between numerical and self-similar 
 solutions also indicate the approximate energy-conserving behavior  of the expanding bubble (as already found by Faucher Giguere 2012, Richings 
 \& Faucher Giguere 2018b), although in the center the large gas densities achieved in the $\alpha=2$ case result into efficient cooling yielding the slower shock velocity $V_{S}$ visible in the central panels of fig 1a compared to the  self-similar solution. 
 
As a further test for our numerical solutions we present in the central panels of fig. 1 the scaling of $R_{S}$, $V_{S}$ and $\dot M_{S,}$ with the AGN bolometric luminosity $L_{AGN}$. The solutions are characterized by increasing normalization for all such quantities for increasing $L_{AGN}$, due to the larger energy injection powering the bubble expansion. Again we can test our numerical results against self-similar analytical solutions yielding 
 $log R_{S}\sim (1/3)\,log L_{AGN}$ (see Faucher Giguere \& Quataert 2012; see also Lapi et al. 2005) for the normalization of the shock expansion, again finding an excellent agreement. 
 
Finally, we study the dependence of our solutions on the total host galaxy mass $M_{gas}$ (right panels of fig. 1), corresponding to varying the normalization $\rho_0$ of our assumed density profile. We find that increasing $M_{gas}$ results into faster shock velocities, and into smaller mass outflow rates,  in agreement with previous works. In this case, self-similar solutions yield $R_{S}\sim \rho_0^{1/(5-\alpha)}\sim M_{gas}^{1/(5-\alpha)}$, again in excellent agreement with our numerical results. 
 
\subsection{Solutions for Exponential Gas Density Profiles} 

Having tested the reliability of our numerical solution, we can proceed toward 
a detailed comparison between the properties of outflows observed in different  galaxies  and the predictions of shock 
models. Toward this aim, we consider a disc geometry (see fig. 2) for the distribution of galactic gas, with a gas density depending only on the galacto-centric distance $x=R/R_v$, but confined within vertical boundaries corresponding to a disc scale height $h$. This is assumed to be constant with radius for a given galaxy (although for $M_{BH}<10^8 M_{\odot}$ models predicts the gravitational bending of the interstellar gas  below 100 pc for due to the black hole gravitational field, see Lamastra et al. 2006), and to increase with the galaxy circular velocity according to the observed average relation 
$h = 0.45\,(V_c/100\,{\rm km\,s^{-1}})$ −- $0.14\,{\rm kpc}$ (see van der Kruit \& Freeman 2011 and references therein). Inside the disc 
(where vertical distance from the plane of the disk $Y$ is smaller than the scale height $h$) we adopt an exponential density profile  
 $\rho(R)=\rho_0\,exp(-R/R_d)$ depending only on the galacto-centric distance $R$ and on a scale length $R_d$. Outside the disc (i.e., for $Y\geq h$) the density is assumed to drop rapidly to zero.
We assume that the processes occurring in the regions reached by the expanding shell (white regions in fig. 2) do not affect the expansion of the shock in the other regions interior to the disc (the red region in fig. 2), a reasonable assumption for supersonic shocks.  

Within the above framework, we numerically solve eqs. 5-8 for the expansion of the shock, and eqs. 10-13 describing the evolution of the shock temperature $T_{S}$ and density $n_{S}$ with the shocked gas shell. We consider a grid of 20 equally spaced values of $0\leq \theta\leq \pi/2$ to derive 
at each time step, the shock radius $R_{S,\theta}(t)$ at different inclinations $\theta$, and update the corresponding value of the surface $S(t)$ and volume $Q(t)$ of the hot bubble, and of the associated geometrical factor  $C(R_{S,\theta})$ entering eqs. 5-8 and 10-13. 
Notice that until the shock radius becomes larger than the disc size $h$, the evolution is isotropic due to our assumed isotropic form $g(x)$ for the 
gas density distribution inside the disc, so that $C=1$. When the shock breaks out of the disc, and the bubble expansion will no longer retain an isotropic shape,  the values of $Q(t)$, $C$ and $S(t)$ are computed numerically.

We first derive our solutions for the expansion of the shock in the plane of the galactic disc. Then we use the properties of such solutions to understand the 
full two-dimensional structure of the outflows. 

\subsubsection{Solutions for shock expansion on the plane of the disc} 

We consider a galactic gas density profile $g(x)= exp{(-x/\xi)}$ where $\xi\equiv R_d/R_v$ is the ratio between the disc 
scale length $R_d$ and the virial radius $R_v$, in turn related to the circular velocity $V_c=10\,H(z)\,R_v$ (Mo, Mao, \& White 1998). We take $\xi=1/60$, a value consistent with  determinations from 
both detailed disc models (Mo, Mao, \& White 1998, assuming a DM angular momentum parameter $\lambda=0.05$) and existing observations 
(Courteau 1996, 1997; for a recent review see Sofue 2018).
The form factor $I$ entering the normalization of the density profile $\rho_0=M_{gas}/4\,\pi\,R_v^{3}\,I$ is obtained 
performing the volume integral $I=4\,\pi\int_0^{1}\,g(x)\,x^{2}\int_{0}^{arctg(h/x\,R_v)}\,cos\theta\,d\theta\,dx$ over the regions interior to the disc (see Sect.  2). 

The results are illustrated in fig. 3 for different values of the AGN luminosity $L_{AGN}$ and galactic gas mass $M_{gas}$, 
 assuming a fixed, fiducial value for DM mass $M=2\,10^{12}\,M_{\odot}$ (corresponding to a circular velocity $V_c=200$ km/s). 
 
The qualitative behavior of the shock velocity and outflow mass  is similar to that obtained for a power-law density profile,  with a direct dependence of $M_S=M_{S,\theta=0}$ on both  AGN luminosity and galactic gas content, while  $V_S=V_{S,\theta=0}$ increases with increasing AGN luminosity and decreases with gas mass $M_{gas}$. 
We note that in this case the outflow rate $\dot M_{S}$ reaches a maximum value and declines at large radii.  This is due to two factors: the first is the rapid drop in the gas density $\rho$, while the second is related to the large volume encompassed by the bubble expansion when it breaks out of the disc, and is expressed by the quantity $C(R_{S,\theta})$ in eqs. 4 and 5. This retains the initial value $C(R_{S,\theta})=1$ until the bubble reaches 
the disc boundary in the vertical direction. The subsequent rapid expansion of the bubble in the vertical direction (due to the low density encountered in this direction, see sect. 3.2.2) reduces the pressure exerted on the portion of the bubble surface contained in the disc, thus reducing the expansion and the mass outflow in the direction parallel to the disc.  In fact, in such direction the volume of the sphere with radius $R_{S,\theta}$ becomes increasingly smaller than the  actual volume of the bubble $Q(t)$ yielding progressively larger values for the quantity $C(R_{S,\theta})$ and resulting into a smaller efficiency 
 of the propulsive effect of the pressure term in eqs.  1, 3, 4, 5. 

As for the properties of the shocked shell, the initial decline of the temperature $T_{S}$ is followed by a sharp drop due to 
fast radiative cooling when $T_{S}\approx 10^{6.5}$ K, and  $T_{S}$ drops rapidly to $10^{4}$ K. Correspondingly, the 
density increases sharply; for a given total gas mass  $M_{gas}$, the density reached by the shocked gas shell depends on the position of the shock at the moment of gas cooling, and can easily reach values as large as $10^{4}\,{\rm cm^{-3}}$ at $R_{S}\lesssim 1$ kpc typical of observed molecular outflows. 
Of course, the temperature drop (and hence the increase in density) of the shocked gas shell is delayed for increasing values of the AGN luminosity (due to the heating term in eq. 9), while it is favored by increasing values of the total gas mass $M_{gas}$, which also correspond to increasing shocked gas 
densities $n_{S}$. 

\vspace{0.cm}\hspace{-1cm}
\scalebox{0.4}[0.4]{\rotatebox{-0}{\includegraphics{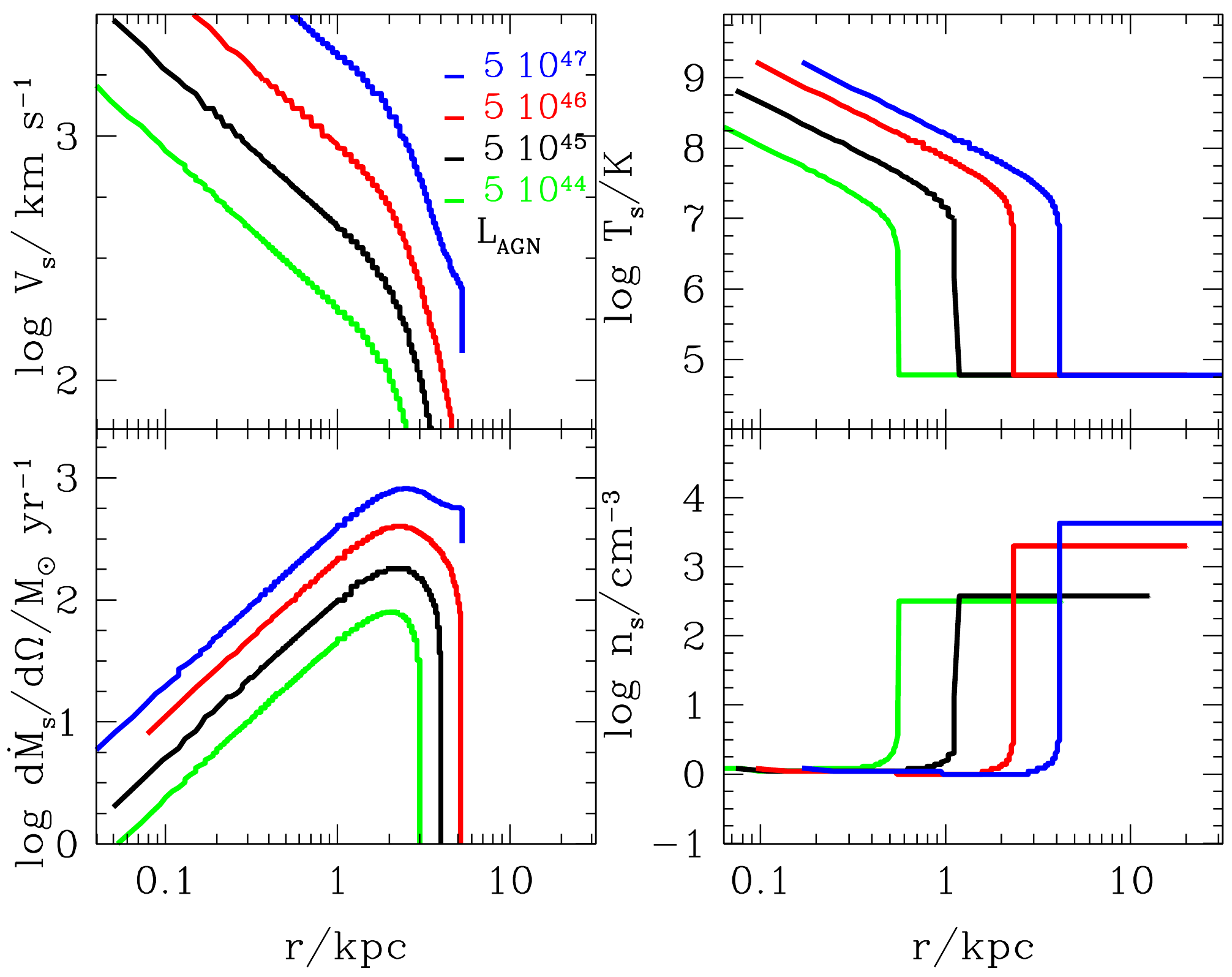}}}\hspace{0.4cm}
\scalebox{0.4}[0.4]{\rotatebox{-0}{\includegraphics{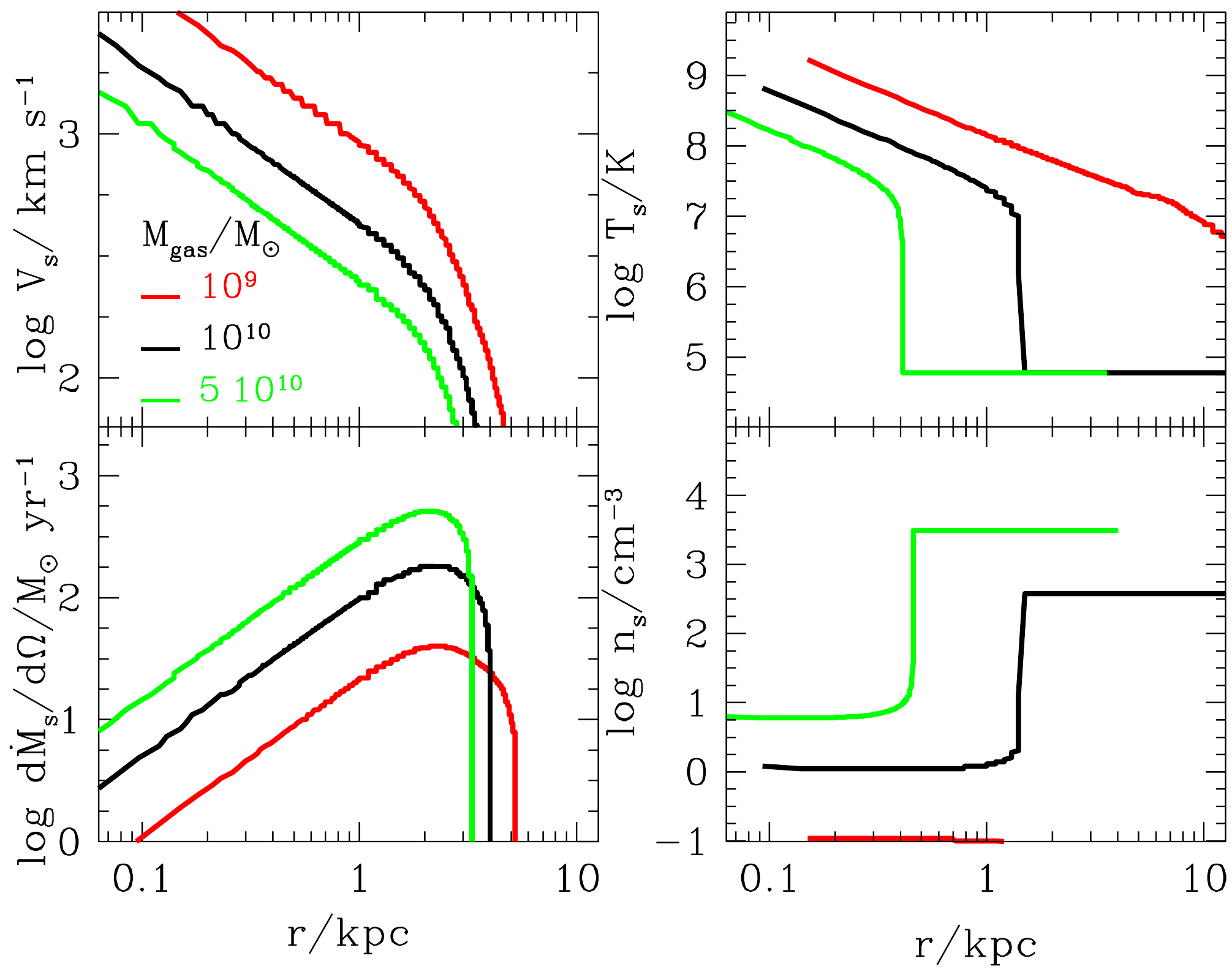}}}\newline
{\footnotesize Fig. 3. For the shock expansion in the plane of the disc ($\theta=0$), we show the dependence of our numerical solutions on the AGN luminosity $L_{AGN}$ (in erg/s, left), and on the total gas mass of the host galaxy $M_{gas}$ (right), for an exponential density profile of the galactic gas.  
For each case, we show the shock velocity $V_{S}$ and mass outflow rate per unit solid angle $\dot M'_{S}$, and the temperature $T_{S}$ and density $n_{S}$ of the shocked gas shell. A dark matter mass $M=2\,10^{12}\,M_{\odot}$ has been assumed, and the black hole mass is derived from $L_{AGN}$ assuming Eddington emission. 
In all panels the black line corresponds to the reference case $L_{AGN}=5\,10^{45}$ erg/s, and $M_{gas}=10^{10}\,m_{\odot}$, and an 
initial wind velocity $V_{in}=3\,10^{4}$ km/s has been assumed in all cases.
}
\vspace{-0.2cm }

\subsubsection{The Two-Dimensional Structure of the outflows} 

We now proceed to compute the full two-dimensional structure of the outflows. In this case, we compute the expansion of the shock in all directions
solving eqs. 4-7, assuming the usual exponential  density profile $\rho(R)=\rho_0\,exp(-R/R_d)$ until the shock position reaches the disc boundary, and a vanishing  density 
 in the regions external to the disc, i.e.,  where the vertical distance $Y$ from the plane of the disc is larger than the scale height $h$. Our approach is similar to that adopted by Hartwick, Volonteri \& Dashyan (2018), although the latter authors adopt a 
scale height $h$ which depends on the position along the disc. 

The solutions for the outflow velocity $V_{S,\theta}$, mass outflow rate $\dot M_{S,\theta}$ and shock position $R_{S,\theta}$ as a function of time, are plotted in fig. 4 for a reference galaxy with DM mass $M=10^{12}\,M_{\odot}$, a gas 
mass  $M_{gas}=10^{10}\,M_{\odot}$, and AGN bolometric luminosity $L=10^{45}$ erg/s. The $X$ coordinate represents the distance from the center in the direction parallel to the plane of the disc, while the $Y$ coordinate corresponds to the distance in the (vertical) direction perpendicular to the disc.  Along the plane of the disc the velocity $V_{S,\theta}$ rapidly decreases with increasing radius (top-left panel), while in the vertical direction the shock decelerates until it reaches the disc boundary $h$, but it rapidly accelerates afterward due to the drop of the gas density outside the disc. The opposite is true for the mass outflow rate (top-right panel), which instead grows appreciably only along  the plane of the disc, where the larger densities allow to reach values $\dot M_{S,\theta=0}\sim 10^{3}$ $M_{\odot}$/yr, as we have seen in the previous section. 

As for the shock expansion radius, this follows the paths of  least resistance (see bottom panel of fig. 4), yielding an elongated shock front in the vertical direction. E.g., inspection of fig. 4 (bottom panel) shows that while in the direction perpendicular to the disc the outflows reaches a distance of 20 kpc in approximatively $10^{7}$ yrs, it takes about $10^{8}$ yrs to reach the same distance in the plane of the disc.
This has important implications for studies of AGN feedback in galaxy formation models. E.g., for an AGN 
 life time$\sim 10^{8}$ yrs this would results into null gas expulsion along the plane of the disc. 
 
We notice that such a behaviour does not depend on the particular choice for the cutoff in initial density distribution outside the disc (i.e., for $Y\geq h$). 
 Indeed, Hartwick, Volonter \& Dashyan (2018) find similar results for a 
radius-dependent scale length and with a different functional form for the cutoff. 
Thus, although the shock expansion follows the paths of  least resistance (see bottom-left panel of fig. 4), yielding an elongated shock front in the vertical direction, it is only in  directions close  to the  plane of the disc that massive outflows ($\dot M_{S}=10^{2}-10^{3}\,M_{\odot}$/yr) can be generated. 
This is shown in detail in the bottom-right panel of fig. 4, where the expansion of the shock position $R_{S,\theta}$ is shown as a function of time 
for both the vertical and the horizontal directions, along with the mass $M_{S,\theta}$ swept out by the outflows in the considered directions. 
 While in the direction perpendicular to the disk the shock expands rapidly to reach 10 kpc in a short time scale $\approx 2\,10^{7}$ yr, 
  the denser medium encountered by the shock in the direction parallel to the disk results into a slower expansion (a distance  
  10 kpc is reached only after $t\approx 10^{8}$ yr). However, the mass swept out by the outflow in the vertical direction saturates to a small value $M_{s \perp}\approx 10^{7}\,M_{\odot}$, while a much larger value $M_{s \parallel}\approx 10^{9}\,M_{\odot}$ is attained along the direction 
  parallel to the disc. 

While outflows 
in the vertical directions can have a significant impact on the expulsion of gas from the disc (due to the large velocities attained on a short time scale),  
 on the escape fraction of UV photons from the galactic center, and possibly on the formation of Fermi bubbles like those detected in our Galaxy
(see Su, Slatyer, and Finkbeiner 2010),  they are of minor importance in determining the observed massive molecular and ionized 
 outflows in galaxies. 
  
\vspace{-0.4cm}\hspace{-1cm}
\scalebox{0.4}[0.4]{\rotatebox{-0}{\includegraphics{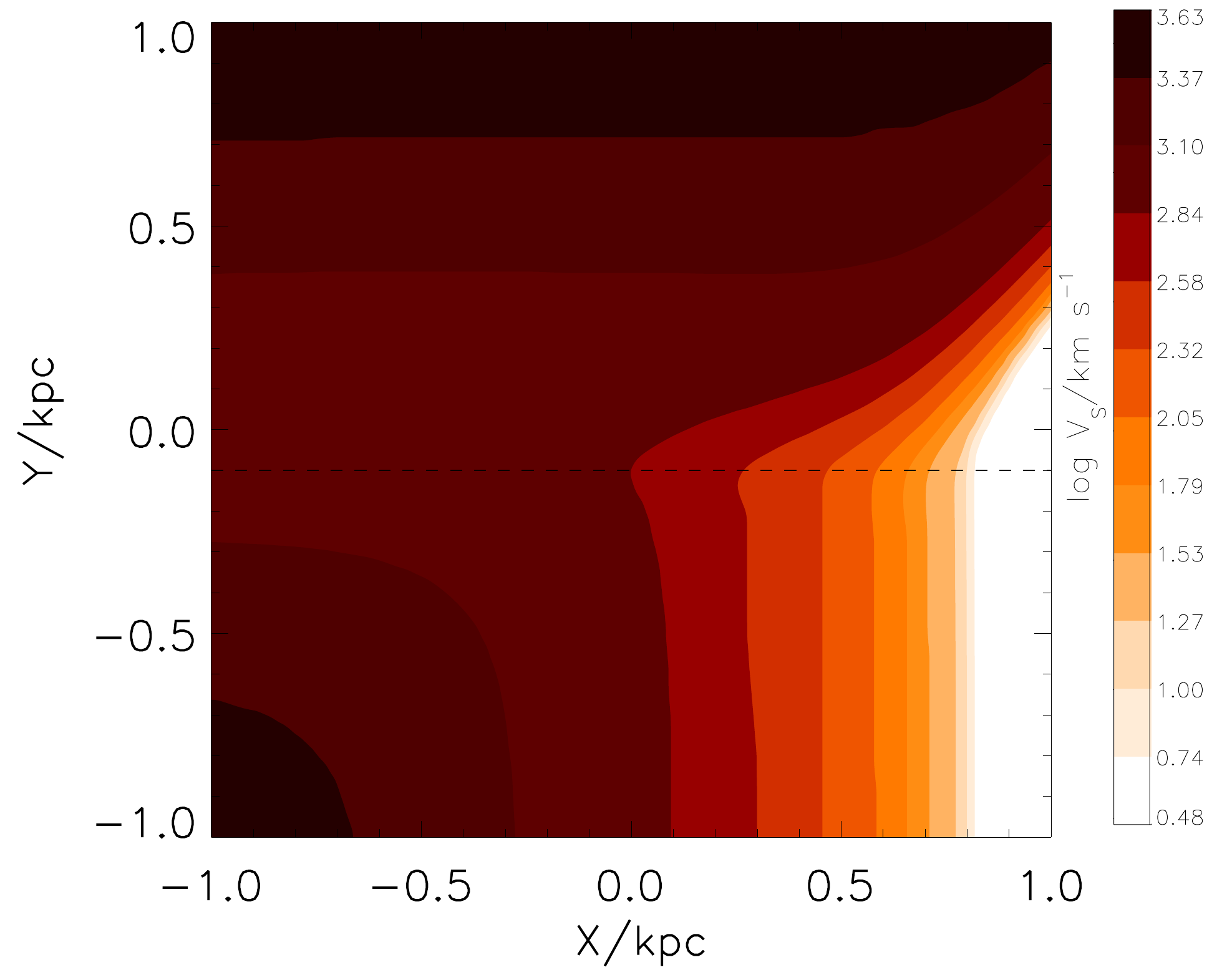}}}\hspace{0.2cm}
\scalebox{0.42}[0.4]{\rotatebox{-0}{\includegraphics{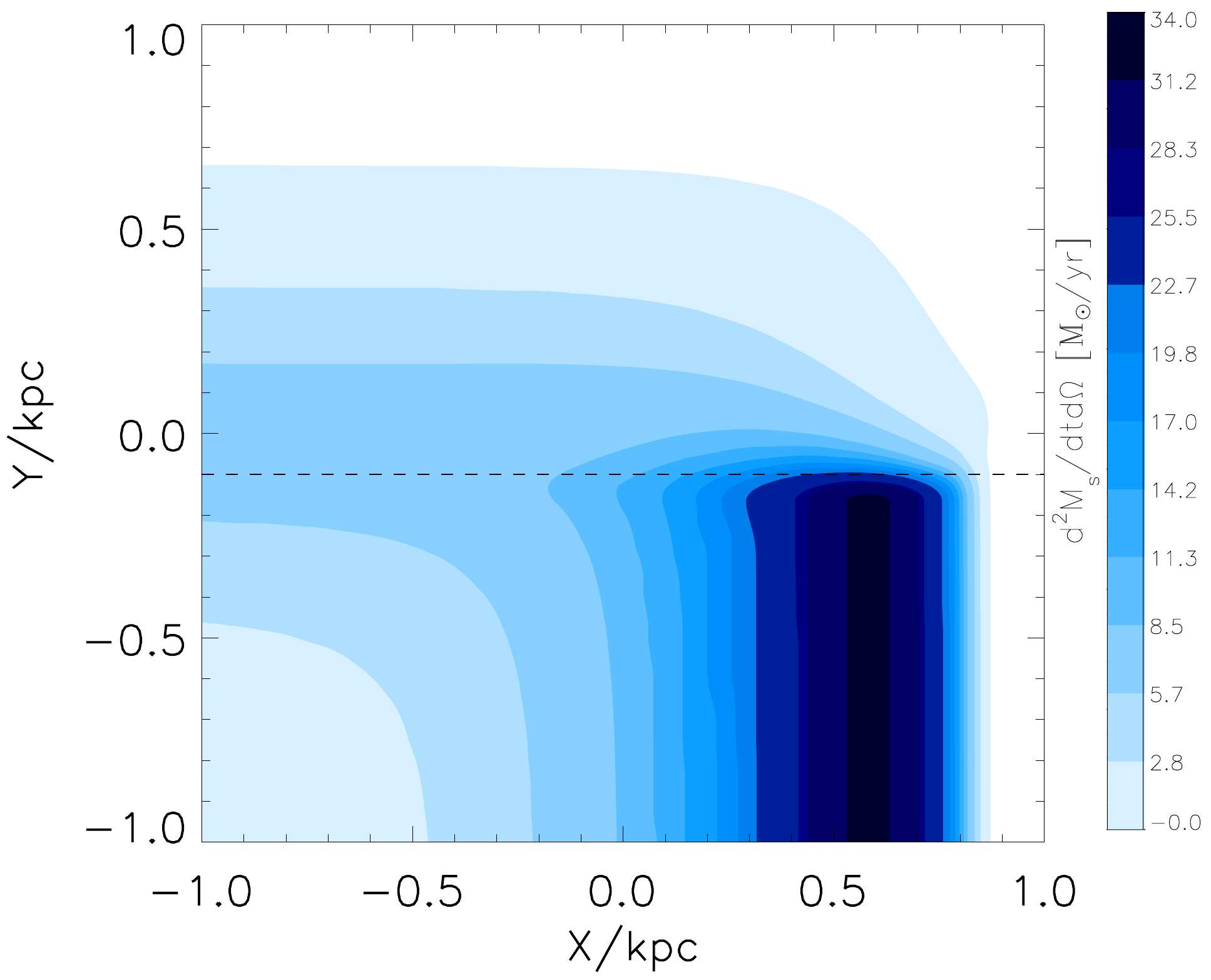}}}
\vspace{0.1cm}
\scalebox{0.4}[0.4]{\rotatebox{-0}{\includegraphics{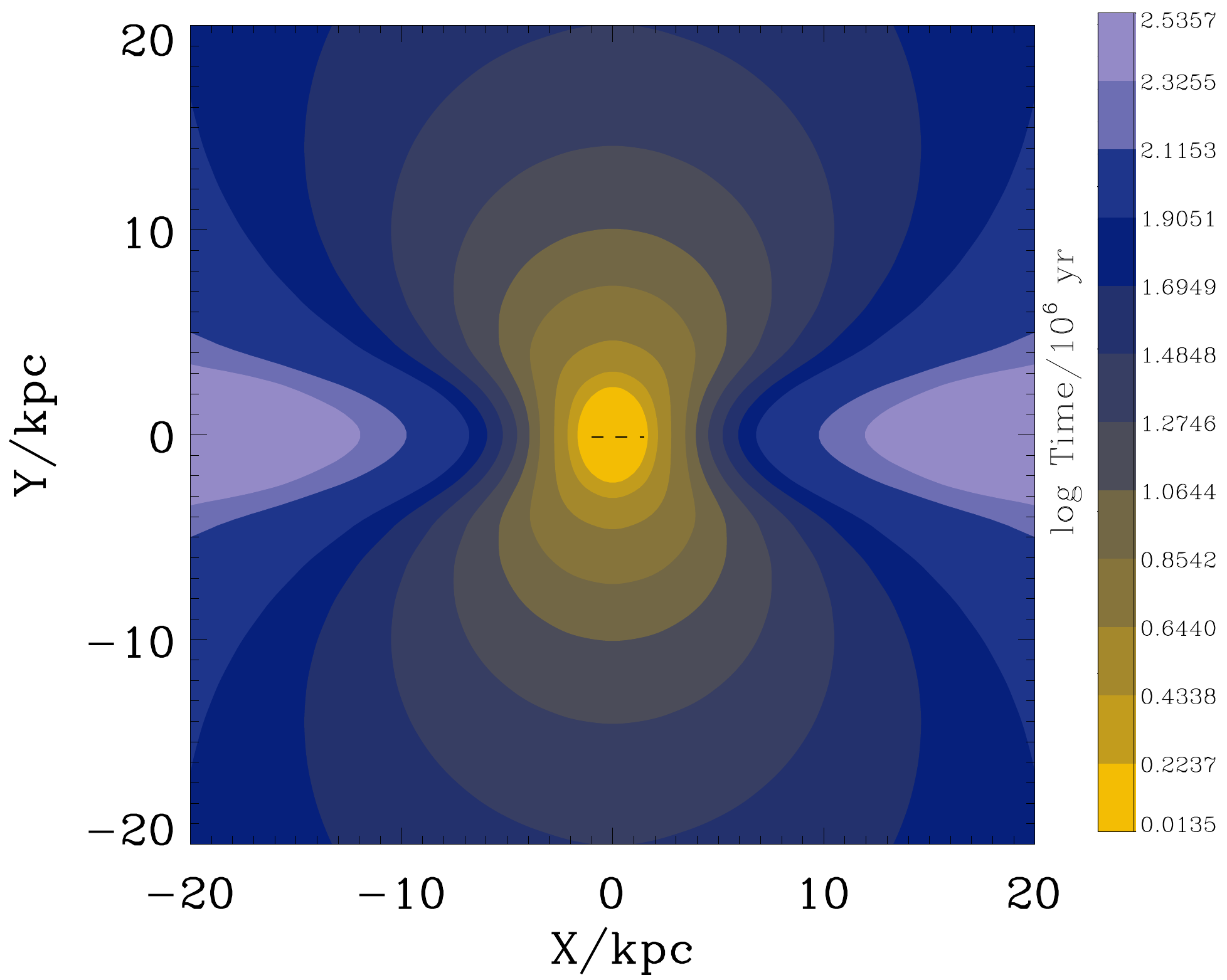}}}\hspace{0.6cm}
\scalebox{0.4}[0.4]{\rotatebox{-0}{\includegraphics{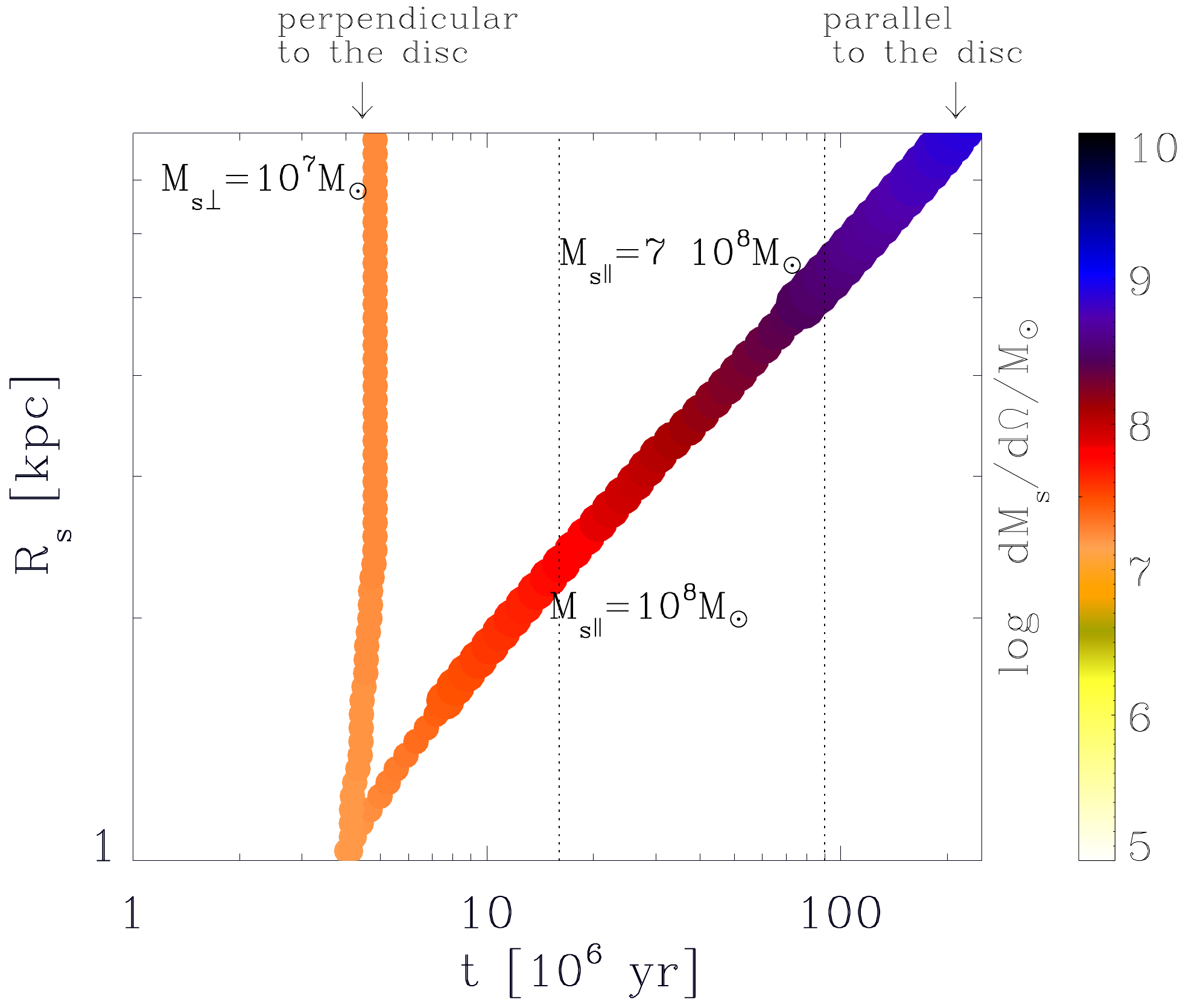}}}
\newline
{\footnotesize Fig. 4. Top panels. The  velocity map (left) and mass outflow rate (per unit solid angle) map (right) for our reference galaxy. The values corresponding to the colored contours are displayed on the bars. The $X$ and $Y$ coordinates correspond to the distance from the 
galaxy center in the directions parallel and perpendicular to the plane of the disc, respectively. Bottom Left Panel. The positions of the shock at the different times represented by different colors and displayed on the right bar. Bottom Right Panel. The time evolution of the shock radius $R_{S,\theta}$ in the direction perpendicular and parallel to the disc. The size and the colors of the dots correspond to the logarithm of the mass of the swept-out gas (per unit solid angle)  in the considered  direction, as shown by the color bar. We also marked the values of the swept-out mass  in the perpendicular ($M_{s\perp}$) and parallel directions ($M_{s\parallel}$) at $t=10^{7}$ yrs and  $t=5\,10^{7}$ yrs.
}

\section{Comparison with Observations}

The above numerical solutions allow us to perform a detailed comparison with available data concerning observed molecular and ionized outflows. 
When key  properties of the host galaxy mass are measured, we can use the observed AGN luminosity $L_{AGN}$, the total gas mass $M_{gas}$,  and the the dynamical mass $M$ (or equivalently the circular velocity $V_c$) as inputs for the model. 
This allows us to compute, for each observed galaxy, the expected expansion properties of the shock, and to compare the results with the observed properties of the outflow in the considered galaxies. To perform a fair comparison, we must take into account how the properties of the outflows are derived from observations. First, present observations are not able to resolve the angular dependence of the outflow quantities (i.e., $R_S$, $V_S$ and $\dot M_S$). 
Thus, when comparing with observations 
we first compute the full two-dimensional solutions in all directions (i.e., $R_{S,\theta}$, $V_{S,\theta}$ and $\dot M_{S,\theta}$), and then we derive their  mass-weighted average over the directions $\theta$, which are then compared with observed values. In the following, for the sake of simplicity, we simply denote with $R_S$, $V_S$, $\dot M_S$ such averaged quantities. A second consideration concerns the measurement of the  mass outflow
 rate, which is observationally derived as $\dot M_S=M_s\,V_S/R_S$ (see Appendix in Fiore et al. 2017 for a discussion). Such a definition is not identical to the ($\theta$-averaged)  mass outflow rate $\dot M_S$ derived from the time-derivative of eq. 8. When comparing with data we shall present the model predictions for both definitions  and we show that, in the regions usually covered by observations, they  are basically equivalent.


\subsection{Comparison with Single Objects}

For molecular outflows, and for a single ionized outflow, observations in the literature have recently led to assemble a sizable sample of objects for which the gas mass and the rotation velocity of the host galaxies have been measured. 
In the following, we compare with the data sample summarized in Table 1, that extends the sample of molecular outflows in  Fiore et al. (2017) to include those in M51, Circinus, 
XID2028, zC400528, APM08279, 3c298 (references are given in the caption). In addition, the data for I11119 have been updated using the 
recent results by Veilleux et al. (2017). 
We use only AGN for which there is not only a measurement of the physical properties of the outflow (the physical size $R_{S}$, the velocity $V_{S}$, and of the mass outflow rate $\dot M_{S}$) but also an estimate of the gas mass within $R_{S}$, of the projected rotation velocity $V_{rot}(<R_{S})\,sin\,i$ within  $R_{S}$, and of the inclination angle $i$; for some objects, the asymptotic rotation velocity $V_{asympt}$ is also available. 
 
 For each object, the observed AGN and host galaxy properties summarized in the left side of Table 1 (left of the vertical line) are used to obtain the input quantities for the model. 
 The most uncertain quantity is the host circular velocity $V_c$. For most objects, we derive a lower limit 
 from the rotation velocity within $R_{S}$ (corrected for the inclination), and explore the effect of changing the assumed value of $V_c$. For objects where the asymptotic rotation velocity has robust estimates, we adopt such a value as  an upper limit (for the two objects where only the asymptotic is available we explore the effect of assuming lower values).
 The input value for the total gas mass $M_{gas}$ is derived 
   by extrapolating the observed value $M_{gas}(R<R_{S})$ out to the virial radius using an exponential profile with disc scale radius $R_d$ related to the circular velocity as explained in Sect. 3.2. 
For each object, the input quantities $L_{AGN}$, $V_c$ and $M_{gas}$ derived as above allow us to compute the corresponding predicted values of outflow velocity and outflow rate. These are compared with the observed values of $V_{S}$ and $M_{S}$  (shown on the right of the vertical line in Table 1). 

\begin{table}[!ht]
\caption{Sample of Observed Outflows}
 \footnotesize

\label{my-label}
\begin{tabular}{c c c c c c c c | c  c c l l}
\hline
\hline
Object    & \scriptsize Redshift & \scriptsize $L_{AGN}$   &$M_{BH}$ &  \scriptsize{log $M_{gas}(<R_{S})$} & \scriptsize {$V_{rot}(<R_{S})\,sin(i)$} & {$i$} & $V_{asympt}$ & $R_{S}$     & $V_{S}$    & $\dot M_{S}$      & Ref.   \\
          &          & \scriptsize{[$10^{45}$ erg/s] }&  \scriptsize $[10^{9}\,M_{\odot}$]&  {[}$M_{\odot}${]}   & [km/s]                       &         deg      & [km/s]                 & [kpc] & [km/s] & {[}$M_{\odot}$/yr{]} &                                                                                            \\ \hline
mark231   & 0.042   & 5       & 0.087        & 8.9                 & 77         & 36      & 340       & 0.3       & 750          & 1000        & \tiny 1, 2, 3, 23 \\ 
mark231   &  0.042  & 5       & 0.087        & 9.3                 & 77         & 36      & 340       & 1          & 850           & 700         & \tiny 1 ,2, 4, 5 , 23     \\ 
n6240      & 0.025   & 0.63   & 0.1            & 9.3                 & 230       & 70     & --          & 0.6       & 500            & 500         &\tiny 6, 7 8, 9  \\ 
n6240      & 0.025   & 0.63   & 0.1            & 9.8                 & 188       &70 	   & -- 	       & 5          & 400            & 120         & \tiny 6, 7, 8, 9 \\ 
I08572     & 0.06     & 4.6     &  --              & 9.1                 & 100       & 75 	   & --          & 1          & 1200          & 1200       & \tiny 9, 10 \\ 
I10565     & 0.04     & 0.65   & 0.02           & 9.3                 & 75        & 20     & 250 	 	& 1.1       & 600            & 300         & \tiny 9, 10, 11, 12 \\ 
I23060     & 0.17     & 11.5   &  --               & 10.4               & 175      & 75     & --       	& 4          & 1100          & 1100       &\tiny 10    \\
I23365     & 0.06     & 0.47   &  0.037         & 9.47               & 130      & 30    & 260    	& 1.2       & 600             & 170        &\tiny 9, 10, 11, 12 \\
J1356      & 0.12     & 1.25   &  0.3             & 8.5                 & 200      & 45     & --         & 0.3        & 500             & 350        &\tiny 13  \\
ngc1068  & 0.03     & 0.087 & 0.01             & 7.8                & 52         & 41    & 270   	& 0.1       & 200             & 120        &\tiny 5, 14, 15          \\
ic5063     & 0.01     & 0.1    &  0.055           & 7.7                & 166       & 74    & --       	& 0.5        & 400            & 22          &\tiny 16, 17, 18, 19 \\
ngc1266  & 0.01     & 0.02   &  0.003          & 8.6                & 110       & 34    & --       	& 0.45      & 360            & 13  		   & \tiny 20, 21 \\
I17208    & 0.04      & 1.3 &  0.05             & 11.13              & 130      & 30      & --        & 1           & 370           & 65   		& \tiny 12, 22, 23 \\
I11119    & 0.19      & 15 &  0.016              & 9.95               & 142      & 30      & --        & 7           & 1000           & 800   		& \tiny 24 \\
M51         & 0.002   & 0.1     &  0.001          & 9.8                &   --        & 22    & 200   		& 0.04      & 100-200           & 11.6   	   & \tiny 25, 26 \\
Circinus    & 0.001   & 0.04   &  0.0017        & 8.46              &  --         & 65   & 220        & 0.45      & 150            & 3.1   		& \tiny 27 \\
XID2028   & 1.6       & 20      &  --                & 10                 & 210       & 30     & 350     & 10         & 700            & 350   		& \tiny 28, 29 \\
zC400528& 2.3       & 1.7     &  --                & 11                 & 250       & 37     & --         & 4.2        & 450            & 768   		& \tiny 30, 31 \\
APM08279& 3.9      & 280    &  10               & 11.15            & 550       & 30     & --         & 0.27      & 1340          &1000   		& \tiny 32, 33 \\
3c298      & 1.43     & 70      &  3.2             & 9.81               & 190      & 54      & --        & 1.6        & 400             & 2300   	& \tiny 34 \\
     &      &   &                  &                &                               &      &         &            &   &   \\
\end{tabular}
{\footnotesize Ref. 1 = Feruglio et al. (2015); 2 = Lonsdale et al. (2003); 3 = Davies et al. (2004), 4 = Veilleux et al. (2009), 5 = Davies et al. (2007); 6 = Feruglio et al. (2013); 7 = Tacconi et al. (1999); 8 = Engel et al. (2010); 9 = Howell et al. (2010); 10 = Cicone et al. (2014); 11 = Dasyra et al. (2006); 12 = Downes \& Solomon (1998); 13 = Sun et al. (2014); 14 = Garcia-Burillo et al. (2014); 15 = Krips et al. (2012); 16 = Morganti et al. (1998); 17 = Morganti et al. (2013); 18 = Woo \& Urry (2002); 19 = Malizia et al. (2007)
20 = Alatalo et al. (2011); 21 = Alatalo et al. (2014); 22 = Veilleux et al. (2013); 23 = Xia et al. (2012); 
24= Veilleux et al. (2017); 25=Querejeta et al. (2016); 26= Shetty et al. (2007); 27= Zschaechner et al. (2016); 28 = Perna et al. (2015a); 29=Brusa et al. (2018); 30=Genzel et al. (2014); 31=	Herrera-Camus et al. (2018); 32=Feruglio et al. (2017); 33=Riechers et al. (2009); 
34=Vayner et al. (2017)}
\end{table}

The comparison between model predictions and observations is shown in fig. 5 for each object in Table 1. The values of 
 $V_c$, $M_{gas}$, and $L_{AGN}$ that have been adopted as inputs for the model are shown in the labels for each objects. 
The model seems to capture the basic dependence of the outflow velocity and mass outflow rate on the AGN luminosity $L_{AGN}$ and host galaxy gas mass $M_{gas}$ for a relatively large range of input parameters
$10^{44}\lesssim L_{AGN}/{\rm erg\,{s^{-1}}}\le10^{46}$ and 
  $10^{8}\lesssim M_{gas}/{M_{\odot}}\le10^{10}$
covered by the data in Table 1. 
For all the objects  the density $n_{S}$ of the shocked gas at the observed outflow positions is close to  the critical threshold for emission from the rotational transition of CO (corresponding to 2700 cm$^{-3}$), although in a few cases the predicted densities are slightly below the critical threshold (for I11119, I10565, ngc1266, J356). However, as noticed in the Introduction, a detailed treatment of the position-dependent ionization properties of the shocked gas and  of its molecular content requires numerical simulations (Richings \& Faucher-Giguere 2018a,b). 


The evolution of  the outflow velocity in the models is characterized by an upturn. This is related to the two-dimensional properties of the shocks discussed above. While initially the mass-weighted average $V_S$ is dominated by the component $\theta=0$ aligned with the plane of the disc (due to the large mass involved in such a direction, see fig. 4) when the shock in the plane of the disc reaches a standstill, the average $V_S$ is mainly contributed  
by components not aligned with the plane of the disc, characterized by an increasing expansion velocity (see upper left panel of fig. 4) related 
to the low gas density encountered in such a direction. The same effect is responsible for the downturn of the average mass outflow rate $\dot M_s$. 
 In fact, this is largely contributed by the gas mass in the disc; however, in the disc direction, the combined effect of large densities and of the drop 
  in the pressure term associated to the bubble expansion as it breaks out of the disc (see Sect. 3.2.1) leads to the drop of the mass outflow rate.

We note that our results are not sensitive to the uncertainties affecting $V_c$ (and hence the extrapolated $M_{gas}$) defining the properties of the host galaxy. This results from the balance between the effect of changing $V_c$ and the correction that relates the observed values $M_{gas}(<R_{S})$ in Table 1 to the overall gas content $M_{gas}$. E.g., increasing $V_c$  tends to shift the predicted curves on the right along the x-axis;  however, such effect is balanced by the larger value of $M_{gas}$ corresponding to the observed $M_{gas}(<R_{S})$. 
 
\hspace{-0.9cm}\scalebox{0.4}[0.4]{\rotatebox{-0}{\includegraphics{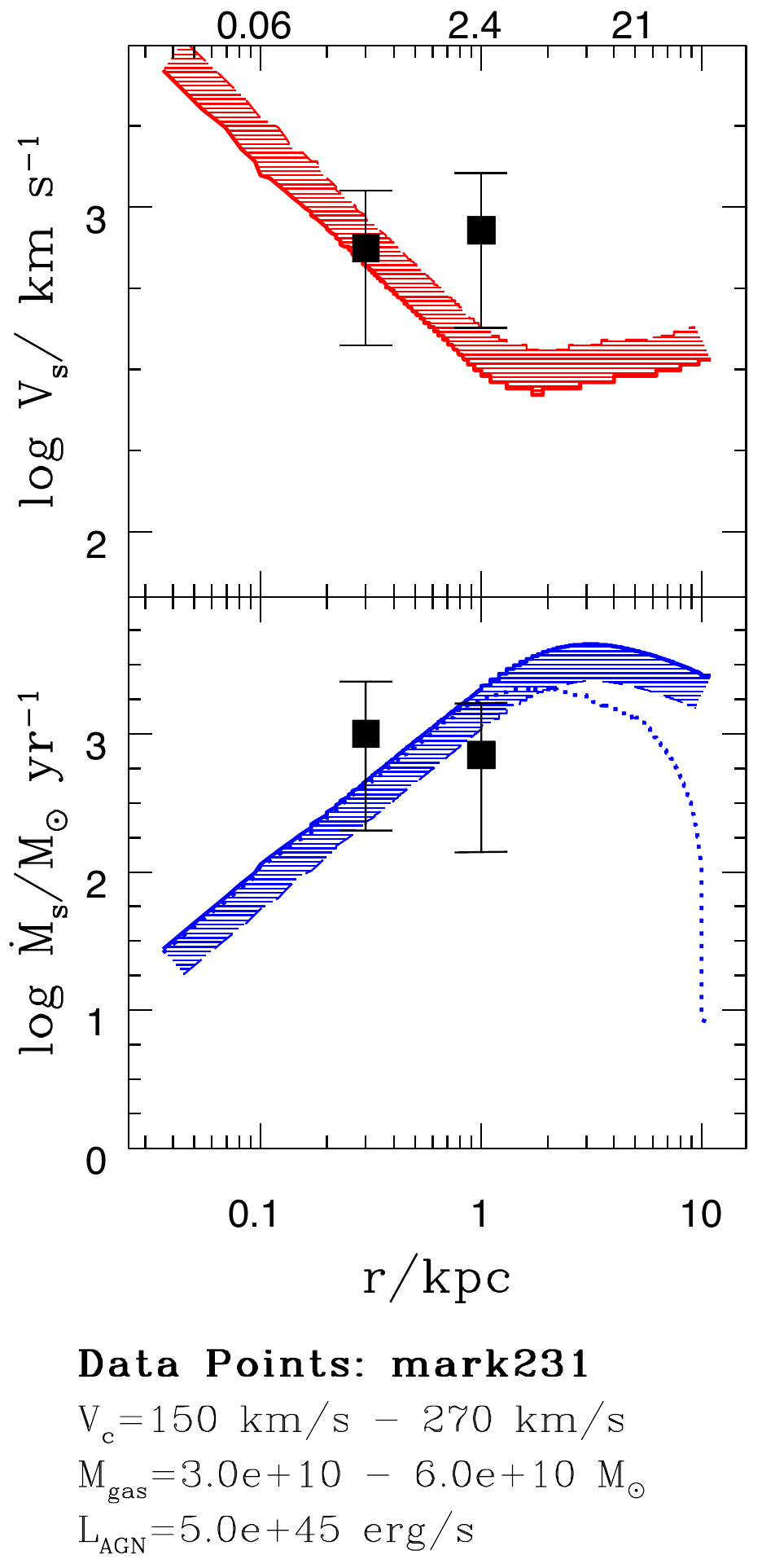}}}
\hspace{0.cm}\scalebox{0.4}[0.4]{\rotatebox{-0}{\includegraphics{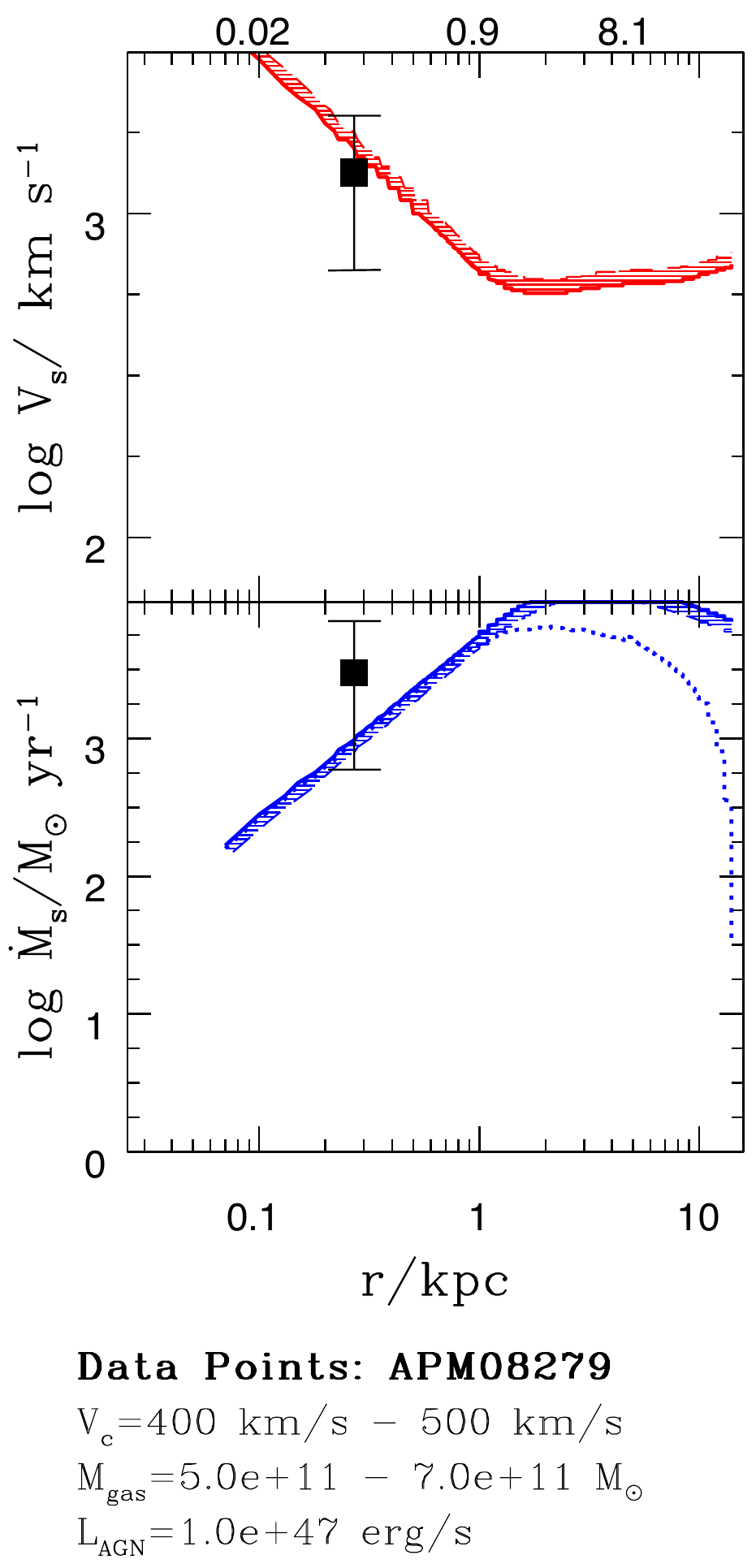}}}
\hspace{0.cm}\scalebox{0.4}[0.4]{\rotatebox{-0}{\includegraphics{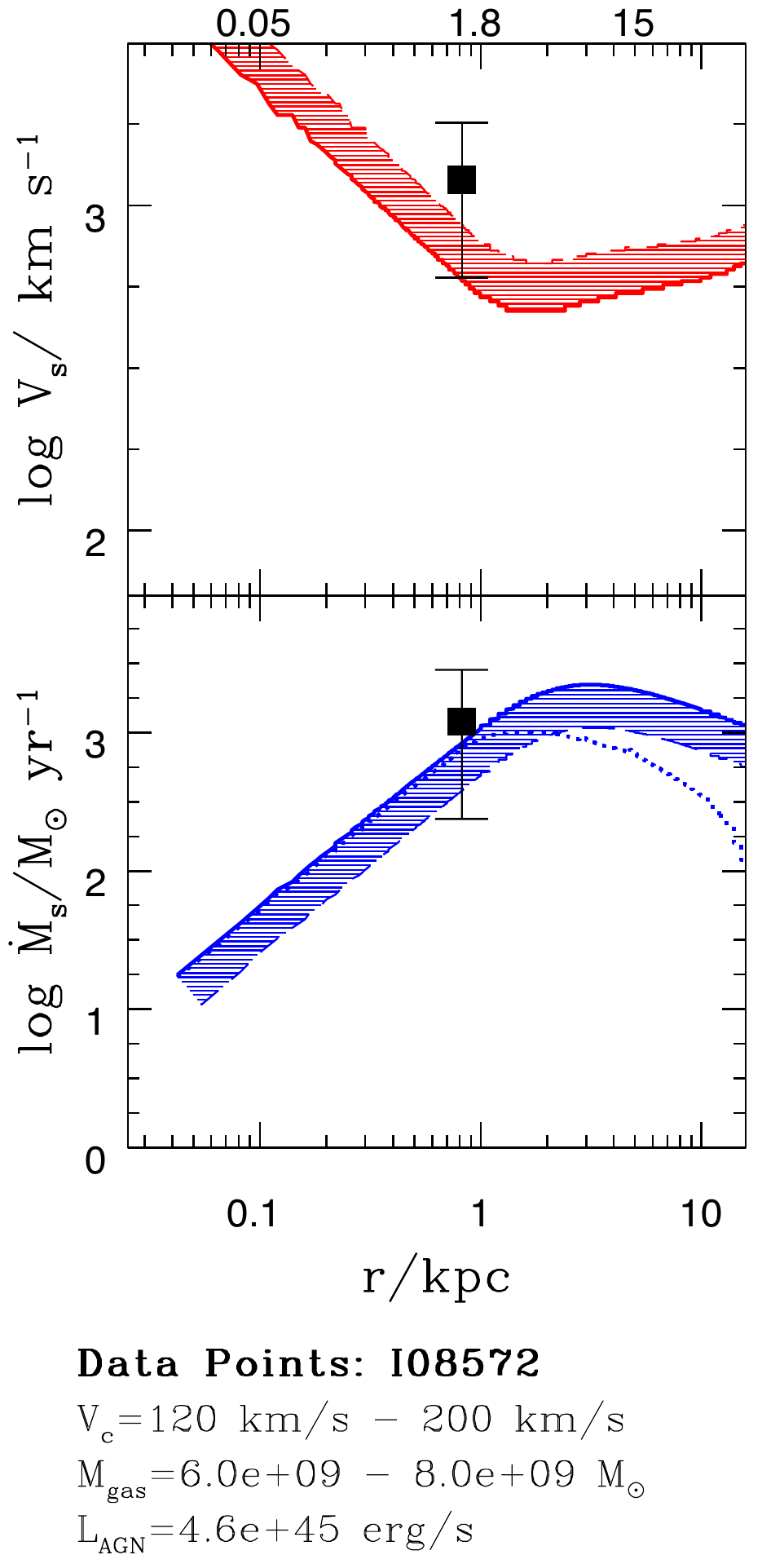}}}
\hspace{0.cm}\scalebox{0.4}[0.4]{\rotatebox{-0}{\includegraphics{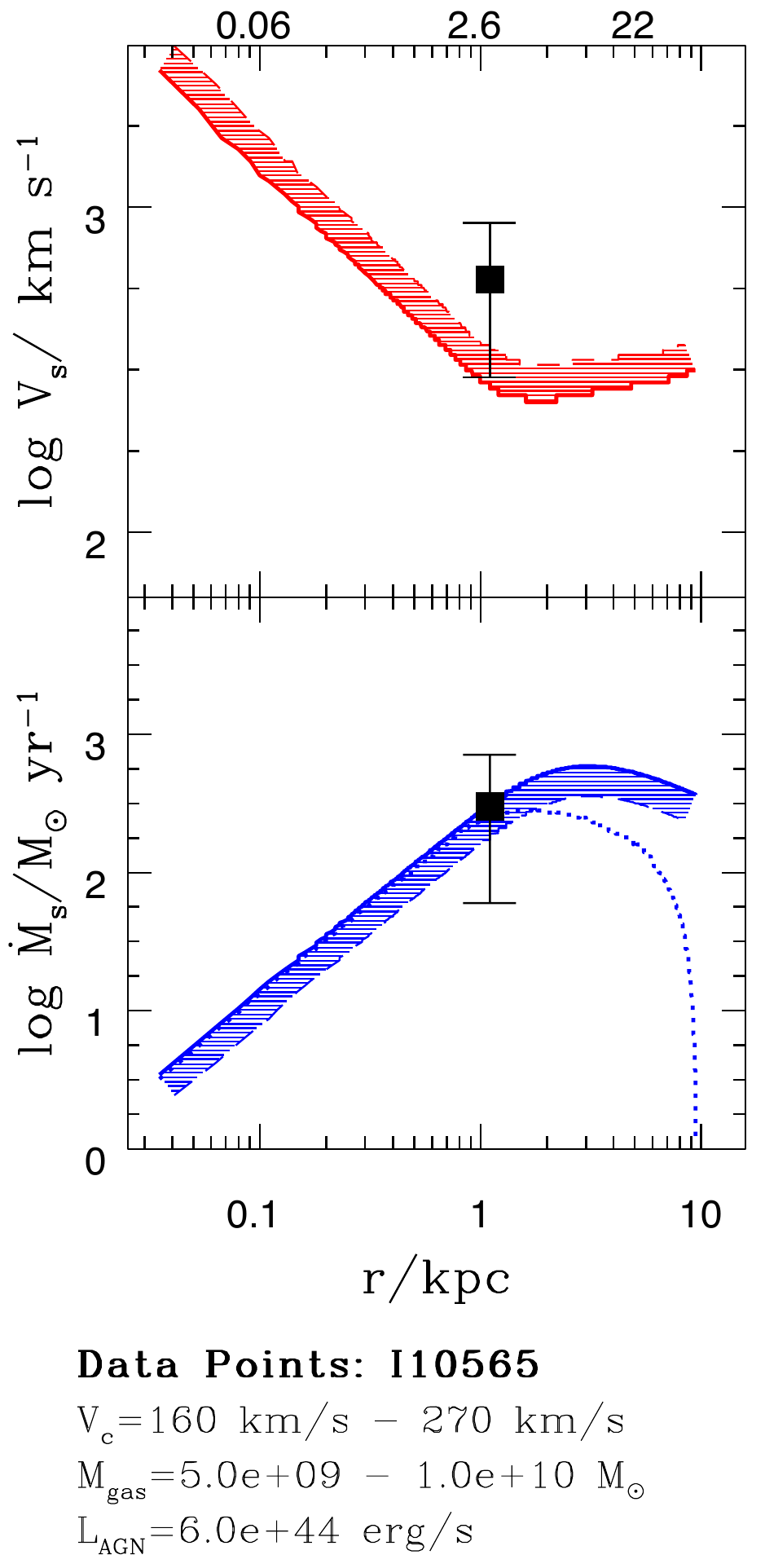}}}
\newline
\hspace{-0.9cm}\scalebox{0.4}[0.4]{\rotatebox{-0}{\includegraphics{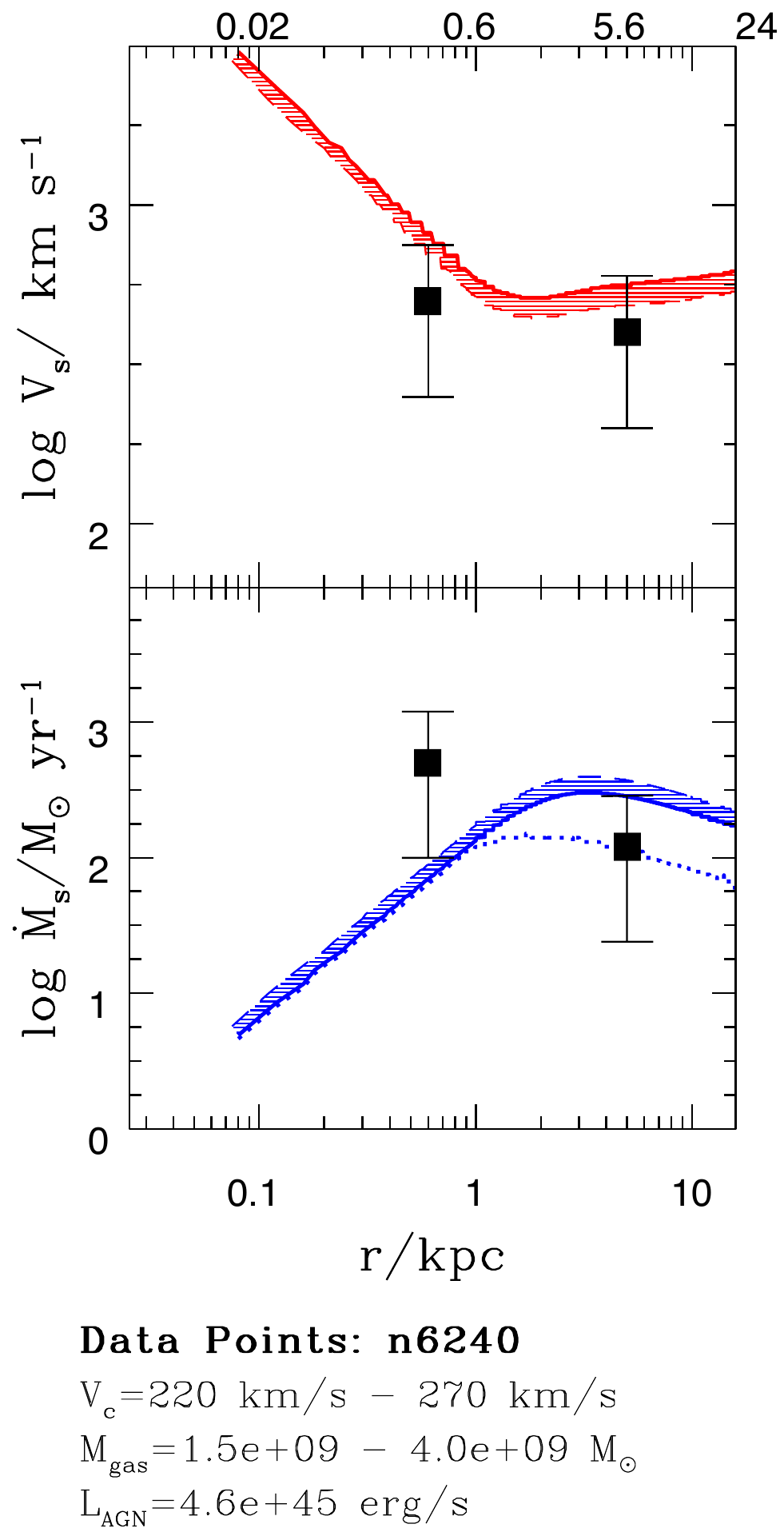}}}
\hspace{0.cm}\scalebox{0.4}[0.4]{\rotatebox{-0}{\includegraphics{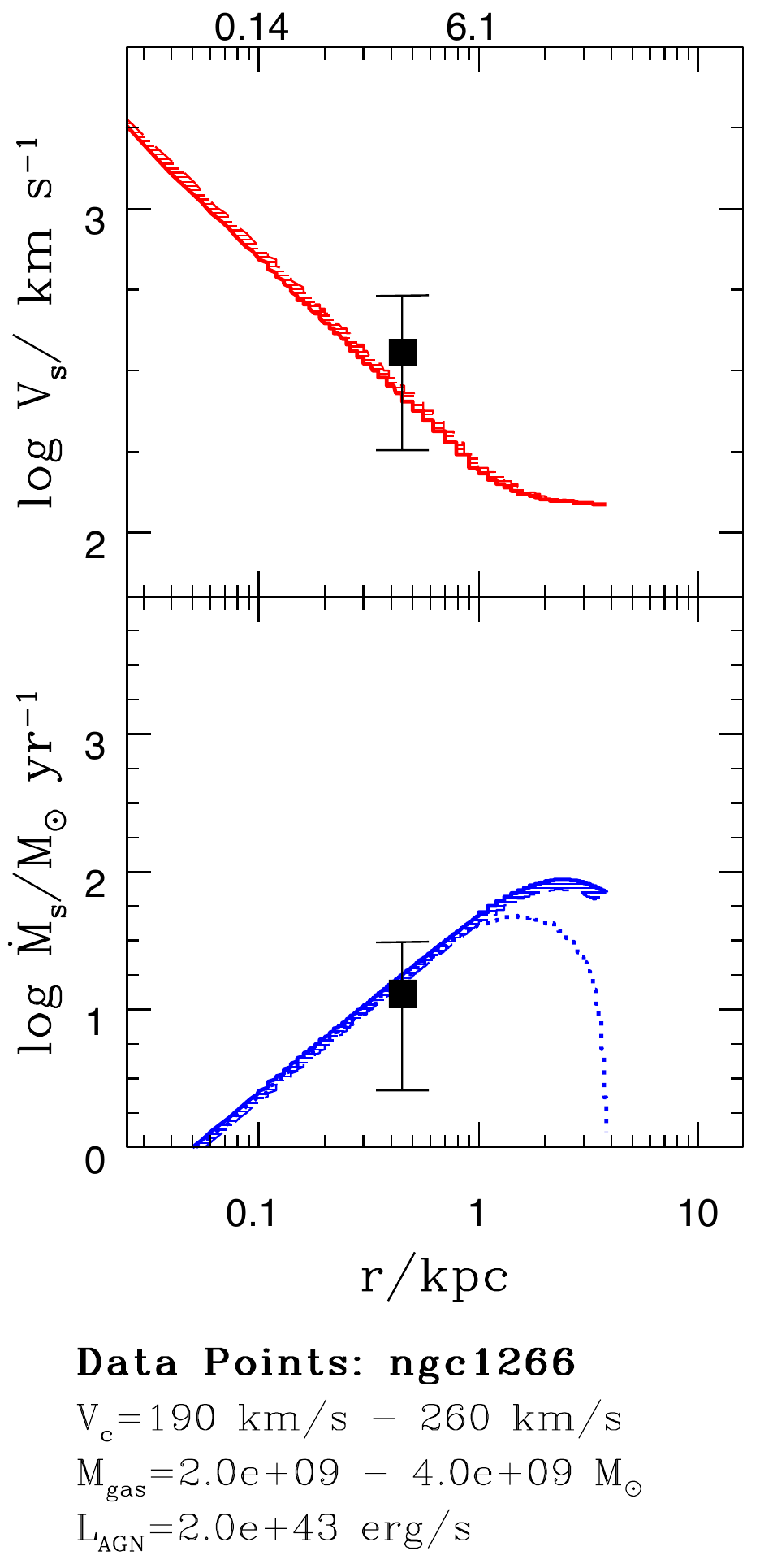}}}
\hspace{0.cm}\scalebox{0.4}[0.4]{\rotatebox{-0}{\includegraphics{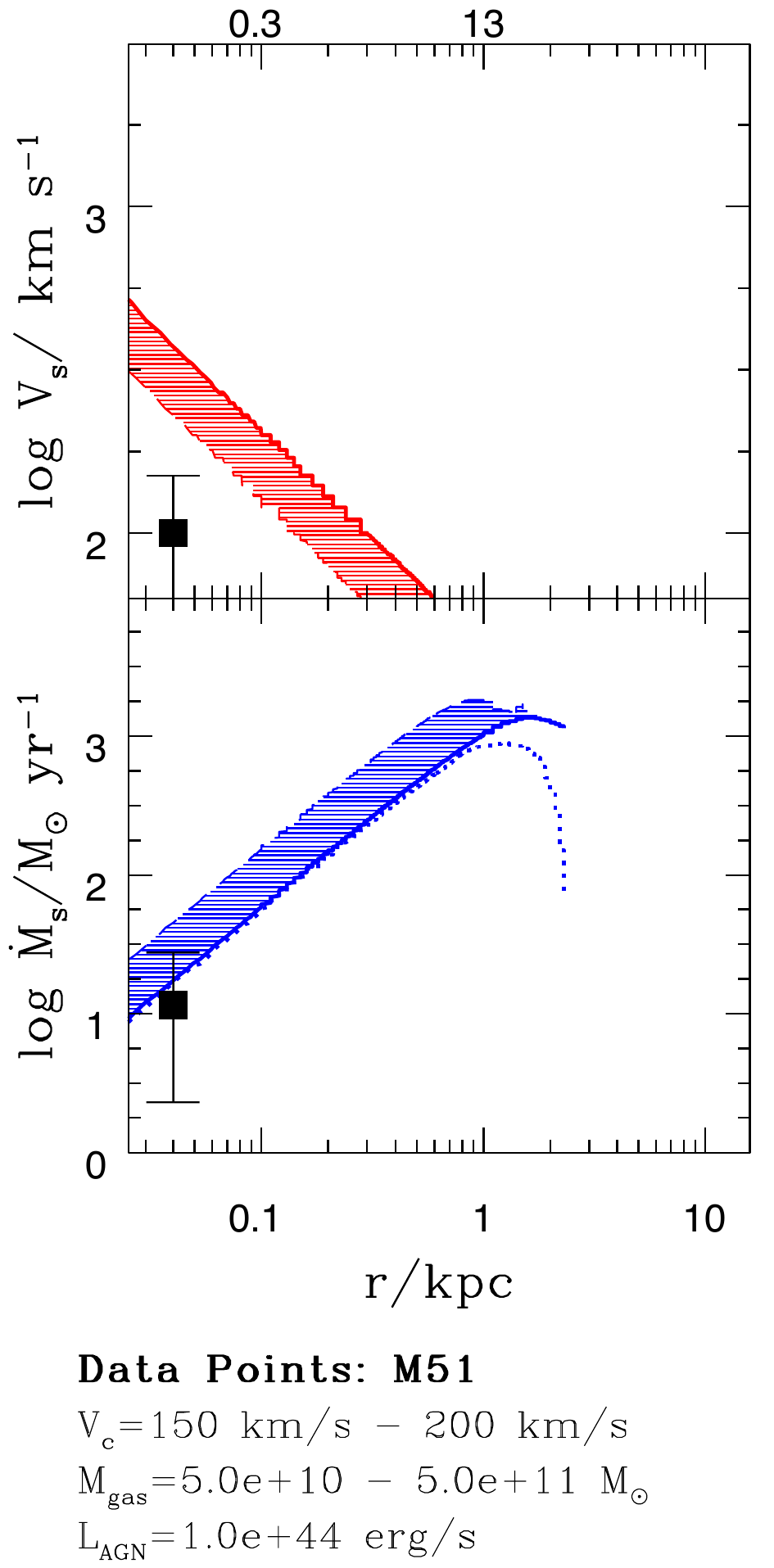}}}
\hspace{0.cm}\scalebox{0.4}[0.4]{\rotatebox{-0}{\includegraphics{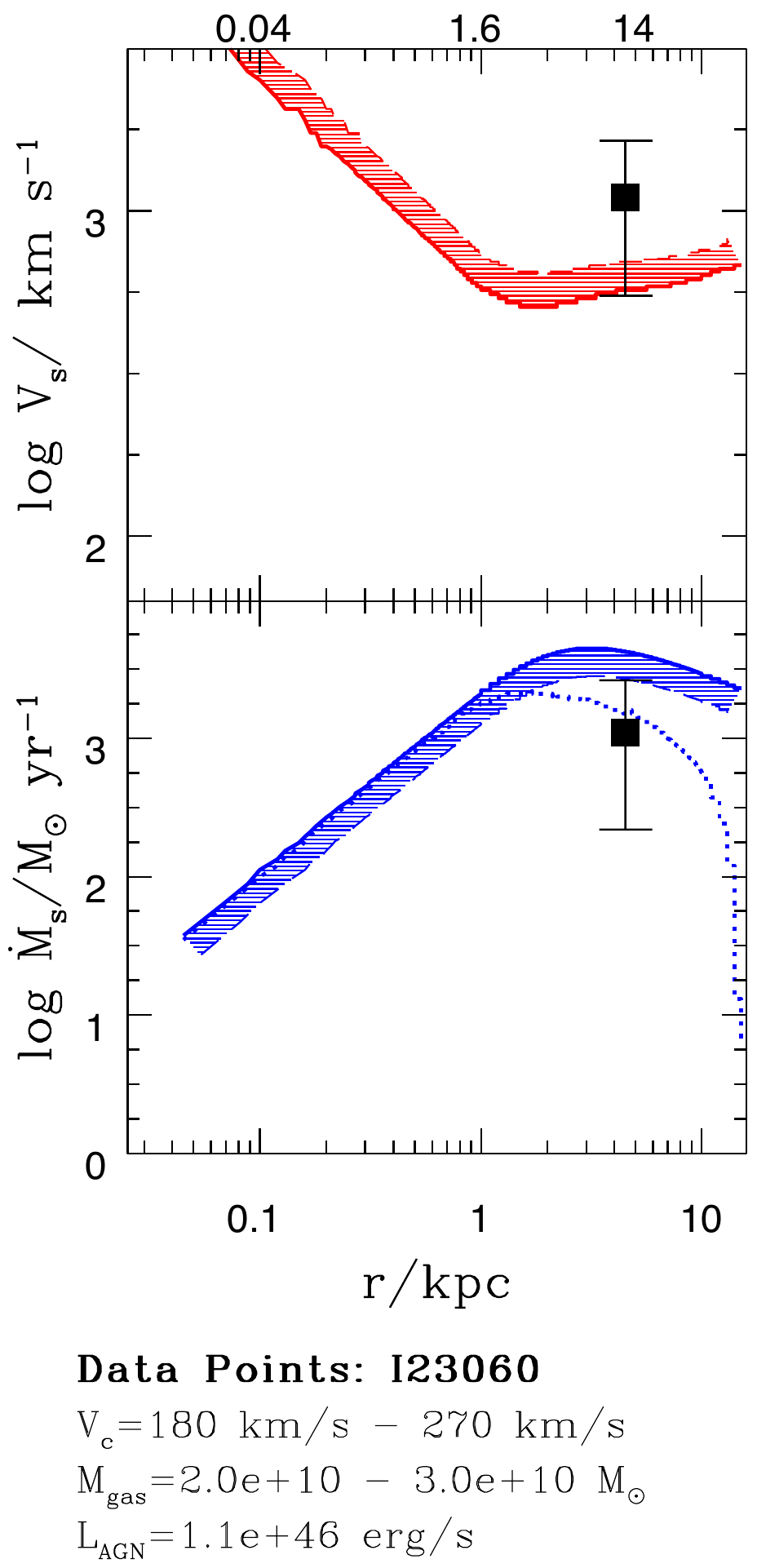}}}
\newline
\hspace{-0.8cm}\scalebox{0.4}[0.4]{\rotatebox{-0}{\includegraphics{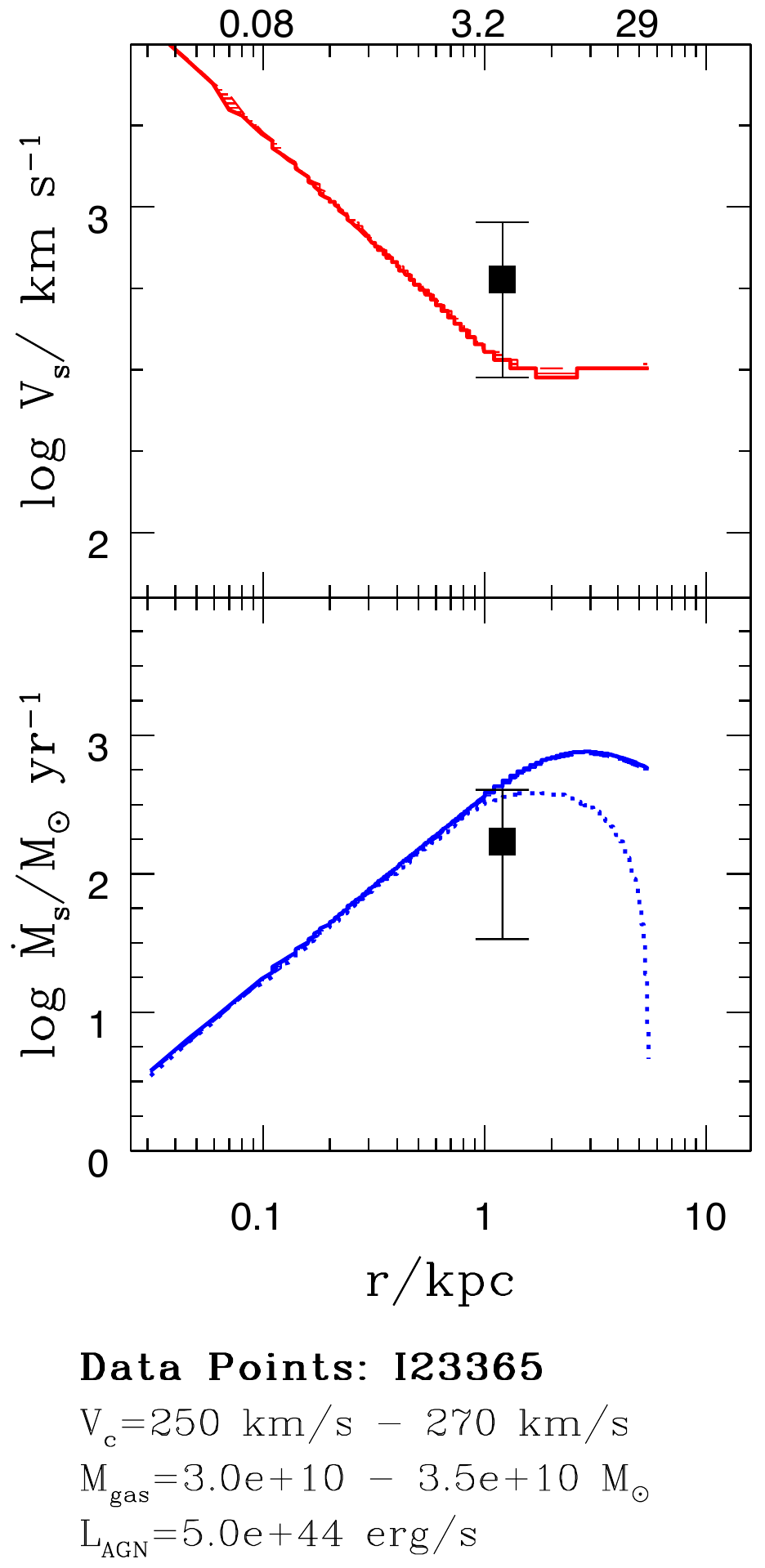}}}
\hspace{0.cm}\scalebox{0.4}[0.4]{\rotatebox{-0}{\includegraphics{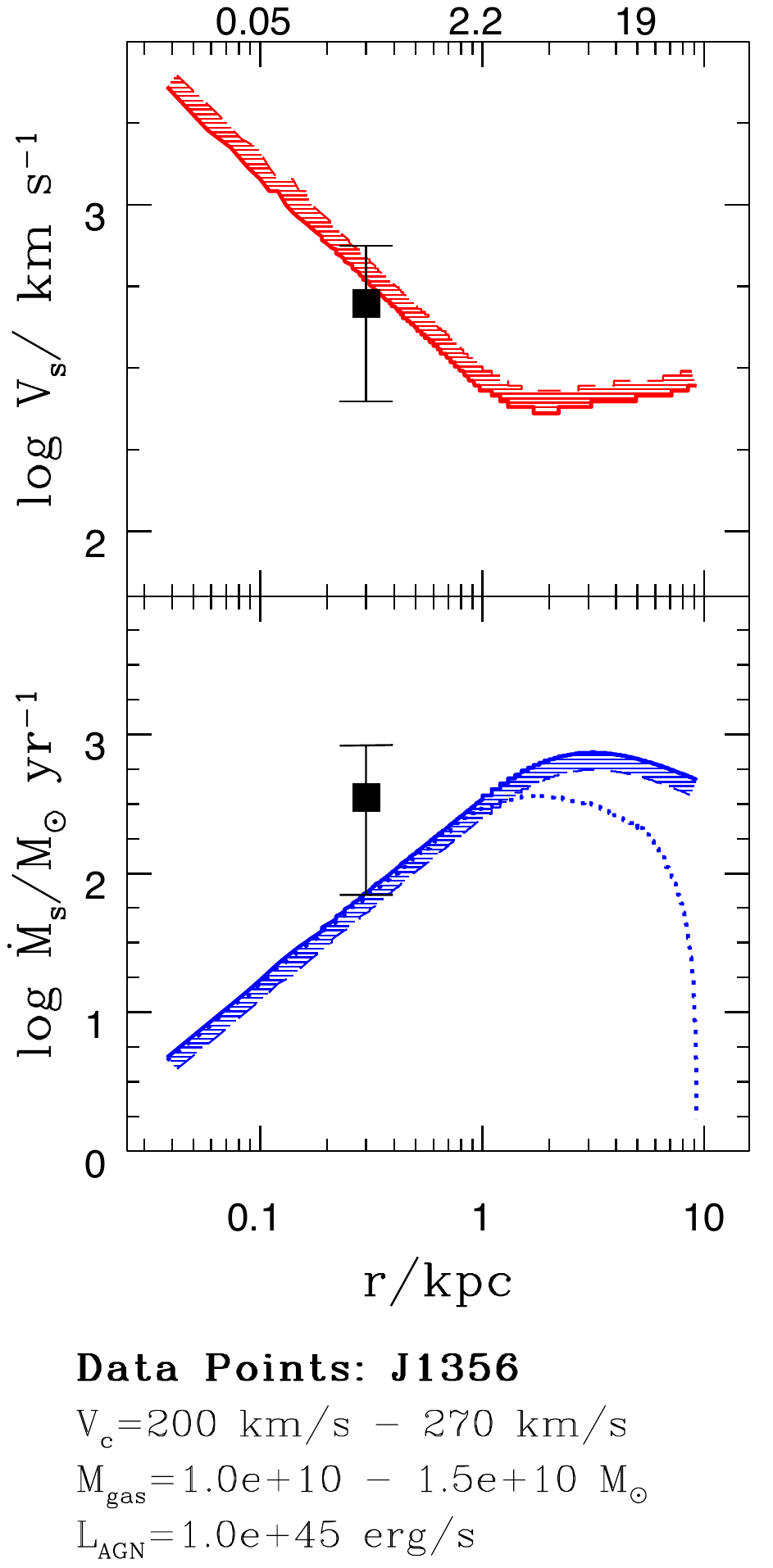}}}
\hspace{0.cm}\scalebox{0.4}[0.4]{\rotatebox{-0}{\includegraphics{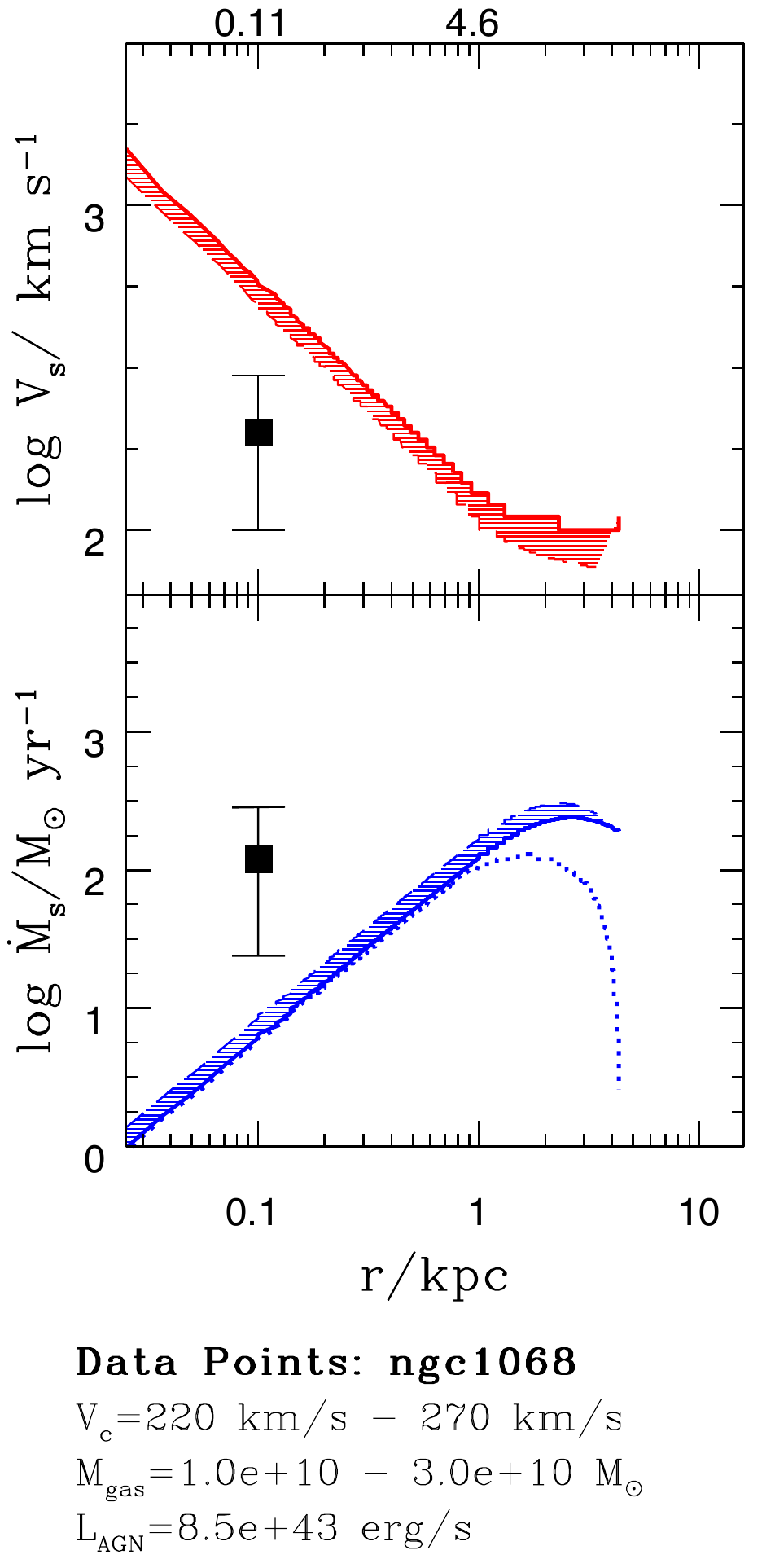}}}
\hspace{0.cm}\scalebox{0.4}[0.4]{\rotatebox{-0}{\includegraphics{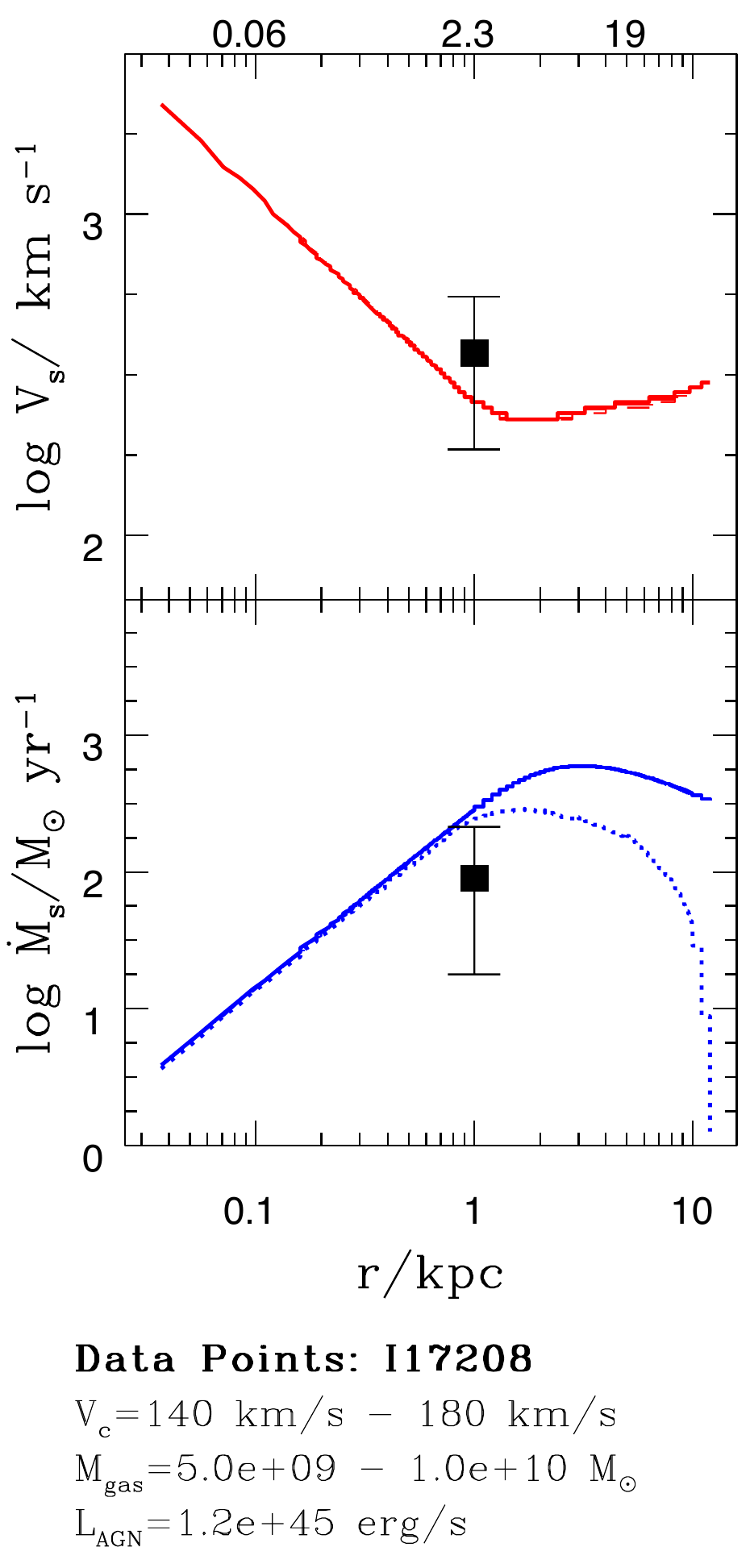}}}
\newline
\hspace{-0.8cm}\scalebox{0.4}[0.4]{\rotatebox{-0}{\includegraphics{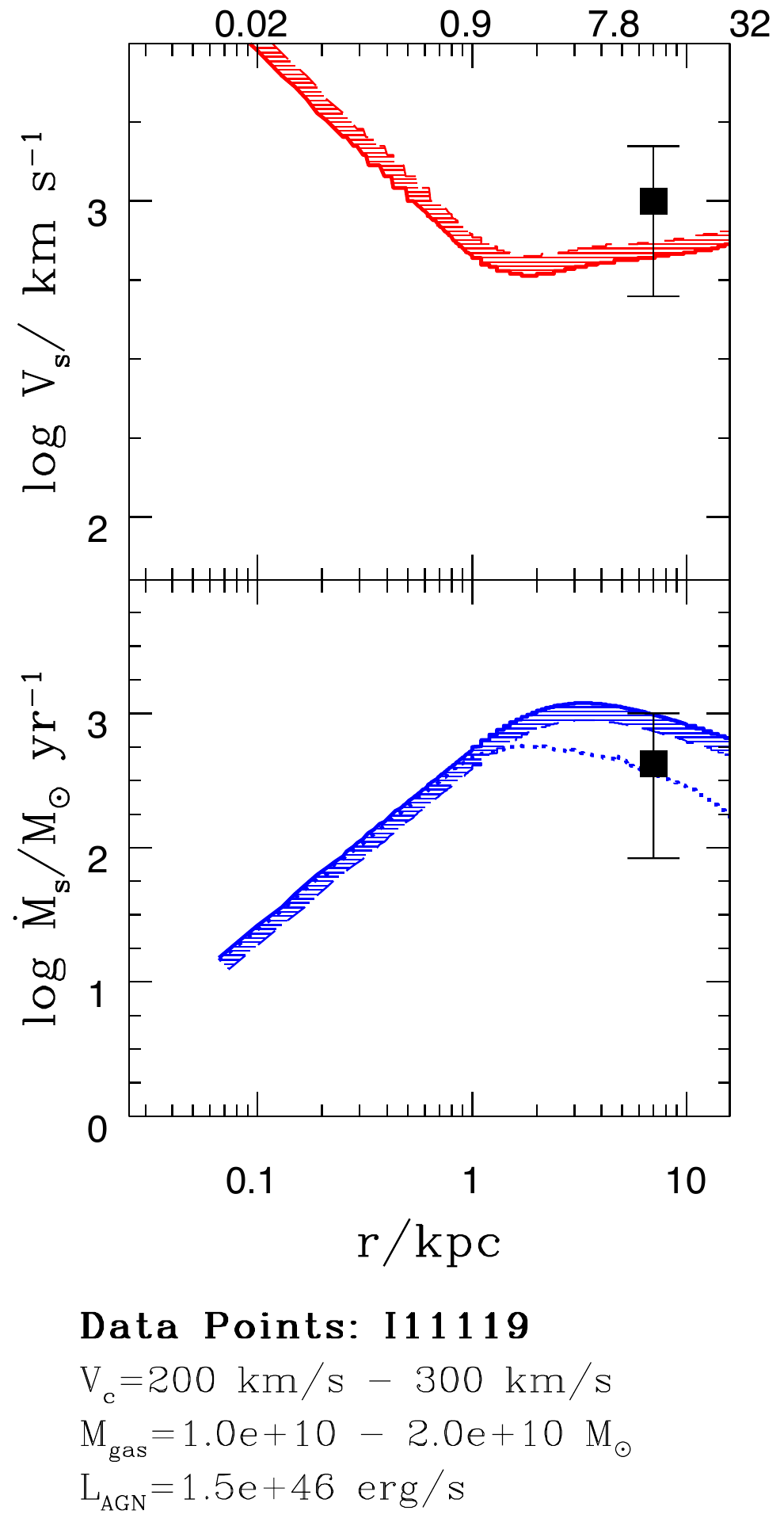}}}
\hspace{0.cm}\scalebox{0.4}[0.4]{\rotatebox{-0}{\includegraphics{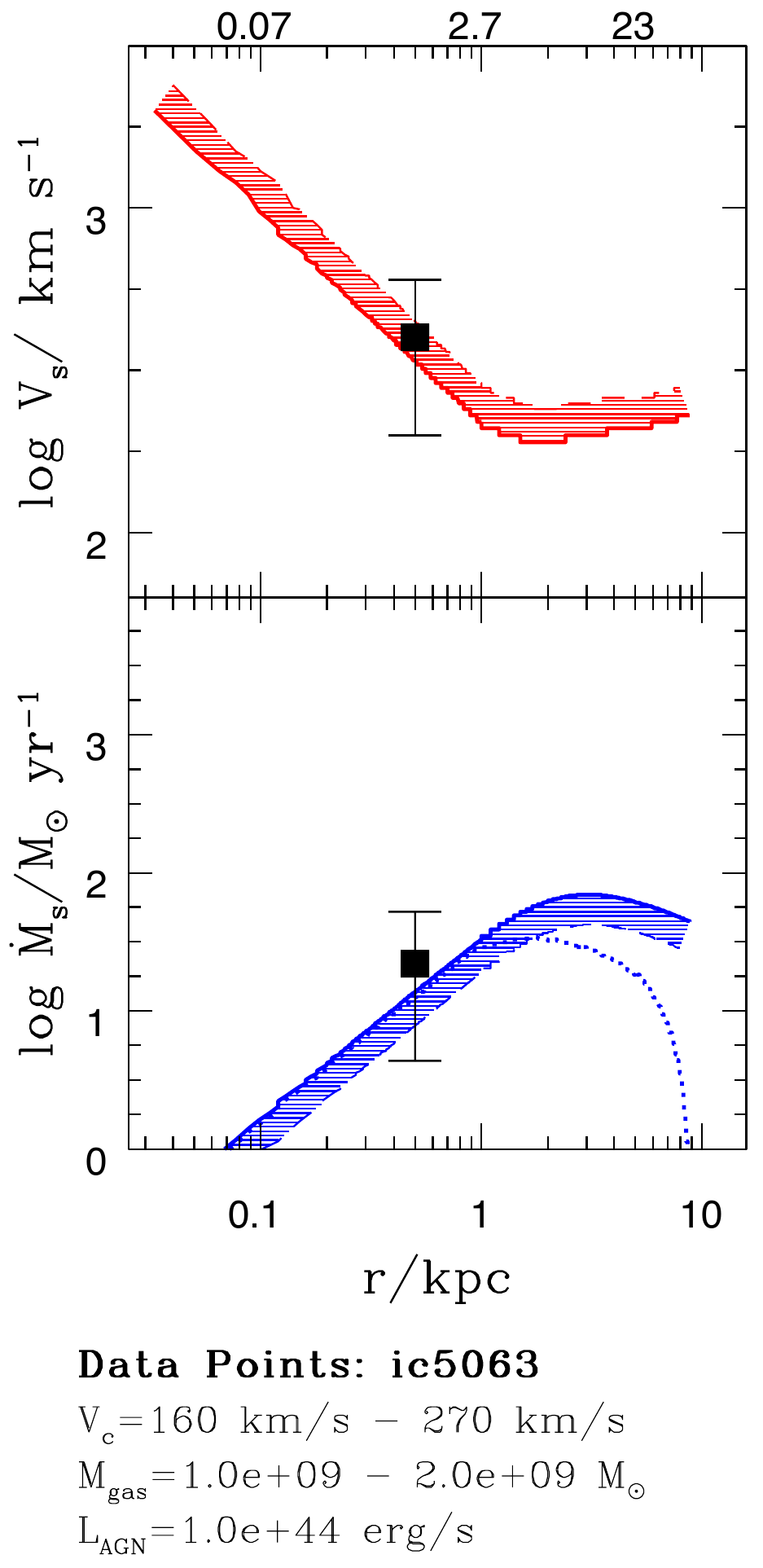}}}
\hspace{0.cm}\scalebox{0.4}[0.4]{\rotatebox{-0}{\includegraphics{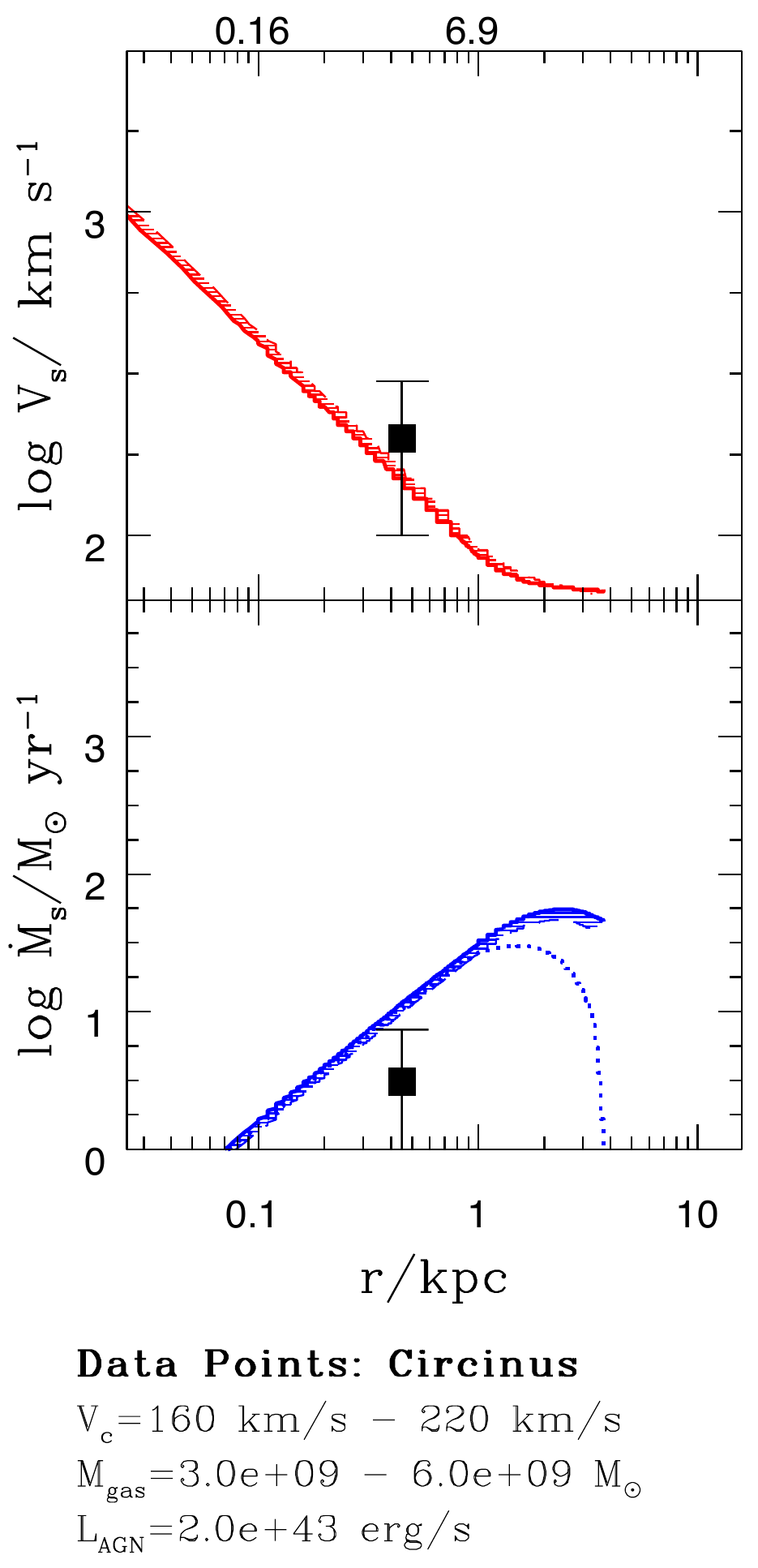}}}
\hspace{0.cm}\scalebox{0.4}[0.4]{\rotatebox{-0}{\includegraphics{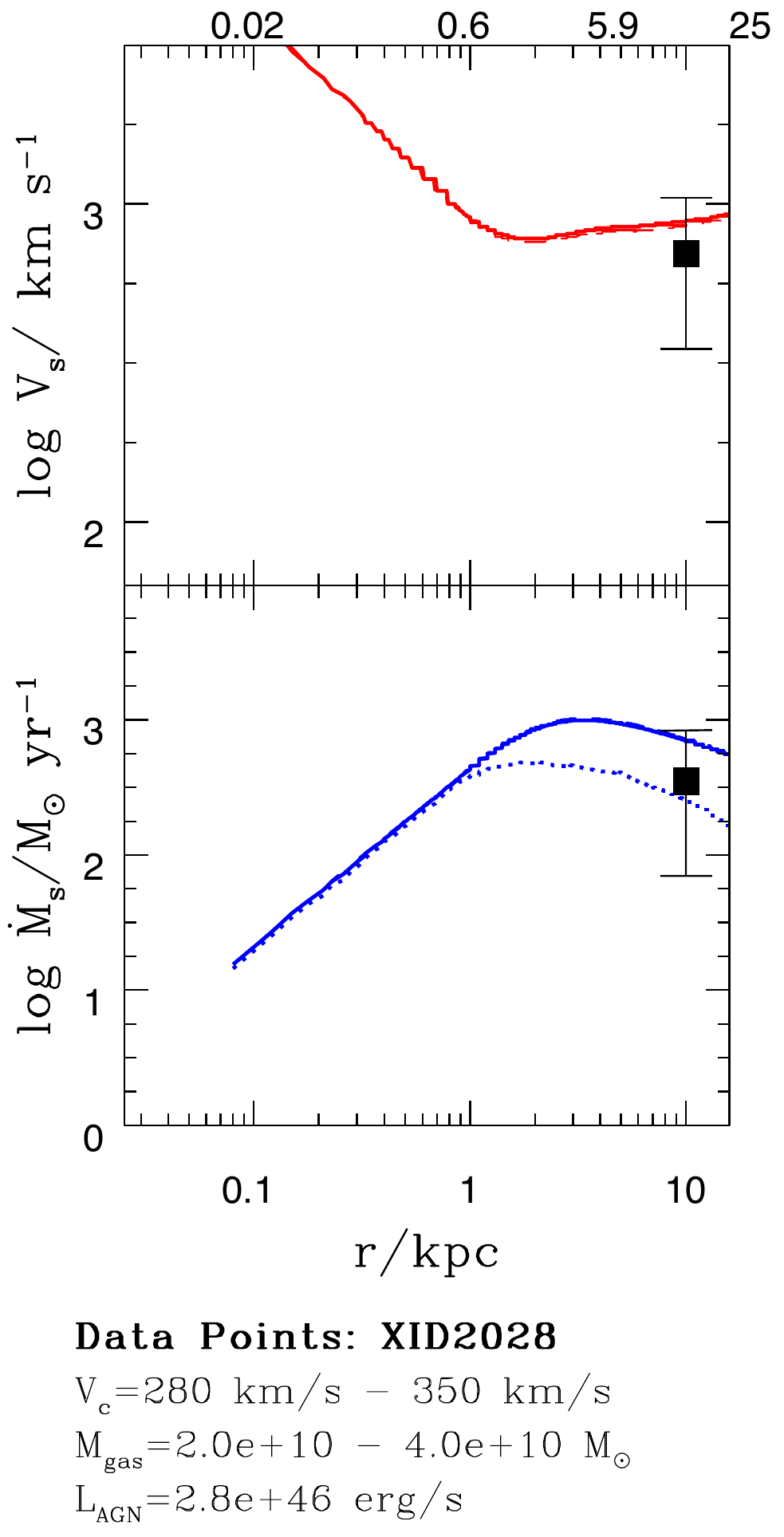}}}
\newline
\hspace{-0.8cm}\scalebox{0.4}[0.4]{\rotatebox{-0}{\includegraphics{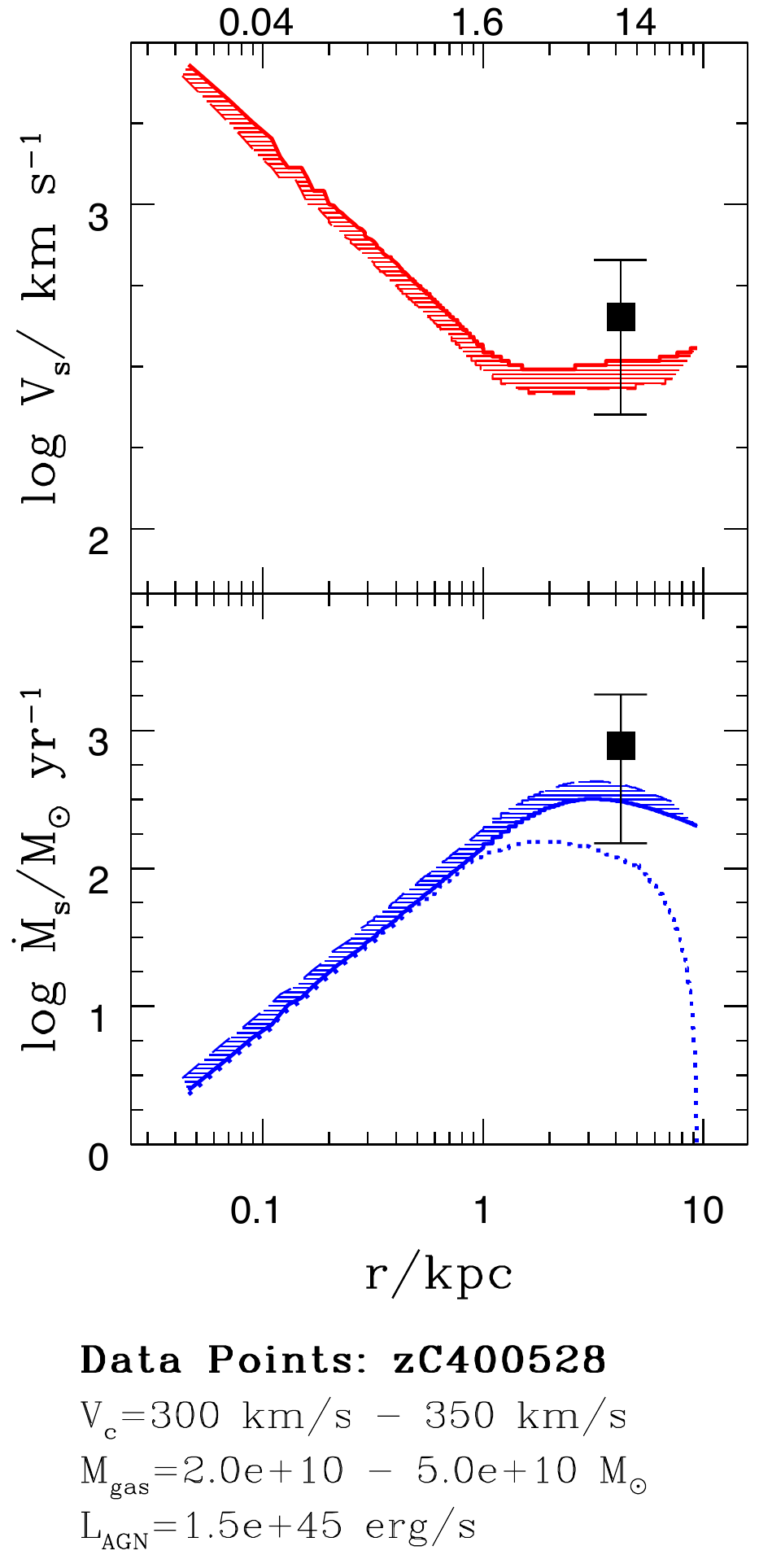}}}
\hspace{0.cm}\scalebox{0.4}[0.4]{\rotatebox{-0}{\includegraphics{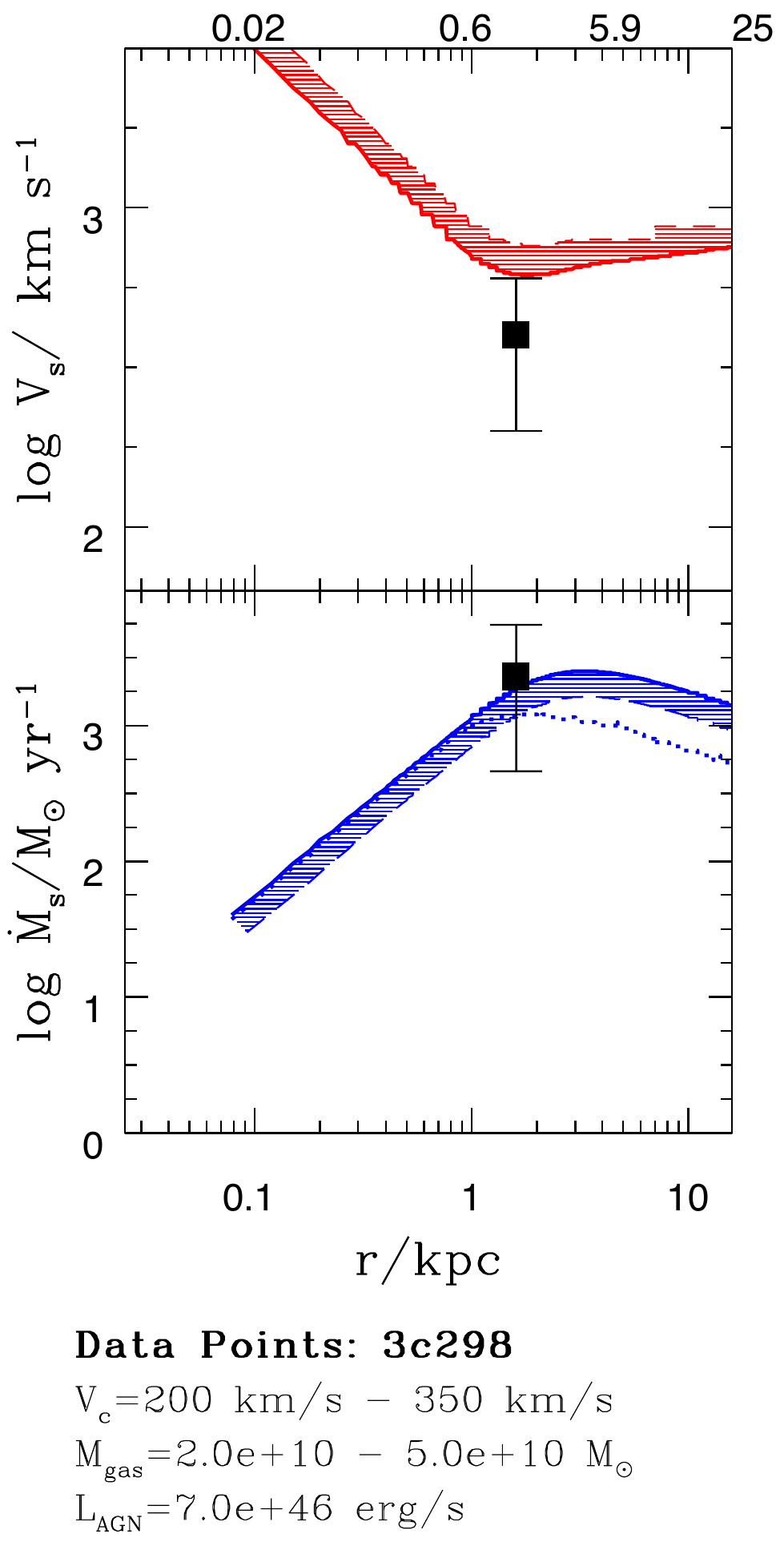}}}
\scalebox{0.4}[0.4]{\rotatebox{-0}{\includegraphics{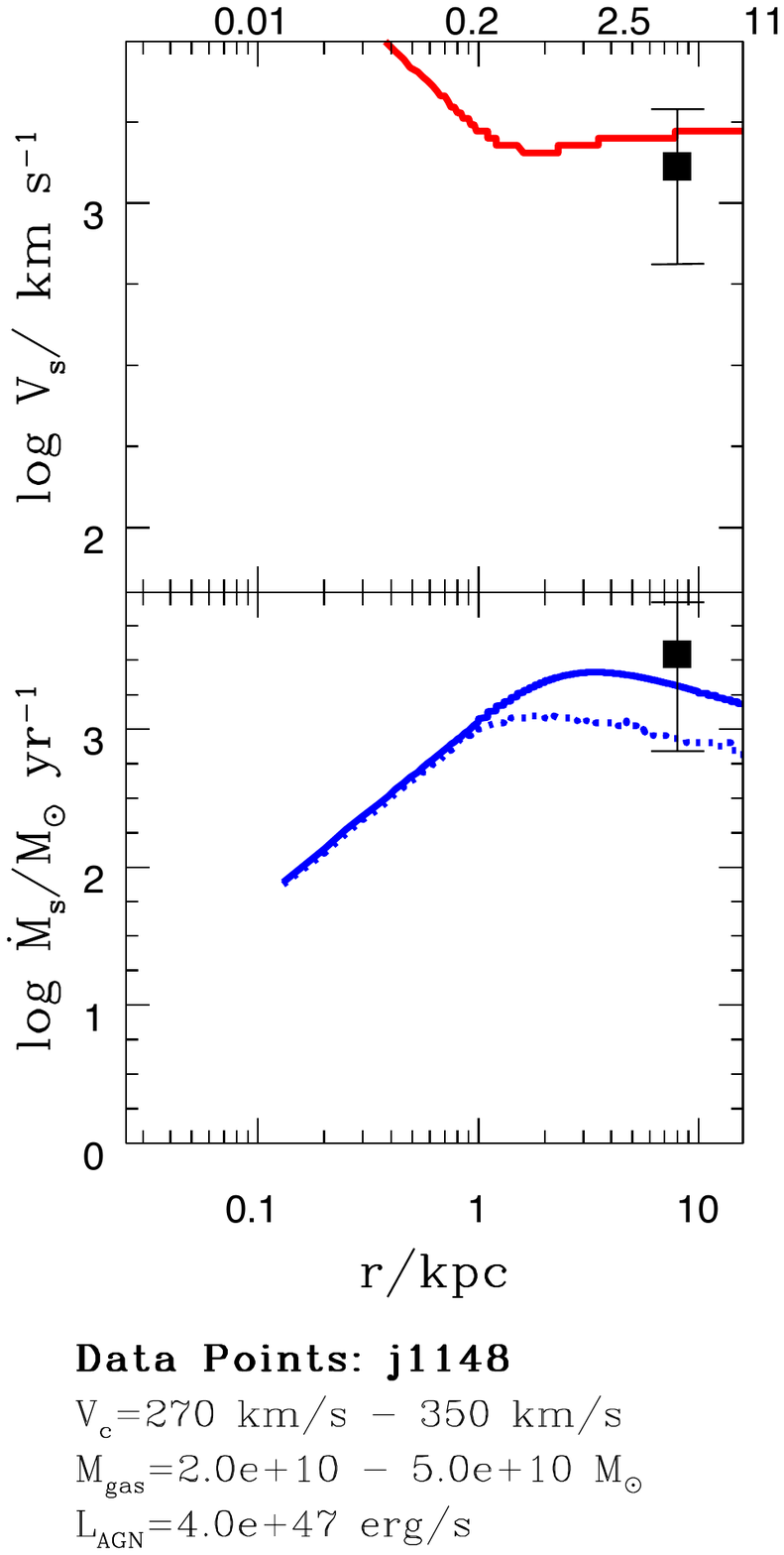}}}\hspace{1.2cm}

{\footnotesize Fig. 5. For all the objects listed in Table 1, we show the predicted shock velocity $V_{S}$ and 
mass outflow rate $\dot M_{S}$ as a function of $R_{S}$, and compare them with observation. All predicted quantities are derived from the 
full two-dimensional model after performing a mass-weighed average over their dependence on the inclination angle $\theta$ with respect to the plane of the disc (see text).
For the mass outflow rate we show both the $\dot M_S$ resulting from our full solutions (the time derivative of eq. 8, dotted line) and the value that corresponds to   $\dot M_S=M_S\,V_S/R_S$ (solid line), the definition adopted to derive the observational points. 
The data points are taken from the references in Table 1. The labels on the top axis show the time (in units of $10^{6}$ yr) corresponding to the 
shock position $R_{S}$ in the x-axis. For each object we also show the input values (derived from the left side of Table 1 as explained in the text) 
that have been used to run the model. The uncertainties in the model predictions due to the adopted range of input values are shown as a shaded region.
For SDSSJ1148 the adopted values for $M_{gas}$ and $V_c$ of the host galaxy (used as inputs for the model) are taken from 
Maiolino et al. (2012; see also Cicone et al. 2015; Fiore 2017). 
}

In the last panel of fig. 5  we present the case of SDSSJ1148 at z=6.4 where Maiolino et al. (2012) and Cicone et al. (2015) reported the detection of a massive [CII] outflow (i.e. associated with the cold gas phase of the ISM) powered by a high luminosity QSO for which the properties of the observed host galaxy (i.e., $M_{gas}$, $V_{rot}$) are known.  Even for this object, characterized by a huge AGN luminosity, by a large $R_{S}$ and by an extreme mass outflow rate, the agreement with the model predictions is excellent. In this case we also obtain a predicted density $n_{S}$ of the shocked gas shell  lower than the critical density for the molecular CO emission, as expected in the case of ionized outflows.

\subsection{Comparison with high-luminosity, ionized outflows}

For all the other ionized outflows present in the literature, detailed measurements of the host galaxy 
$M_{gas}$ and $V_{rot}$ are not available. In Table 2 we report the values of $V_{S}$ and $\dot M_{S}$  for a large sample of objects taken from  
Fiore et al. (2017), where objects are listed in order of increasing bolometric luminosity $L_{AGN}$. From the sample in Fiore et al. (2017) we have excluded I10565 due to the ambiguous interpretation of the outflow (possible earlier bubbles due to previous ejection episodes, see Rupke \& Veilleux 2013), mark231 (the analysis of Rupke \& Veilleux 2013 excludes  a part of the nuclear emission), and J1339 (its identification with an AGN is uncertain, see Harrison et al. 2014). 

Since the properties of the observed host galaxies are not available, we cannot perform a detailed on-by-one comparison with the model as we did for molecular outflows. Thus, we  have divided the observed AGN luminosity range in different bins. For each luminosity bin, the input quantities for the model, i.e., the total mass $M$ (or the circular velocity $V_c$) and the gas mass $M_{gas}$ of the host galaxy, are related to $L_{AGN}$ through available average scaling relations. The total mass $M$, we adopt the central value of the relation $log(M_{BH}/M_{\odot}) = (1.55 \pm 0.02)\,log(M/M_{\odot})$ −- $(11.26 \pm 0.20)$ found from analysis of the Illustris N-body simulations by Mutlu-Pakdil et al. (2018) 
such a relation is consistent with a wide set of observational data (see references in the above paper). An average $M-L_{AGN}$ relation is then derived 
after converting the black hole mass in bolometric luminosity assuming Eddington emission; assuming an Eddington ratio peaked at 0.3 (as indicated by some observations, see, e.g., Kauffmann \& Heckman 2009; Shankar, Weinberg, Miralda-Escud\'e  2013) does not change appreciably our results. Although, observationally, the large scatter of the relation at small galaxy masses makes the  correlation weak  (see, e.g., Kormendy \& Bender 2011; Sabra et al. 2015), the scatter reduces appreciably for large halos with $V_c\gtrsim 200$ km/s  and large black hole masses $M_{BH}\gtrsim 10^{8}\,M_{\odot}$ like those corresponding to the objects in Table 2. To derive the other input quantity $M_{gas}$ we use the relation $log(M_{gas}/M)\approx -2-[[log(M/M_{\odot})-12]$ approximating the scaling found from abundance matching by Popping, Behroozi, Peebles (2015) for galaxies with $12\leq log M/M_{\odot}\leq 13$, the range covered by the masses corresponding to the luminosities in Table 2.  With such an approximation, 
the gas mass stays constant to  $M_{gas}\approx 10^{10}\,M_{\odot}$ over the whole interval of interest $M=10^{11}-10^{13}\,M_{\odot}$ (corresponding to $V_c=150-400$ km/s). 

\begin{table}[]
\caption{Sample of Observed ionized outflows: objects are sorted by increasing AGN luminosity}
 \footnotesize
\label{my-label}
\begin{tabular}{l c c  c c c l}
\hline
\hline
Object    & Redshift & \footnotesize $L_{AGN}$ {[$10^{45}$ erg/s] }  & $R_{S}$   [kpc]  & $V_{S}$ [km/s]   & $\dot M_{S}$  {[}$M_{\odot}$/yr{]}    & Ref.   \\ \hline
SDSSJ0958   &   0.10   &    45.0 	&    2.6		& 	866		&  1.1  & 1    \\ 	
SDSSJ1356   &   0.12   &    45.1 	&    3.1		&	1049	&  1.6  & 1   \\ 	 
SDSSJ1130   &   0.13   &    45.1 	&    2.8		&	616     &  0.3  & 1    \\   	
SDSSJ1125   &   0.17   &    45.2 	&    2.9		&	1547    &  0.75 & 1  	\\	
SDSSJ1430   &   0.08   &    45.3    &	 1.8		&	999	    &  1.7  & 1     \\ 	
SDSSJ1316   &   0.15   &    45.4 	&    3.1 		&	1216	&  1.48 & 1 	\\
SDSSJ0945   &   0.13   &    45.5 	&    2.7		&	1511    &  1.62 & 1   \\ 	
SDSSJ10100  &   0.10   &    45.6    &	 1.6		&	1267    &  1.46 & 1   \\ 	
GS3-19791   &   2.22   &    45.6    &	 1.3		&	530		&  3.23 & 2    \\ 	
SDSSJ1000   &   0.15   &    45.7 	&    4.3		&	761     &  1.16 & 1    \\   	
SDSSJ1355   &   0.15   &    45.7 	&    3.5		&	797		&  0.57 & 1    \\ 	
GS3-28008   &   2.29   &    45.9	&    1.3		&	300		&  2.34 & 2   \\ 	
XID5395     &   1.47   &    45.9 	&    4.3		&	1600	&  2.65	& 4	\\
SDSSJ10101  &   0.20   &      46 	&    3.9		&	1523    &  1.82 & 1   \\ 	
SDSSJ1100   &   0.10   &      46 	&    1.9		&	1192    &  1.65 & 1    \\
SDSSJ0210   &   0.54   &    46.1 	&    7.5		&	560		&  2.62	& 5, 6, 7, 8	\\
SDSSJ1040   &   0.49   &    46.2 	&    7.6		&	1821	&  3.16	& 5, 6, 7, 8		\\
COS11363    &   2.10   &    46.2 	&    1.3		&	1240   	&  2.83 & 2    \\ 	
SMMJ1636    &   2.38   &    46.3 	&    7			&	1054    &  1.44 & 9   \\	
MRC0406     &   2.44   &    46.3 	&    9.3		&	960		&  3.82	& 10	\\
XID5321     &   1.47   &    46.3 	&    11			&	1950	&  1.84	& 11, 12	\\
RGJ0302     &   2.24   &    46.3 	&    8			&	1234    &  1.48 & 9   \\	
SDSSJ0319   &   0.62   &    46.4 	&    7.5		&	934		&  2.32	&  5, 6, 7, 8		\\
SDSSJ0321   &   0.64   &    46.5 	&    11			&	946		&  2.30	&  5, 6, 7, 8		\\
SDSSJ0841   &   0.64   &    46.5 	&    6.4		&	675		&  2.60	& 5, 6, 7, 8		\\
MIRO20581   &   2.45   &    46.6 	&    4.8		&	1900	&  2.29	& 13\\
MRC1138     &   2.20   &    46.6 	&    20			&	800		&  2.39	& 14\\
MRC0828     &   2.57   &    46.6 	&    9			&	800		&  3.87	& 10	\\
SMMJ1237    &   2.06   &    46.7 	&    7			&	1200    &  1.48 & 9    \\	
SMMJ0943    &   3.35   &    46.7 	&    15			&	1124    &  1.57 & 9   	\\
SDSSJ0842   &   0.56   &    46.8 	&    9.			&	522		&  2.59	& 5, 6, 7, 8		\\
HB8905      &   2.48   &    46.8 	&    1.3		&	500		&  2.65 & 9   \\
SDSSJ1039   &   0.58   &    46.9 	&    5.8		&	1046	&  2.81	& 5, 6, 7, 8		\\
SDSSJ0149   &   0.57   &    46.9 	&    4.1		&	1191	&  2.60	& 5, 6, 7, 8		\\
SDSSJ0858   &   0.45   &    47.2 	&    5.6		&	939		&  2.79 & 5, 6, 7, 8		\\	
HB8903      &   2.44   &    47.3 	&    1.9     	&	1450 	&  1.76 & 13 	\\
SDSSJ0759   &   0.65   &    47.3 	&    7.5		&	1250	&  2.87	& 5, 6, 7, 8		\\
HE0109      &   2.40   &    47.4 	&    0.4		&	900		&  3.14	& 5, 6, 7, 8		\\
LBQS0109    &   2.35   &    47.4 	&    0.4		&	1850	&  2.84	& 13	\\
SDSSJ1326   &   3.30   &    47.6	&  	 7			&	2160	&  3.81 & 14, 15   \\	
SDSSJ1201   &   3.51   &    47.7 	&    7			&	1850	&  3.50 & 14, 15     \\ 	
SDSSJ1549   &   3.30   &    47.8 	&    7			&	1380	&  3.42 & 14, 15     \\	
SDSSJ0900   &   3.30   &    47.9 	&    7			&	2380	&  3.52 & 14, 15    \\ 	
SDSSJ0745   &   3.22   &    48.0 	&    7			&	1890	&  3.76 & 14, 15    \\ 	
\end{tabular}

{\scriptsize Ref. 1 =Harrison et al. (2014); 2 = Genzel et al. (2014) , assuming $H_{\alpha} /H_{\beta}$  = 2.9, extinction corrected; 3 = Cicone et al. (2014), 4= Brusa et al. (2016); 5 = Liu et al. (2013a); 6 = Liu et al. (2013b), extinction corrected; 7 = Wylezalek et al. (2016); 8 = Reyes et al. (2008); 
9 = Harrison et al. (2012); 10 = Nesvadba et al. (2008); 11 = Brusa et al. (2015a); 12 = Perna et al. (2015a); 13 = Perna et al. (2015b); 14 = Nesvadba et al. (2006); 
13= Carniani et al. (2015); 14 = Bischetti et al. (2017); 15 = Duras et al. (2017)}
\end{table}

With the above approximations for the input values of $M$ and $M_{gas}$, we computed the model predictions for the outflow velocity $V_{S}$, outflow mass rate $\dot M_{S}$, and shocked gas density  $n_{S}$ for different bins of $L_{AGN}$, and compare with the observed values taken from Table 2 for each $L_{AGN}$ bin.  The results are shown in fig. 6 for AGN  luminosities ranging from 
 $10^{45}$ erg$s^{-1}$ to $10^{48}$ erg$s^{-1}$. To account for uncertainties in the input values of $M_{gas}$ derived from the scaling law in Popping, Behroozi, Peebles (2015),  in each bin we show the effect of assuming an input value of $M_{gas}$ differing from the average relation by a factor three above and below the mean value. In all panels the model predictions are computed at $z=0$. Since we cannot perform a one-by-one comparison with data points, the model predictions are all computed at $z=0$. However, objects at high redshifts are more compact (see, e.g., Mo, Mao  \& White 1998), and all sizes are expected to scale accordingly. Thus, to compare with data corresponding to objects with different redshifts in the same plot, we have rescaled all the observed sizes to $z=0$ according to the expected evolution of the disc radius $r_d(z)/r_d(z=0)=(\Omega_{\Lambda}+\Omega_0\,(1+z)^{3})^{-0.5}$ (Mo, Mao \& White 1998, with density parameters $\Omega_{\Lambda}=0.7$ and $\Omega_{0}=0.3$ for the dark energy and matter, respectively). 
 
Within the unavoidable uncertainties due to the derivation of the input quantities from the above average relations, the model predictions are in general in agreement with the observations, the agreement becoming excellent for the highest luminosity bins.  Also, for all objects the predicted shocked gas density  is below the value required for the CO emission, as appropriate for 
ionized outflows. Notice that in the vast majority  of cases (although not in all them), the shocked gas has reached the cooling radius at the observed shock position, so the expected temperatures 
for the shocked shell are $T_{S}\sim 10^{4}$ K. Nevertheless, the large AGN luminosities push the cooling radius to the outer regions, where the lower gas densities yield values for $n_{S}$ smaller than 
the threshold for CO emission. It is also noticeable the huge range spanned by the model predictions when $L_{AGN}$ is changed, with mass outflow rates ranging from $\dot M_{S}\sim 10\,M_{\odot}$ yr$^{-1}$ for the lowest luminosity bin to $\dot M_{S}\sim 10^{3}\,M_{\odot}$ yr$^{-1}$ for $L_{AGN}\sim 10^{48}$ erg/s. The agreement of model predictions with observations over such a large range of input and output quantities provides  a strong support to the model predictions. On the other hand, in the lowest luminosity bin $45\leq log L_{AGN}/erg\,s^{-1}\leq 45.4$ , the model over-predicts the mass outflow rates. However, as discussed in Fiore et al. (2017), measured ionized mass outflow in low-luminosity objects are likely to represent only a fraction of the total mass outflow. In particular, from the comparison with model predictions we  expects a correction factor $\sim 10-50$ in this luminosity 
 range. Thus, observational determinations of the fraction of mass outflow in ionized winds in low-luminosity objects will constitute an important  consistency check for the model predictions in this regime. 

\hspace{-0.8cm}
\scalebox{0.38}[0.38]{\rotatebox{-0}{\includegraphics{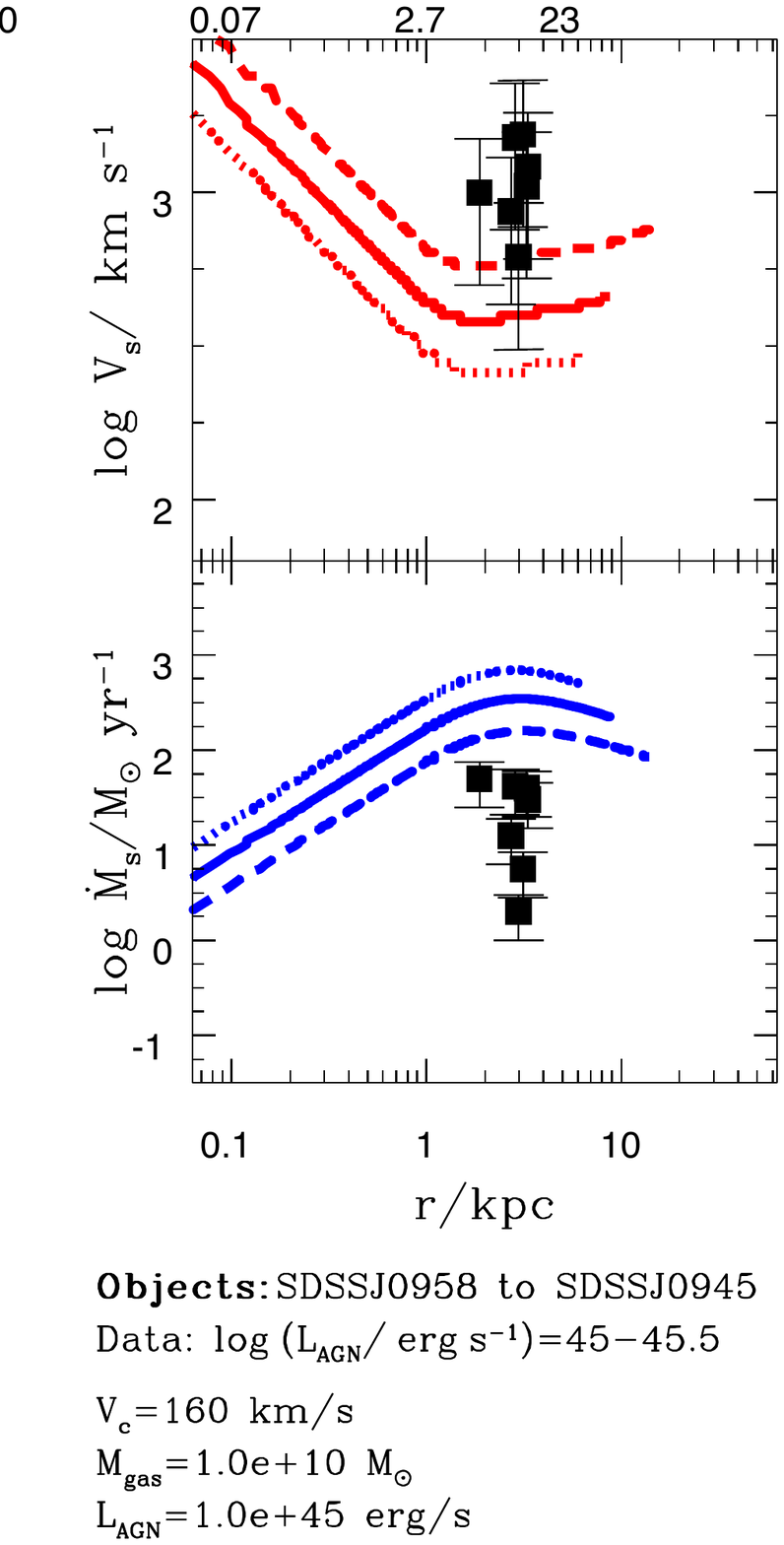}}}
\scalebox{0.38}[0.38]{\rotatebox{-0}{\includegraphics{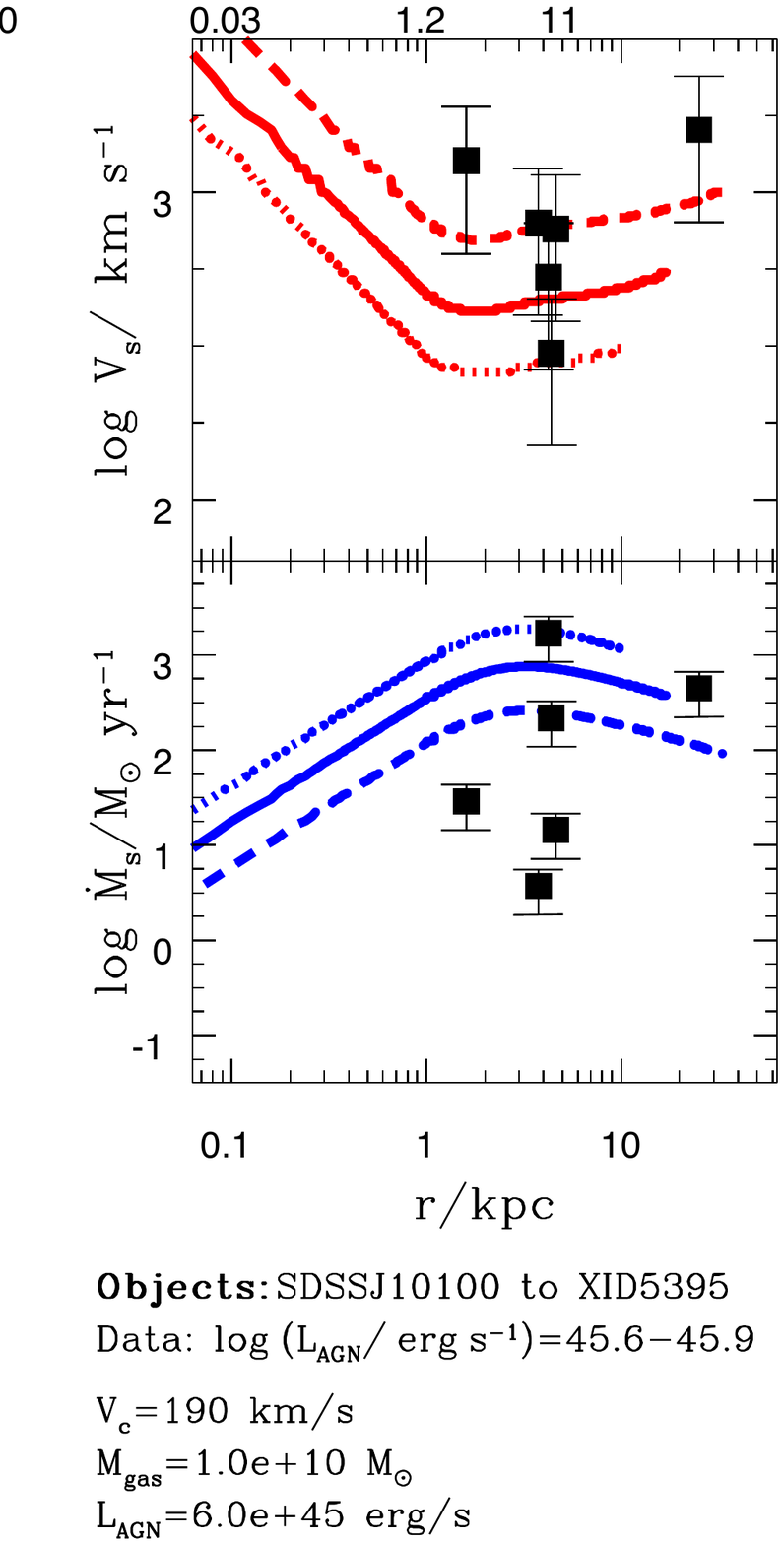}}}
\scalebox{0.38}[0.38]{\rotatebox{-0}{\includegraphics{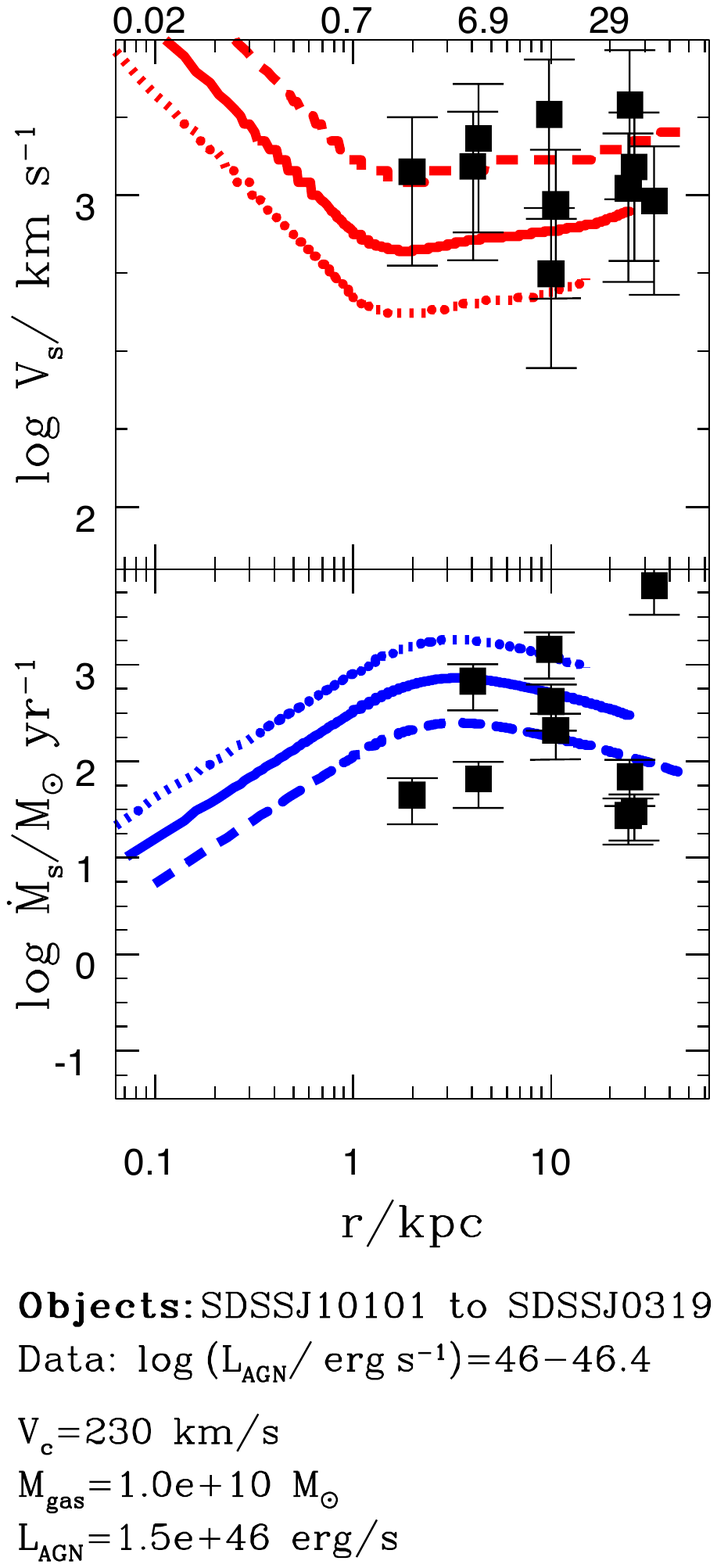}}}
\newline
\scalebox{0.38}[0.38]{\rotatebox{-0}{\includegraphics{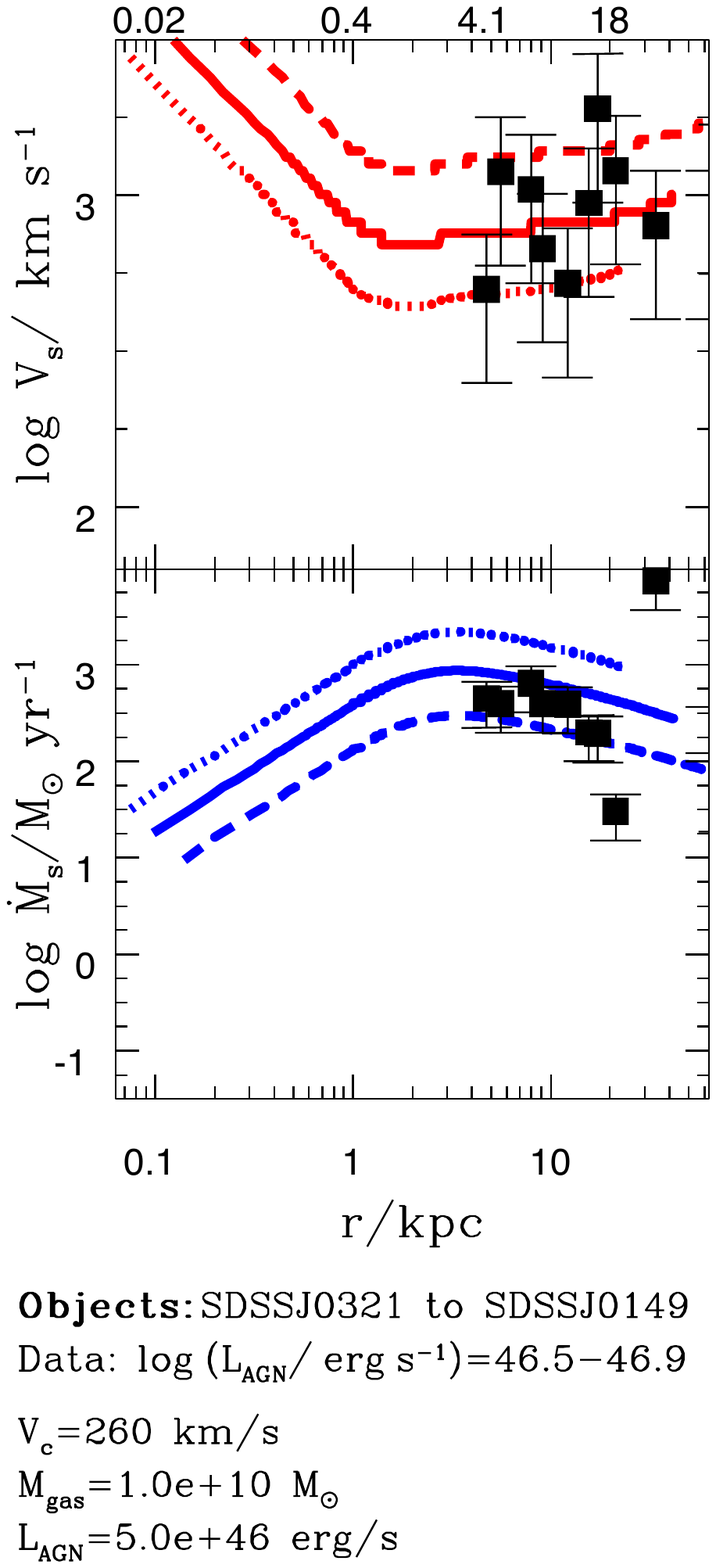}}}
\scalebox{0.38}[0.38]{\rotatebox{-0}{\includegraphics{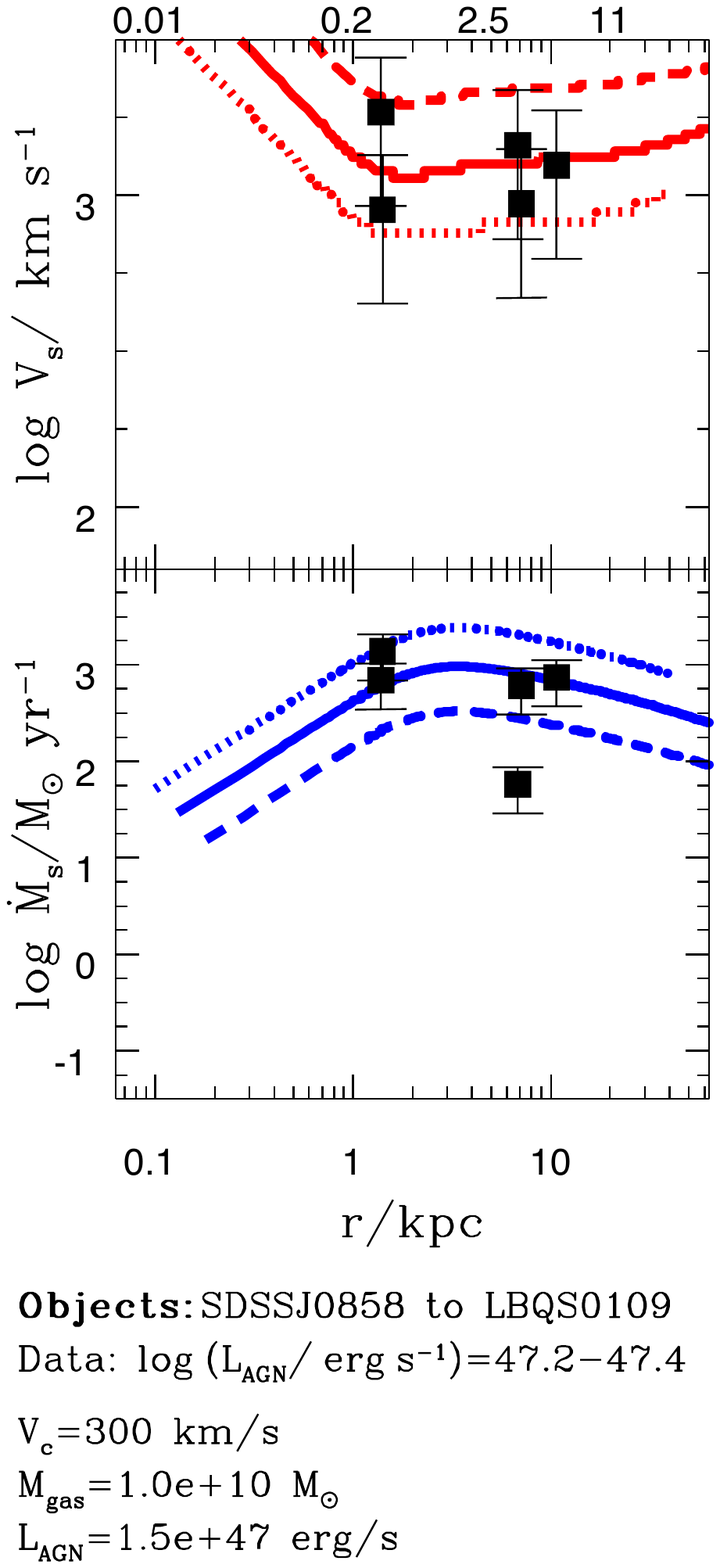}}}
\scalebox{0.38}[0.38]{\rotatebox{-0}{\includegraphics{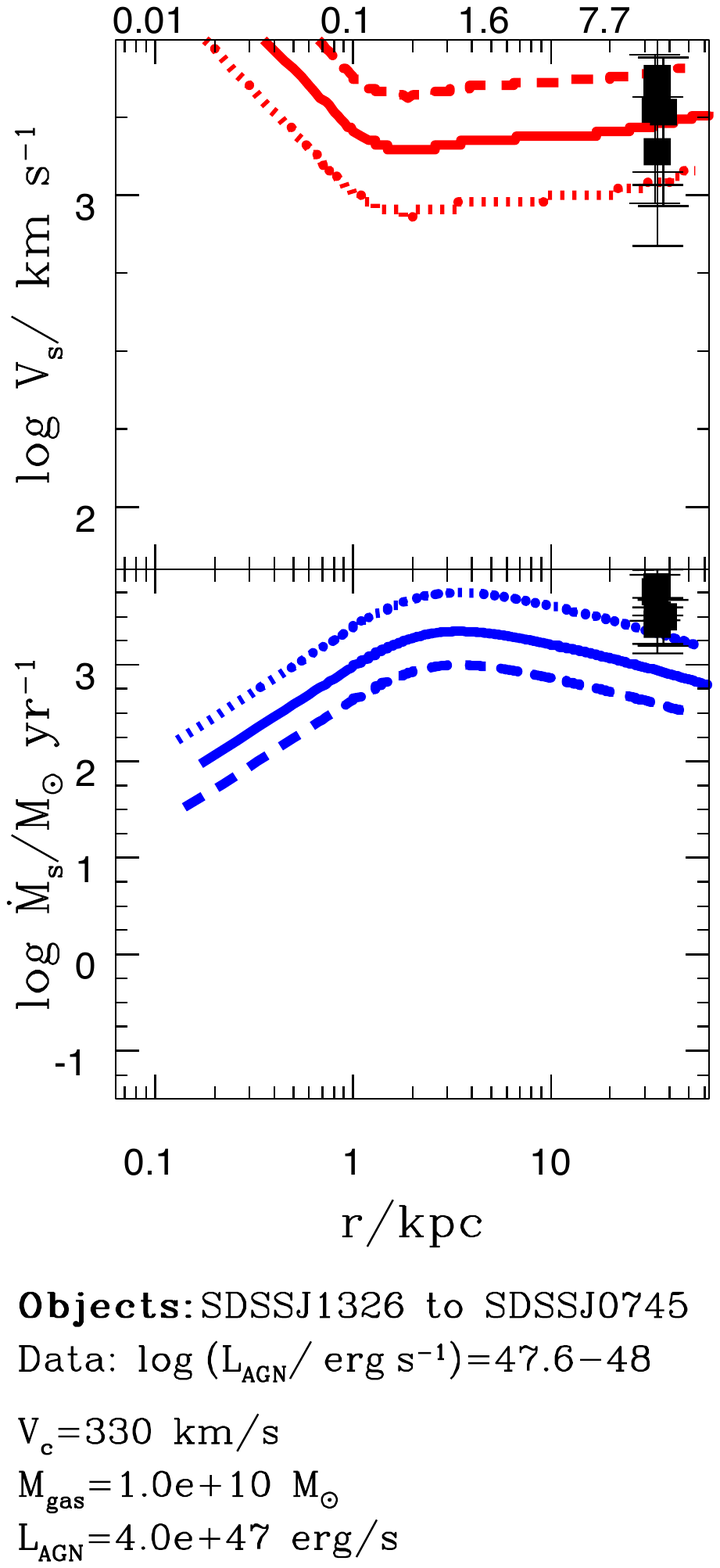}}}
\vspace{0.2cm }
\newline
{\footnotesize Fig. 6. Predicted shock velocity $V_{S}$ and 
mass outflow rate $\dot M_{S}$ as a function of $R_{S}$ for the different AGN luminosities shown in the legend. 
All quantities are derived from the full two-dimensional model after averaging over their dependence on the inclination angle $\theta$ with respect to the plane of the disc. The circular velocity $V_c$ used in the computation is derived as explained in the text and shown in the legends. The input gas mass is  $M_{gas}=10^{10}\,M_{\odot}$ (solid line); the results for $M_{gas}=3\,10^{10}\,M_{\odot}$ and $M_{gas}=3\,10^{9}\,M_{\odot}$ are shown as dotted and dashed lines, respectively. The results are compared with the
observed ionized outflows shown in the legends (the object names are referred to Table 2 where they are sorted by increasing AGN luminosity), with the AGN luminosity range shown in the labels. To display the data corresponding to objects with different redshifts on the same plot, we have rescaled the measured outflow radius according to the expected evolution of the disc radius (see text). The labels on the top axis show the time (in units of $10^{6}$ yr) corresponding to the 
shock position $R_{S}$ in the x-axis. 
}

\section{Scalings}

Finally, we focus on the scaling properties of the outflow quantities $V_{S}$ and $\dot M_{S}$. 
In fig. 7 we show the dependence of both quantities on the AGN luminosity for observed outflows  at  small ($R_{S}\leq 1$ kpc) and 
large ($R_{S}\geq 1$ kpc) distances from the galaxy center. The observational data points are compared with the model predictions for $V_{S}$ and $\dot M_{S}$ at different 
luminosities  computed at $R_{S}=0.5$ kpc (upper panels) and $R_{S}=7$ kpc (lower panels). In all cases, the assumed value for $V_c$ has been computed following the procedure described in Sect. 5, while we considered three equally spaced values for $M_{gas}$ in the range $0.3-3\cdot \,10^{10}\,M_{\odot}$ as done in Sect. 5 and in fig. 6. 
Thus, we do not perform a one-by-one comparison between the data and the model predictions since the latter are computed only at particular values of $R_{S}$ and $M_{gas}$.

The predicted $V_{S}$ scale as $V_{S}\sim L_{AGN}^{0.35}$ at small radii and as  $V_{S}\sim L_{AGN}^{0.37}$ at larger radii $R_{S}\geq 1$ kpc. The correlation is  consistent with  the best fit to the data $V_{S}\sim L_{AGN}^{0.3\pm 0.4}$ given in Fiore et al. (2017), although we stress that the model slope is computed at fixed $M_{gas}$ and $R_{S}$ 
while the observed points in the $V_{S}-L_{AGN}$ plane are characterized by different values of gas mass and shock position. 

The corresponding predicted scaling of the mass outflow rate $\dot M_{S}\sim L_{AGN}^{0.3}$ for $R_{S}=0.5$ kpc, and $\dot M_{S}\sim L_{AGN}^{0.35}$  for $R_{S}=7$ kpc. 
In this case the observed overall correlations show a sensibly stronger dependence ($\dot M_{S}\sim L_{AGN}^{0.76\pm 0.06}$  and 
$\dot M_{S}\sim L_{AGN}^{1.29\pm 0.38}$ for molecular and ionized outflows, respectively), although with some variance depending on the observational sample (see, e.g., Bischetti et al. 2019)
However, besides the scatter in $R_{S}$ and $M_{gas}$ of the observed points, the observed steep correlation for ionized winds is  largely determined by the points with small $\dot M_{S}$ at large radii $R_{S}\geq 1$ kpc. Indeed, most of the ionized outflows in the lower-right panel of fig. 7 are below the value expected by our model at low  low $L_{bol}\lesssim 10^{46}$ erg s$^{-1}$. 
This is due to the fact that at low $L_{bol}$ ionized outflows represent only a small fraction of the total mass outflow rate. 
At high $L_{bol}$ the mass outflow rates of ionized and molecular outflows are about similar, meaning that both can be used as
relatively good tracers of the  total mass outflow rate  (see Fiore et al. 2017). 

In any case, the above comparison is largely affected by the fact that the model predictions are computed at particular values of $R_{S}$ and $M_{gas}$, while the observed points 
correspond to objects with a wide range of $R_{S}$ and $M_{gas}$. Thus, the steeper logarithmic slope of the observed correlation can be either due to a true inadequacy  of the model 
in describing ionized outflows in low-luminosity objects, or to large biases affecting the  determination of the total mass outflow rate from the observation of ionized outflows in low-luminosity AGN, or to the intrinsic scatter in $R_{S}$ and $M_{gas}$ of the observational data points. 

While present data are too sparse and incomplete to allow for detailed determination of the  $\dot M_{S}- L_{AGN}$ and $V_{S}- L_{AGN}$ relations in different bins of $M_{gas}$ and $R_{S}$, a step forward to test the predicted correlations can be performed scaling the observed values of $V_{S}$ and $\dot M_{S}$  (corresponding to different outflows with different AGN luminosity and gas mass and circular velocity) to a reference value of $V_c$, $M_{gas}$ and $L_{AGN}$ using the predictions of the model. 
The run of the rescaled observed quantities with $R_{S}$ can be then compared with the model predictions for $V_{S}(R_{S})$ and $\dot M_{S}(R_{S})$ computed at the reference 
values for $V_c$, $M_{gas}$  and $L_{AGN}$.

To this aim, we first express the model predictions in a compact form. In fact, the scaling properties of our solutions with the main input quantities of the model at different shock radii $R_{S}$ can be approximated by the following  power- law relations: 
\begin{eqnarray}
log\Bigg ( {V_{S}\over {\rm km  s^{-1}}}\Bigg )=2.65+0.35\,log\Bigg (  {L_{AGN}\over 10^{45} {\rm erg s^{-1}}}\Bigg )-{1\over 3} log \Bigg({M_{gas}\over 10^{10}\,M_{\odot}}\Bigg)+log \Bigg ({V_c\over 200 {\rm km s^{-1}}}\Bigg )-0.65\,log \Bigg({R_{S}\over 1 {\rm kpc}}\Bigg)\\
log\Bigg ( \dot {M_{S}\over M_{\odot} yr^{-1}}\Bigg )=1.9+ 0.35\,log\Bigg (  {L_{AGN}\over 10^{45} {\rm erg s^{-1}}}\Bigg )+0.65\,log \Bigg({M_{gas}\over 10^{10}\,M_{\odot}}\Bigg)-1.7\,log \Bigg ({V_c\over 200 {\rm km s^{-1}}}\Bigg )+1.3\,log \Bigg({R_{S}\over 1 {\rm kpc}}\Bigg)
\end{eqnarray}
We stress that the above eqs. constitute a valid approximation (within 10 \%) only in the inner region  ($< 5$ kpc for most cases) where the scaling of the solutions with $R_{S}$ is approximatively a power-law. At larger radii,  our solutions (as described in Sect. 3) cannot be fitted with a single power-law, and  are characterized by a turn over which depends on the 
input quantities $V_c$, $M_{gas}$ and $L_{AGN}$. 

\begin{center}
\vspace{-0.1cm}\hspace{-0.5cm}
\scalebox{0.4}[0.4]{\rotatebox{-90}{\includegraphics{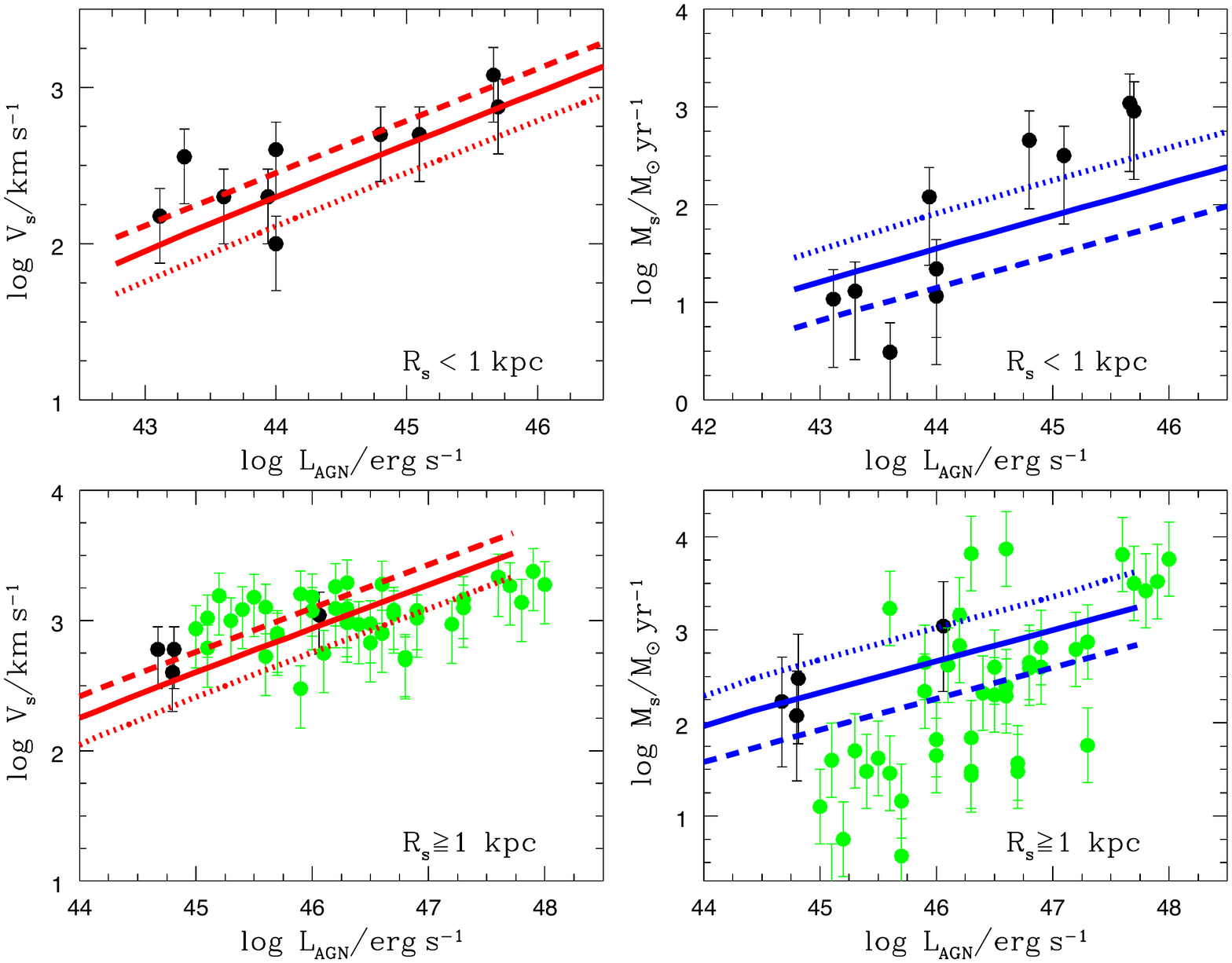}}}\hspace{1.2cm}
\vspace{0.1cm }
\end{center}
{\footnotesize Fig. 7. We show the overall scaling of the predicted shock velocity $V_{S}$ (left panels) and mass outflow rate $\dot M_{S}$ (right panel) on the AGN bolometric luminosity $L_{AGN}$. All predicted quantities are derived from the full two-dimensional model  after performing a mass-weighed average over their dependence on the inclination angle $\theta$ with respect to the plane of the disc. 
The data points have been grouped so that all data corresponding to $R_{S}\leq 1$ kpc are shown in the 
 top panels, while all the remaining points with $R_{S}> 1$ kpc are shown in the bottom panel.
The curves are computed at $R_{S}=0.5$ kpc (upper panels) and $R_{S}=7$ kpc (lower panels). Continuous lines correspond to an assumed gas mass $M_{gas}=10^{10}\,M_{\odot}$, 
 while dashed and dotted lines to $M_{gas}=3\,10^{10}\,M_{\odot}$ and $M_{gas}=0.3\, 10^{10}\,M_{\odot}$, respectively. The data points  show the observational determinations 
 summarized in Table 1 and 2 for molecular (black circles) and ionized (green circles). 
}

The above fitting formulas allow us to test 
the typical dependencies of our model on the input quantities $V_c$, $M_{gas}$ and $L_{AGN}$ against observations. We first re-scale the data 
(corresponding to different outflows with different AGN luminosity and gas mass and circular velocity) to a  reference value of $V_c$, $M_{gas}$ and $L_{AGN}$ 
through the dependencies in eqs. (14) and (15). This leaves in evidence the dependence of the data points on the shock radius $R_{S}$ which can be compared with the  predictions. 
Such a test is performed in fig. 8, where the all points (corresponding to objects in Table 1 for which measurements of $M_{gas}$, $R_{S}$ and $V_{S}$ are available) have been rescaled to the same reference value $L_{AGN}=10^{45}$ erg/s, $M_{gas}=10^{10}\,M_{\odot}$ and 
$V_c$=200 km/s after eqs. 14 and 15, and the resulting dependence on $R_{S}$ is compared with the model solutions for the reference input value of $V_c$, $M_{gas}$ and $L_{AGN}$. 

\begin{center}
\vspace{-0.2cm}\hspace{-0.5cm}
\scalebox{0.5}[0.5]{\rotatebox{-90}{\includegraphics{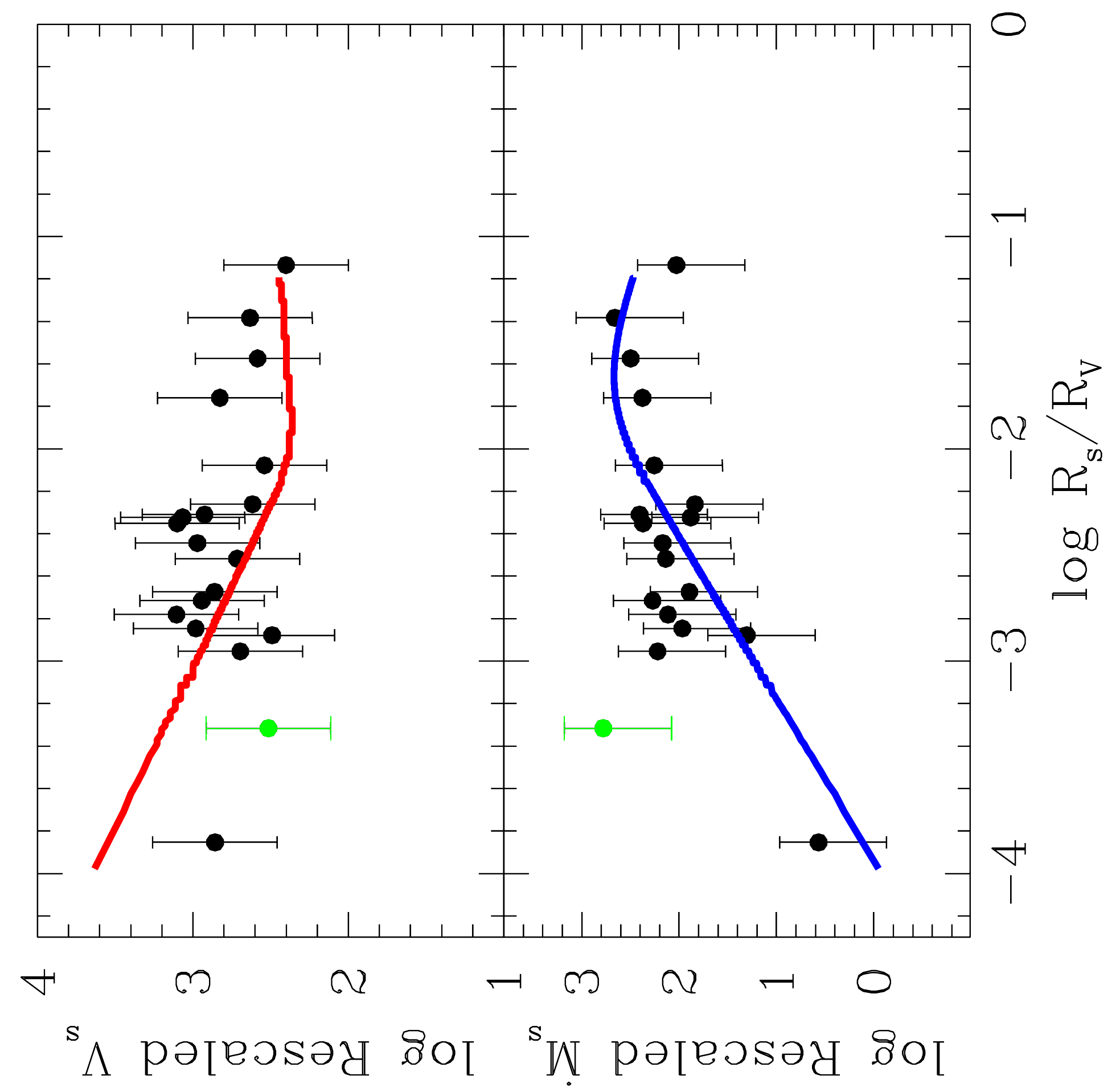}}}\hspace{1.2cm}
\vspace{-0.8cm }
\end{center}
{\footnotesize Fig. 8. We show the predicted dependence on $R_{S}$ of our model solutions $V_{S}$ and $\dot M_{S}$ for a reference case $L_{AGN}=10^{45}$ erg/s, $M_{gas}=10^{10}\,M_{\odot}$ and $V_c$=200 km/s. 
All predicted quantities are derived from the full two-dimensional model after performing a mass-weighed average over their dependence on the inclination angle $\theta$ with respect to the plane of the disc. We compare it with the data points taken from Table 1 where -for each object - the different measured values of $V_{S}$ and $M_{S}$ are rescaled for the 
different values of $L_{AGN}$, $M_{gas}$ and $V_c$ according to eqs. (13) and (14). Since Table 1 includes objects with different redshifts,  
the values of $R_{S}$ are normalized to the virial radius to account for size evolution. The green points corresponds to NGC1068 (see text).}

The correspondence of model predictions with the observed scalings of $\dot M_{S}$ is actually very good, despite the simple assumptions of the model and the large uncertainties affecting the data. Only for one object (NGC1068, green points in the figure) the model yields significant deviations from the observed 
values of $V_{S}$ and $\dot M_{S}$ (see also fig. 5). However, this galaxy is characterized by  radio jets  not aligned with the line of sight  (Gallimore et al. 1996, Crane \& Van der Hulst 1992), a property shared by Circinus and M51. Thus, in this case, the observed molecular outflow could be determined (or affected) by the interaction of the jet with the interstellar medium (not described by our model). 

While the comparison of model predictions with present data is encouraging, we stress that at present several uncertainties affect the data we are comparing with. Measured mass outflow rates depend on the adopted conversion from 
  CO luminosities into $H_2$ masses, and on the estimated size of the outflow. In particular, the conversion factor $\alpha_{CO}=0.8\,M_{\odot}$ 
 (K km s$^{-1}$ pc$^{-2}$)$^{-1}$  
  adopted in Fiore et al. (2017)  can be a function of density, metallicity and gas distribution (see the discussion in Fiore et al. 2017; see Bolatto et al. 2013, for a review), while the size of the outflow is based on the maximum radius up to which high velocity gas is detected (baseline method, but alternative methods proposed in the literature, see Carniani et al. 2015). Both these uncertainties will likely greatly be reduced by future higher resolution observations with the Atacama Large Millimiter Array (ALMA)  and Northern Extended Millimeter Array (NOEMA). 

\section{Discussion and Conclusions}

We computed the two-dimensional expansion of outflows driven by Active Galactic Nuclei (AGN) in galactic discs as a function of the global  properties of the host galaxy and of the luminosity of the central AGN. We derived the expansion rate, the mass outflow rate, and the density and temperature of the shocked shell in the case of an exponential profile for the disc gas, for different expansion directions $\theta$ with respect to the plane of the disc. 
Having expressed our model results in terms of global properties of the host galaxies, we  compared our predictions to a large sample of 19 outflows (mostly molecular, except for one object)  in galaxies with measured AGN luminosity and gas mass, and with estimated total 
mass. This allowed us to perform  a detailed, one-by-one comparison with the model predictions,  to assess to what extent the  present status of the modeling is consistent with the existing observational distribution of outflow properties.

We find - in the vast majority of cases - an encouraging agreement for a wide range of gas mass and AGN bolometric luminosity (including the hyper luminous quasars with $L_{AGN}\approx 10^{47}$ erg/s, see Vietri et al. 2018).  The model yields - for each considered galaxy and at the observed outflow radii - 
values of velocity and mass outflow rates (averaged over the directions $\theta$) that are in good agreement with observations. The predicted densities of the shocked shell are consistent with the observed molecular emission  of the outflows in the vast majority of cases. Significant deviations from the model predictions are found only  for NGC1068 (concerning both the  shock velocity $V_{S}$ and the mass outflow rate $\dot M_{S}$) and for the ic5063 and Circinus (concerning the gas density needed to produce the observed molecular outflow). 
 However, these three objects are all characterized by weak radio jets not aligned with the line of sight  (Gallimore et al. 1996, Elmouttie et al. 1998, Crane \& Van der Hulst 1992, Morganti et al. 2015)  which  could be at the origin of (or contribute to) the observed outflow, a situation  outside the reach of our model. 

We notice that some features characterizing the predicted expansion and mass outflow rates  are specific of the exponential density profile and of the two-dimensional geometry that we consider. This includes the upturn of the expansion rate and the drop of the mass outflow rate at large radii $\approx 0.1\,R_v$ (see figs. 3 and 8).  Interestingly, signatures of such a typical behavior at the expected distance from the galaxy center seem to be already present in the considered sample of measured outflows (fig. 8).

 Then we considered a larger sample of 48 outflows (mostly) ionized outflows in galaxies with no reliable measurements of the gas and dynamical mass, and we perform an approximate comparison of the model predictions for different bins of AGN luminosities assuming different  reference values for the  gas mass and dynamical mass derived from average scaling relations. Within the unavoidable uncertainties due to the derivation of the input quantities from  average relations, the model predictions are in general in agreement with the observed outflow properties, the agreement becoming excellent for the highest luminosity bins.  Also, for all objects the predicted shocked gas density  is below the value required for the CO emission, as appropriate for  ionized outflows.
Notice that, in the lowest luminosity bin, the model over-predicts the mass outflow rates measured in ionized winds. However, as discussed in Fiore et al. (2017), measured ionized mass outflows {\it in low-luminosity objects} are likely to represent only a fraction of the total mass outflow, 
 while our model predictions concern the total mass outflow rate. 
When comparing with observations, we need to assess whether the observed quantities are good tracers of the total outflow rate. The much lower ionized outflow rates with respect to molecular rates found in the past, in particular at low bolometric luminosity, suggest that only the latter are good tracers of the total outflow rate, with the former probably a good tracer only at high $L_{bol}$ (Carniani et al. 2016; Fiore et al. 2017). In particular, comparing our model predictions 
for the total mass outflow rates with\ the observed rates in ionized winds for low luminosity AGN 
$45\leq log L_{AGN}/erg\,s^{-1}\leq 45.4$  we  expects that ionized winds trace only a fraction  correction factor $\sim 0.1$ of the total mass outflow rate . Observational determinations of the fraction of mass outflow in ionized winds in low-luminosity objects will constitute an important  consistency check for the model predictions in this regime. 

While the encouraging, quantitative agreement of the model predictions with a wide set of existing observations constitutes a baseline for the interpretation of forthcoming data, and for a more detailed treatment of AGN feedback in galaxy formation models, we stress that the comparison with observations is still affected by large uncertainties related to the data. 
These  mainly affect the estimates of the mass outflow rate $\dot M_{S}$ and of the shock position $R_{S}$.  In fact, the adoption of different approaches in the measurement of the outflow velocity $V_{S}$ (velocity peak of broad emission lines vs. width of the emission lines at the 80\% of the line flux)  - while resulting in a mild error $\approx 4\%$ on $V_{S}$ itself -  produce 
 uncertainties $\sim 35\%$ in the associated estimates of $\dot M_{S}$. For molecular outflows, the latter is also affected by the uncertainties related to the conversion of  CO luminosities into $H_2$ masses. While Fiore et al. (2017) adopted a constant conversion factor $\alpha_{CO}=0.8$, this can  actually be a function of density, metallicity and gas distribution (see Bolatto et al. 2013, for a review). For ionized winds the estimates of mass of outflowing gas depend linearly on the assumed gas temperature $T$ and 
 inversely on the density $n$.  While the  data we base on are derived for $T=10^{4}$ K and $n=200$ cm$^{-3}$, uncertainties up to a factor 2 can affect both quantities (see Fiore et al. 2017 and references therein). As for the size of the outflow $R_{S}$, in most cases this is taken as the maximum radius up to which high velocity gas is detected (baseline method). On the other hand, Carniani et al. (2015), evaluate a size of the ionized wind systematically lower than all other cases, because they adopt a different astrometric procedure. However, 
 this uncertainty will likely  be reduced by future higher resolution ALMA and NOEMA observations.
  While the above uncertainties in present data do not allow to unambiguously determine the effectiveness of the model in providing a full description of the 
   expansion of outflows, the typical dependence of the mass outflow rate $\dot M_{S}$ on $R_{S}$ in fig. 9 (bottom) could be tested in detail to probe the 
   position of the upturn when a larger sample of outflows at large $R_{S}\gtrsim 5 $ kpc will be detected.  Also, testing  dependencies on the properties of the host galaxy (summarized in eqs. 13 and 14) on a solid statistical ground will require a large sample of outflows with associated measurements of the 
   galaxy properties.  The above observational goals actually characterize the SUPER (Survey for Unveiling the Physics and Effect of Radiative feedback) ongoing ESO’s VLT/SINFONI Large Programme (see Circosta et al. 2018). SUPER will perform the first systematic investigation of ionized outflows in a sizeable and blindly-selected sample of 39 X-ray AGN at $z\approx 2$, linking the outflow properties to a number of AGN and host galaxy properties. 
   The large sample of outflow in AGN with high bolometric luminosities (up to $L_{AGN}\approx 10^{47}$ erg $s^{-1}$) will enable to trace the 
   evolution of the outflows up to large  distances from the central AGN providing a larger leverage to test the specific predictions of the model at large 
 values of $R_{S}$. 
 
On the model side, 
our results are in excellent agreement with previous studies in overlapping cases. When we test our model solutions for a spherical initial gas distribution
 with a power-law decline with radius, we find results that agree with existing analytical scaling laws in the limit of energy-driven winds 
 (see, e.g., Cavaliere, Lapi, Menci 2002; King, Zubovas \& Power 2011, Faucher-Giguere \& Quataert 2012) and with numerical solutions in 
 Faucher-Giguere \& Quataert (2012). When an  exponential disc  distribution is assumed for the unperturbed galactic gas, the two-dimensional structure 
 that we obtain is similar to that obtained by Hartwick, Volonteri \& Dashyan (2018), and is 
 characterized by a shock expansion that follows the paths of  least resistance (see bottom panel of fig. 4) with an elongated shock front in the  direction perpendicular to the disc. In such a direction, the velocity field is characterized by a decline with increasing distance $Y$  from the disc, followed by a 
 strong increase for large distances $Y>h$. However, massive outflows ($\dot M_{S}=10^{2}-10^{3}\,M_{\odot}$/yr) can be generated only 
in the plane of the disc. 

In the present paper we have focused on the comparison with available observations of massive outflows, 
which do not resolve the spatial structure of the shock. Thus,  when comparing with data, we did not fully exploited the full two-dimensional description 
of our model, since the predicted quantities (outflow radius, velocity and mass rate) compared with data have been averaged over the angular direction relative to the plane of the disc.
 However, two-dimensional spectroscopic maps obtained by present and upcoming integral field units (IFU) facilities will allow more detailed comparison of the 
mass outflow rate and velocity maps with those predicted by the present model.  For example the ongoing MAGNUM (Measuring Active Galactic Nuclei Under MUSE Microscope, Venturi et al. 2018) survey  by the MUSE instrument at the Very Large Telescope (VLT),  aimed at mappping the ionized outflows from local AGN, and observations with ALMA and NOEMA, will provide a crucial sample to test the two-dimensional picture described by our model.
In the next future the NIRSPEC IFU facility at the James Webb Space Telescope (JWST) will allow us to extend any comparison to higher redshifts. As for the molecular fraction of the AGN outflows, millimeter facilities (like ALMA) will increase both sensitivity and resolution of the comparison sample.

At the same time, our two-dimensional description of outflows can be applied  to a number of different investigations. E.g.,  the model can  provide a detailed estimate  of the escape fraction of ionizing photons in active galaxies, an issue relevant for  studies on cosmic  re-ionization. 
In fact, the fast motion of the outflow in the direction perpendicular to the disc  can effectively sweep out the interstellar gas. Computing such an effect as a function of the  AGN power and lifetime will enable to  estimate the amount of HI ionizing photons escaping from the population of AGN host galaxies, and to provide a refined computation of the contribution AGN to the ionizing background, extending and improving the results derived from the analytic blast-wave model (Menci et al. 2008) adopted in Giallongo et al. (2012). Our model can also be applied to compute the acceleration of particles in the shock front of AGN outflows and the ensuing generation of gamma-rays 
and neutrinos, to compute in detail the contribution of AGNs to the extragalactic gamma-ray and neutrino backgrounds (Lamastra et al. 2017). 
Besides, the two-dimensional description developed here can contribute to provide a quantitative description of the origin of Fermi bubbles (see Su et al. 2010). As  suggested by earlier authors (see 	Zubovas, King, Nayakshin 2011, Zubovas \& Nayakshin 2012, Lacki 2014), these are connected to the 
 anti-correlation between the outflow speed and the gas density which rapidly decreases in the direction perpendicular to the disc. The consideration of the azimuthal dependence of the gas density in our two-dimensional  description will then allow for a detailed, quantitative comparison between the observed properties of the Fermi-LAT lobes and the prediction of the model: 
 we plan to address this point in a next paper.


 Finally, the model can be provide a refined description of the AGN feedback in galaxy formation models by enabling to compute  
 the ejected gas mass, i.e., the gas mass that passes the virial radius with a velocity larger than the escape velocity.  Indeed, most 
analytic and numerical cosmological models of black hole growth have included mostly "thermal-like" AGN feedback recipes (e.g., Dubois, Pichon, Devriendt et al. 2013), which are based on approximatively  isotropic injections of thermal energy into the gas surrounding the black hole. 
 Barausse, Shankar, Bernardi et al. (2017) showed that current implementations of AGN (quasar mode) feedback in comprehensive galaxy evolution models fall drastically short in reproducing the observed strong dependence of black hole mass with velocity dispersion at fixed stellar mass. A kinetic-like feedback, as the one discussed in this work, may provide a stronger coupling to velocity dispersion in view of the possibly more efficient removal of gas in lower mass systems. 
 In this case, the full two-dimensional  description of the outflows plays a relevant role, since the  larger velocities attained in the vertical direction (perpendicular to the disc)  easily exceed the escape velocity at the virial radius in a short time scale, while the slower expansion of the shock in the plane of the disc can prevent the escape of gas in this direction  within the life time of the AGN. Inspection of fig. 4 (bottom panel) shows that while in the direction perpendicular to the disc the outflows reaches a distance of 20 kpc in approximatively $10^{7}$ yrs, it takes about $10^{8}$ yrs to reach the same distance in the plane of the disc. For the AGN with life time$\sim 10^{8}$ yrs this would results into null gas expulsion along the plane of the disc. 
 The description of the shock expansion that we provide in terms of global galactic quantities allows for a fast implementation of the above description  in semi-analytic models of galaxy formation.   We plan to investigate the above  issues in  next papers. 
 
\begin{acknowledgements}
We acknowledge support from INAF under PRIN SKA/CTA FORECaST and PRIN SKA-CTA-INAF ASTRI/CTA Data Challenge. FS acknowledges partial support from the H2020 AHEAD project (grant agreement n. 654215) and a Leverhulme Trust Research Fellowship. EP acknowledges financial support from ASI and INAF under the contract 2017-14-H.0 ASI-INAF. We thank the referee for the constructive comments that significantly helped to improve the paper. 
 \end{acknowledgements}

\end{document}